\documentclass{aa}  
\usepackage{natbib}
\usepackage{bm}
\usepackage{adjustbox}
\usepackage{blindtext}
\usepackage{ulem}
\usepackage{color}
\usepackage{graphicx}
\usepackage{txfonts}
\usepackage{gensymb}
\usepackage{booktabs}
\usepackage{threeparttable}
\usepackage[colorlinks,
        linkcolor=blue,
           citecolor=blue]{hyperref}
\usepackage{varwidth}

\def\nh3{$\rm{NH_3}$}
\def\NH3{$\rm{NH_3}$}

\def\msun{\,$M_\odot$}

\def\kms{\,km/s}
\def\cm2{\,$\rm{cm^{-2}}$}
\def\cm3{\,$\rm{cm^{-3}}$}

\begin{document} 

\title{The role of turbulence in high-mass star formation: Subsonic and transonic turbulence are ubiquitously found at early stages}

   \author{Chao Wang \inst{1,2}, Ke Wang\inst{2}\fnmsep\thanks{The only corresponding author email: kwang.astro@pku.edu.cn}, Feng-Wei Xu \inst{2,1}, Patricio Sanhueza \inst{3,4},  Hauyu Baobab Liu \inst{5}, Qizhou Zhang \inst{12}, Xing Lu \inst{6}, F. Fontani \inst{8}, Paola Caselli \inst{7},  Gemma Busquet \inst{13,14,15},  Jonathan C. Tan \inst{22,23}, Di Li \inst{17,18,19}, J. M. Jackson \inst{20,21}, Thushara Pillai \inst{9}, Paul T. P. Ho\inst{10,11}, Andrés E. Guzmán \inst{16}, Nannan Yue \inst{2}   
          }

   \institute{Department of Astronomy, School of Physics, Peking University, Beijing, 100871, People's Republic of China
   \and
    Kavli Institute for Astronomy and Astrophysics, Peking University, Beijing 100871, People's Republic of China\\
    \email{kwang.astro@pku.edu.cn}
    \and
    National Astronomical Observatory of Japan, National Institutes of Natural Sciences, 2-21-1 Osawa, Mitaka, Tokyo 181-8588, Japan
    \and
    Department of Astronomical Science, SOKENDAI (The Graduate University for Advanced Studies), 2-21-1 Osawa, Mitaka, Tokyo 181-8588, Japan
    \and
    Department of Physics, National Sun Yat-Sen University, No. 70, Lien-Hai Road, Kaohsiung City 80424, Taiwan, R.O.C.
    \and
    Shanghai Astronomical Observatory, Chinese Academy of Sciences, 80 Nandan Road, Shanghai 200030, P.\ R.\ China
    \and
    Max Planck Institute for Extraterrestrial Physics, Giessenbachstr. 1, 85748 Garching, Germany
    \and
    INAF Osservatorio Astrofisico di Arcetri, Largo E. Fermi 5, 50125, Florence, Italy
    \and
    Boston University Astronomy Department, 725 Commonwealth Avenue, Boston, MA 02215, USA
    \and
    East Asian Observatory, 660 N. A'ohōkū Place, University Park, Hilo, HI 96720, US
    \and
    Institute of Astronomy and Astrophysics, Academia Sinica, 11F of Astronomy-Mathematics Building, AS/NTU No. 1, Sec. 4, Roosevelt Road, Taipei 10617, Taiwan, Republic of China
    \and
    Center for Astrophysics $|$ Harvard \& Smithsonian, 60 Garden Street, Cambridge, MA 02138, USA
    \and
    Departament de Física Quàntica i Astrofísica (FQA), Universitat de Barcelona (UB), c. Martí i Franquès, 1, 08028 Barcelona, Spain
    \and
    Institut de Ciències del Cosmos (ICCUB), Universitat de Barcelona (UB), c. Martí i Franquès, 1, 08028 Barcelona, Spain
    \and
    Institut d’Estudis Espacials de Catalunya (IEEC), c. Gran Capità, 2-4, 08034 Barcelona, Spain
    \and
    National Astronomical Observatory of Japan, National Institutes of Natural Sciences, 2-21-1 Osawa, Mitaka, Tokyo 181-8588, Japan
    \and
    National Astronomical Observatories, Chinese Academy of Sciences, Beijing 100101, China
    \and
    Computational Astronomy Group, Zhejiang Lab, Hangzhou, Zhejiang 311121, China
    \and
    University of Chinese Academy of Sciences, Beijing 100049, China
    \and
    SOFIA Science Center, NASA Ames Research Center, Moffett Field, CA 94 045, USA
    \and
    Green Bank Observatory, PO Box 2, Green Bank, WV 24 944, USA
    \and
    Department of Astronomy, University of Virginia, Charlottesville, VA 22904-4235, USA
    \and
    Department of Space, Earth \& Environment, Chalmers University of Technology, SE-412 96 Gothenburg, Sweden
         }

   \date{Received xx xx, 2023; accepted xx xx, 2023}

 
  \abstract
{Traditionally, supersonic turbulence is considered to be one of the most likely mechanisms slowing the gravitational collapse in dense clumps, thereby enabling the formation of massive stars. However, several recent studies have raised differing points of view based on observations carried out with sufficiently high spatial and spectral resolution.
These studies call for a re-evaluation of the role turbulence plays in massive star-forming regions.}   
{Our aim is to study the gas properties, especially the turbulence, in a sample of massive star-forming regions with sufficient spatial and spectral resolution, which can both resolve the core fragmentation and the thermal line width.}
{We  observed NH$_3$ metastable lines with the Very Large Array (VLA)
to assess the intrinsic turbulence.
}  
{
Analysis of the turbulence distribution histogram for 32 identified \nh3\ cores reveals the presence of three distinct components. Furthermore, our results suggest that (1) sub- and transonic turbulence is a prevalent (21 of 32) feature of massive star-forming regions and those cold regions are at early evolutionary stage. This investigation indicates that turbulence alone is insufficient to provide the necessary internal pressure required for massive star formation, necessitating further exploration of alternative candidates;
and (2) studies of seven multi-core systems indicate that the cores within each system mainly share similar gas properties and masses. However, two of the systems are characterized by the presence of exceptionally cold and dense cores that are situated at the spatial center of each system. Our findings support the hub-filament model as an explanation for this observed distribution. }
   {}

  \titlerunning{Subsonic turbulence is ubiquitously found in high-mass star formation}
  \authorrunning{Wang et al.}

   \maketitle
%

\section{Introduction}

Massive stars ($M_*$ $>8$\msun) play a major role in the energy budget of galaxies via their radiation, winds, and supernova events, yet the picture of their formation remains unclear. Specifically, assuming a conversion efficiency from the core to the star is at 50\%, a massive star of 10\,M$_\odot$ would require a core of at least 20\,M$_\odot$ \citep{2021ApJ...912..156K}. However, under typical conditions (e.g., sound speed of $c_s=0.2$\,\kms\ and gas number density $n_{\mathrm{H}}=10^5$\,cm$^{-3}$), the Jeans mass is less than 1\,M$_\odot$ \citep[e.g.][]{2021ApJ...912..156K,Xu2023SDC335}. Therefore, it is unclear how a core with more than 100 Jeans masses would survive fragmentation, rather than giving rise to hundreds of low-mass cores.
To address this, \cite{2003ApJ...585..850M} proposed the turbulent core accretion model (hereafter, TCA), where highly turbulent gas provides additional support against gravitational collapse. As a consequence, the equivalent ``turbulent Jeans mass'' enables the formation of massive stars \citep[e.g.,][]{2011ApJ...735...64W,2014MNRAS.439.3275W}. This theoretical model has been supported by following observations \citep[e.g.,][]{2012ApJ...745...61L, 2013ApJ...770...44L, 2013MNRAS.432.3288S, 2013ApJ...779...96T,2018ApJ...867...94K}. However, several studies have revealed thermal fragmented cores, which are consistent with the Jeans length in high-mass star-forming regions \citep[e.g.][]{2005AJ....129.2281B,2018A&A...617A.100B}. Similarly, 
\cite{2015MNRAS.453.3785P} and \cite{2019ApJ...871..185L} found that fragmentation in massive protostellar cores is dominated by thermal Jeans fragmentation rather than turbulence levels, which is further proven in \cite{2019ApJ...886..102S,Morii2023ASHES}. This fact aligns with the competitive accretion model \citep{2001MNRAS.323..785B,2004MNRAS.349..735B}. This model posits that molecular clouds thermally collapse into Jeans cores. Then, cores that are located at the center of the gravitational potential can grow into high-mass stars through the gas inflow. In this model, supersonic turbulence is not needed.

Since dense cores with such high mass are typically located at a few kiloparsecs, interferometers are necessary to reveal their gas properties in detail. Recent studies have used NH$_3$ to study samples of the massive star-forming regions and discovered that supersonic turbulence is universal in regions of massive star formation \citep{2013MNRAS.432.3288S,Lu_2014, Lu_2018}. However, it is important to note that \citet{2014MNRAS.440.2860H, 2017ApJ...841...97S, Lu_2018, 2018A&A...611L...3S, 2022ApJ...936..169R} have found that the turbulence became transonic on the scale of cores (at 0.1\,pc) with NH$_3$ as well N$_2$H$^+$. 

The change of the supersonic to transonic turbulence from those observations suggests that the former consequence may have resulted from limited resolution. With sufficient resolving power, the intrinsic turbulence may be revealed as transonic or even subsonic. To achieve this sufficient resolution, we require: (1) high spectral resolution capable of resolving thermal line widths (typically 0.2\kms\ at 15\,K); (2) high spatial resolution that can resolve dense cores (0.1\,pc or $4''$ at 5\,kpc); and (3) high mass sensitivity that can resolve thermal Jeans mass (typically 1\,$M_\odot$).

Previous VLA observations in NH$_3$ \citep[e.g.,][]{Lu_2014} used a correlator setup with a channel width of 0.6-0.7\,\kms, which was much larger than the thermal width, making it impossible to detect the potential subsonic turbulence. In contrast, \cite{WangKe2012} combined VLA-C configuration \nh3\ image cubes at 0.2\kms\ resolution with VLA-D configuration data at 0.6 \kms\ resolution to study clump P1 in the infrared dark cloud (IRDC) G28.34+0.06. They found that all dense cores coincided with a reduced line width compared to the general clump, indicating dissipation of turbulence from the clump to the core scales \citep[cf.][]{2008ApJ...672L..33W}.  

In addition, \citet{2018ApJ...861...77M} have found the existence of the subsonic turbulence with the VLA observation under about  0.3\kms\ in the Orion molecular cloud. Later, \cite{2021RAA....21...24Y} have found a trend of turbulence dissipation, where the turbulence changed from transonic to subsonic as the spatial resolution increased from $10^4$ au to $10^3$ au. In the study of IRDC G035.39-00.33, \cite{2018A&A...611L...3S,Sokolov2017} found that more than a third of all the pixels, which coincide with star-forming dense cores, show subsonic non-thermal motions.
Similarly, \cite{2020ApJ...896..110L, 2022ApJ...926..165L} have found the turbulence of the filaments and cores in NGC6334S are sub- or transonic. They credited this result to the high spatial and spectral resolutions (0.02 pc and 0.2\kms) of the observations by VLA and ALMA.  

These studies all suggest that the turbulence is resolution-dependent, highlighting the possibility of finding an intrinsic subsonic turbulence. To reveal the turbulence properties in massive star-forming regions, observations with sufficient spatial and spectral resolution are needed. In addition, the targets should be at the early evolutionary stage (e.g., IRDCs) so that the intrinsic turbulence can avoid being affected by star-formation activities. As for evolved sources at the same resolution, our pilot work in the case of G35.20 \citep{2023A&A...674A..46W} has shown that Mach number decreases from $\sim 6$ on the scale of 0.1\,pc to $\sim 2$ towards the scale of 0.01\,pc. If the intrinsic turbulence is commonly found to be subsonic in massive star-forming regions, the role of the turbulence  should be re-evaluated. In TCA, other internal pressure candidates such as magnetic fields could play a more important role in high-mass stars, which depends both on the scale and the evolutionary stage under the overestimated velocity dispersion \citep[e.g.,][]{2013ApJ...779...96T,2018A&A...615A..94F,2021ApJ...915L..10S}. Otherwise, other theories may explain the formation of high-mass stars better. For example, the competitive accretion model or the global hierarchical collapse model \citep[GHC hereafter;][]{2019MNRAS.490.3061V}. GHC has similar performance in smaller cores with a few solar masses which can replace the role of the magnetic field. Besides, \cite{2021MNRAS.504.1219P} and \cite{2020ApJ...900...82P} also proposed different models that better explain the observed phenomena.

To investigate the turbulence properties, we selected 13 massive star-forming regions based on the following criteria: (1) the distance of the selected regions should be less than about 5 kpc, which allows the core identification at about 0.05 pc; (2) the hydrogen column density of the selected regions should be higher than $10^{22}$ cm$^{-2}$ and the mass should be larger than ${5} \times {10}^3 M_{\odot}$, allowing the formation of the massive stars (similar as the criteria in \cite{2019ApJ...886..102S}); and (3) the evolutionary stage of the regions should be earlier than the evolved IRDC (e.g., G35.20), which avoids the influence of the feedback from stellar activities. We selected 13 regions mainly from the APEX telescope large area survey of the galaxy (ATLASGAL; \citealt{2009A&A...504..415S}). We also add some typical IRDCs out of the range of ATLASGAL as supplementary. Besides, we chose several well-studied  filaments from the Galactic Cold Filaments \citep{2015MNRAS.450.4043W} to study the potential effect of the spatial distribution to the star-forming regions. The basic parameters of the selected sources are presented in Table \ref{tab:sample} of the corresponding cited papers. Figure \ref{ob} shows the infrared environment of the sample with the \textit{Spitzer} \citep{2021yCat.2368....0S} infrared three-color images (blue: 3.6\,$\mu$m; green: 4.5\,$\mu$m; red: 8\,$\mu$m) as the background and the white contours from ATLASGAL \citep{2009A&A...504..415S} 870\,$\mu$m continuum emission.

In this work, we present the results of VLA observations towards 13 massive star-forming regions and extract 32 NH$_3$ cores in total. Based on gas properties of those cores, we study their evolution and dynamics. 
The paper is structured as follows. 
In Section \ref{observe} we present our observations and the data reduction.
Section \ref{analysis} introduces the identification, the fitting and the method.
Section \ref{results} studies the parameters of the cores fitted from the NH$_3$ lines and the implications of the turbulence of star formation based on the Mach number.
Section \ref{diss} discusses the separation and the selection effect.
In Section \ref{sum}, we summarize the main conclusions.

\section{Observations}\label{observe}


\subsection{Sample description}

\begin{figure*}[htb!]
    \centering
    \includegraphics[width=0.75\textwidth]{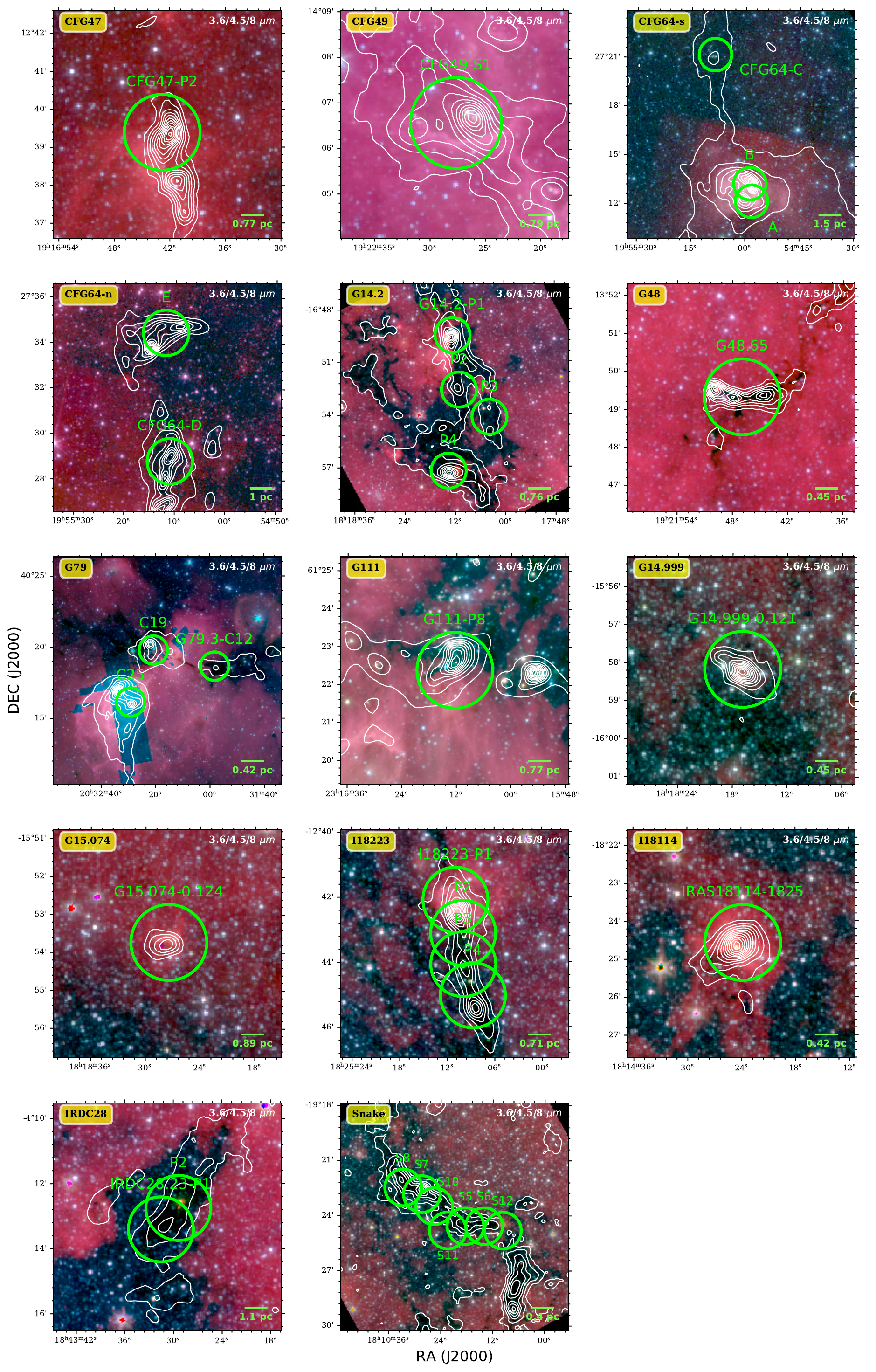}
    \caption{Infrared environment of the sample. The background shows the \textit{ Spitzer} \citep{2021yCat.2368....0S} infrared three-color images (blue: 3.6\,$\mu$m; green: 4.5\,$\mu$m; red: 8\,$\mu$m). White contours are ATLASGAL \citep{2009A&A...504..415S} 870\,$\mu$m continuum emission, with levels starting from $5\sigma$ increasing in steps of $f(n)=3\times n^p + 2,$ where $n=1,2,3,...N$. If no ATLASGAL data, then \textit{Herschel} 250\,$\mu$m continuum emission is used. The green circles with labels mark the VLA field of view (FoV) of $\simeq2$\,arcmin.}
    \label{ob}
\end{figure*}

As shown in Table \ref{tab:sample}, all clumps which are selected from Galactic Cold Filaments, ATLASGAL \citep{2009A&A...504..415S} and other typical IRDCs are massive enough with the order of magnitude at $10^4 M_\odot$. As seen from the three-color images of the regions presented in Figure \ref{ob}, based on previous observations by \textit{Spitzer} \citep{2021yCat.2368....0S} \footnote{https://irsa.ipac.caltech.edu/irsaviewer/}, most clumps are IR-dark with a few IR-bright regions. Within the sample, G11.11, CFG47, CFG49, CFG64, and IRDC28.23 \citep{2012ApJ...756...60S,2013ApJ...773..123S} are representative filaments \citep{2015MNRAS.450.4043W}. G11.11 (also named ``Snake'') is a well-studied S-shaped IRDC \citep{2014MNRAS.439.3275W} with several ongoing star-forming regions (e.g., masers, outflows) and CFG49 is near an HII region \citep{1958BAN....14..215W}. G14.99, G15.07 and G15.19 are compact sources from ATLASGAL \citep{2013A&A...549A..45C}. But G15.19 is presented as a supplementary in appendix due to its much larger distance compared to other sources. The last two have weak IR emission but G14.99 has a maser and a possible young stellar object (YSO) \citep{2012A&A...544A.146W}. G14.2 is another cloud with filamentary structure. Its cold dense cores \citep{2016ApJ...819..139B,2016ApJ...833..209O} and the strong magnetic field \citep{2013ApJ...764L..26B, 2020A&A...644A..52A} may indicate the existence of a possible subsonic turbulence. Other sources in the sample have at least one identified YSO. G48.65 is a cold IRDC with several YSOs in very early stages \citep{2006ApJ...653.1325S}. G79.3 is an IRDC in Cygnus-X with at least five YSOs and still forms protostars \citep{2014A&A...564A..21R}.  G111-P8 is an active star-forming clump with the maser \citep{2019MNRAS.485.2895E}. IRAS18114 has strong outflows and a Class I YSO \citep{2012A&A...540A..95Y}. YSOs in I18223 may be the result of the cloud-cloud collision \citep{2014A&A...565A.101T}.

As presented in Table \ref{tab:sample}, the distance range of the clumps in the selected sample is from 0.9 to 5.4\,kpc, which allows us to investigate the role of turbulence on various spatial scales. Also, the different locations in the Milky Way help to confirm the generality of the conclusion. According to the TCA model proposed by \cite{2003ApJ...585..850M}, the strength of turbulence may be affected by the evolutionary stage of the source. Therefore, the selected sample includes both infrared (IR) dark and bright clumps to ensure that those clumps are at different evolutionary stages. For the filamentary clumps, we adopt the value of the size of the main-axis measured in \cite{2015MNRAS.450.4043W}. Other clumps sizes are roughly measured from the dust maps of the ATLASGAL.

\begin{table*}[htp]
    \centering
    \linespread{2.}
    \caption{Massive star-forming regions observed in NH3 with the VLA.}
    \label{tab:sample}
    \begin{tabular}{llllllllll}
         \toprule
         Name & \multicolumn{2}{c}{Equatorial (J2000)} & \multicolumn{2}{c}{Galactic} & D & Mass & size & IR  & alter-name \\
         \cmidrule(r){2-3} \cmidrule(r){4-5}
          & R.A. & DEC & $l$ & $b$ & (kpc) & ($10^4 M_{\odot}$) & (pc)  & & \\
         \hline
         G14.2&18h18m10.0s&$-$16d53m24.5s&14.17\degree&-0.53\degree&1.7&4.7&29.5&dark&G14.225-0.506\\   
         I18223&18h25m10.1s&$-$12d43m48.6s&18.63\degree&-0.07\degree&3.5&2.2&13.2&dark&IRAS18223-1243\\
         G14.99&18h18m17.9s&$-$15d58m12.5s&14.99\degree&-0.12\degree&2.6&0.7&10.9&dark&G14.999-0.121\\   
         G48.65&19h21m46.1s&+13d49m12.6s&48.65\degree&-0.29\degree&2.6&0.6&18.3&dark&MSXDCG048.65-00.29\\ 
         G111-P8&23h16m12.4s&+61d22m12.2s&111.78\degree&0.58\degree&4.4&-&-&dark&G111.78+00.58\\ 
         G11.11&18h10m34.0s&$-$19d24m00.1s&11.11\degree&-0.10\degree&3.5&1.1&37$\times$0.6&dark&G11.11-0.12, Snake\\
         CFG64&19h55m05.5s&+27d21m00.1s&64.26\degree&-0.41\degree&3.6&0.5&51$\times$2.7&dark&CFG064.27-0.42\\         
         IRDC28.23&18h43m12.7s&$-$04d13m12.4s&28.23\degree&-0.17\degree&4.9&5.5&60$\times$1.6&dark&-\\  
         G79.3&20h32m14.6s&+40d18m00.3s&79.30\degree&0.34\degree&1.4&1.7&22.5&dark&G79.3+0.3\\
         IRAS18114&18h14m22.3s&$-$18d24m36.7s&12.40\degree&-0.46\degree&2.4&0.5&9.1&bright&IRAS18114-1825\\
         G15.07&18h18m29.4s&$-$15d54m00.0s&15.07\degree&-0.12\degree&5.1&0.6&8.7&bright&G15.074-0.124\\         
         
         G15.19&18h18m48.2s&$-$15d48m36.0s&15.18\degree&-0.16\degree&11.6&0.7&16.6&bright&G15.185-0.158\\ 
         CFG47&19h16m43.7s&+12d39m36.3s&47.06\degree&0.25\degree&4.4&2.0&78$\times$3.0&bright&CFG047.06+0.26\\   
         CFG49&19h22m29.2s&+14d06m36.8s&49.00\degree&-0.31\degree&5.4&8.3&85$\times$1.6&bright&CFG049.21-0.34\\

         \hline
    \end{tabular}
    \begin{tablenotes}
    \footnotesize
    \item VLA Observation of sources in F
    Figure \ref{ob} with their basic parameters from other studies. From the left to the right are R.A. (J2000), Decl. (J2000), $l$, $b$, distance (kpc), mass ($10^4 M_{\odot}$), size (pc), IR dark/bright, alter-name. For 5 filamentray IRDCs, we adopt their sizes from \cite{2015MNRAS.450.4043W}, others' sizes are measured from the dust maps of the ATLASGAL. References are: IRAS18114 \citep{2012A&A...540A..95Y}, G14.2 \citep{2011ApJ...733...25X}, I18223 \citep{2014A&A...565A.101T}, G14.99, G15.07 and G15.19
    \citep{2013A&A...549A..45C}, G79.3 \citep{2014A&A...564A..21R}, G48.65 \citep{2006ApJ...653.1325S}, G111-P8 \citep{EB2015BGPSdistance,EB2015BGPSPhy}, IRDC28.23 \citep{2012ApJ...756...60S,2013ApJ...773..123S}, other sources such G11.11, CFG47, CFG49, and CFG64 are from \cite{2015MNRAS.450.4043W}. Note that G15.19 has a much larger distance than other sources, so it is excluded from analysis in the main text. We present its fitting results in the appendix.
    \end{tablenotes}
\end{table*}

\subsection{VLA Observations}

All the observations were executed with the VLA in its D/DnC configuration at K band from 2013 to 2014 (project ID: 13A-373, PI: Ke Wang; 14A-272, PI: Patricio Sanhueza). The settings of the observations are listed in Table \ref{tab:set}. We specially configured the correlator to cover the frequency range of 18-26.5~GHz, containing several typical lines which trace the dense gas (e.g., \nh3 inverse transition lines from (1,1) to (7,7), details can be found in Table \ref{tab:ob}). The spectral resolution of those observations is 15.625 kHz (about 0.2\kms), which is similar to \cite{2018A&A...611L...3S, 2020ApJ...896..110L,2023A&A...674A..46W}. 
The largest recoverable angular scale in D configuration at this frequency is approximately $60''$ in $\sim 2'$ primary beam. Due to a incomplete u-v coverage in a snapshot observation, the largest recoverable angular scale can be even smaller \footnote{https://science.nrao.edu/facilities/vla/docs/manuals/oss/performance/resolution}. Thus the line width may be underestimation because of the missing flux and leads to a higher fraction of the weak turbulence \citep{2013ApJ...768L...5L,2021RAA....21...24Y}. Even though, since the resolved \nh3 core size in this work are much smaller than the largest recoverable angular scale, the measured \nh3 line width should not be severely affected by the more diffuse velocity components from larger scale (typically 1 pc). 
Besides, similar studies lacking single-dish data also estimated line width biases through simulations, finding that the effect is not significant enough to alter conclusions, for instance, \cite{2021RAA....21...24Y}(about 20\%) and \cite{2023ApJ...949..109L}(3\%-10\%).  
The maximum reduction of line width found in those studies is about 20\%, leading to a change in the Mach number by 13\% in our results, which is not enough to significantly alter conclusions.

\begin{table*}[htp]
    \centering
    \caption{Observation setting of sources}
    \label{tab:set}
    \begin{tabular}{llllllll}
         \toprule
         Name&N$_p$&date&gain&bandpass&flux&beam size&rms (mJy beam$^{-1}$)\\
         \hline
         IRAS18114&1&2013 May&J1733-1304&3C454.3&3C48&$2.6''\times4.0''$&3.6\\
         G14.2&4&2013 May&J1832-1035&3C454.3&3C48&$2.9''\times4.1''$&3.8\\   
         I18223&4&2013 May&J1832-1035&3C454.3&3C48&$2.9''\times4.3''$&4.1\\
         G14.99&1&2013 May&J1832-1035&3C454.3&3C48&$3.0''\times4.4''$&6.2\\         
         G15.07&1&2013 May&J1832-1035&3C454.3&3C48&$2.8''\times4.1''$&6.3\\         
         G15.19&1&2013 May&J1832-1035&3C454.3&3C48&$2.9''\times4.2''$&4.0\\ 
         G79.3&3&2013 May&J2015+3710&3C454.3&3C48&$2.5''\times3.9''$&2.0\\
         G48.65&1&2013 Apr&J1924+1540&J1925+2106&3C48&$2.6''\times4.1''$&1.9\\ 
         G111-P8&1&2013 Apr&J0014+6117&3C454.3&3C147&$2.8''\times4.1''$&2.2\\ 
         G11.11&7&2014 Sep&nrao530&3C454.3&3C48&$2.9''\times4.4''$&5.3\\
         CFG47&1&2014 Sep&J1925+2106&3C454.3&3C48&$3.0''\times4.5''$&9.7\\   
         CFG49&1&2014 Sep&J1925+2106&3C454.3&3C48&$2.9''\times4.3''$&5.6\\         
         CFG64&5&2014 Sep&J2023+3153&3C454.3&3C48&$2.9''\times4.4''$&8.7\\         
         IRDC28.23&2&2014 Sep&J1851+0035&3C454.3&3C48&$2.5''\times3.9''$&7.2\\           
         \hline
    \end{tabular}
    \begin{tablenotes}
    \footnotesize
    \item VLA Observation settings of sources in Figure \ref{ob}. They are listed as the number of the pointing observation (N$_p$), the observation date, the standard calibrators of the gain, bandpass and flux, the beam size in NH$_3$ (1,1), and the background noise in mJy beam$^{-1}$.
    \end{tablenotes}
\end{table*}

\subsection{Data reduction}

The VLA antenna baseline and the atmospheric opacity have first been corrected using Common Astronomy Software Applications (CASA) 4.7.2 \citep{2007ASPC..376..127M}. Then the standard calibration solutions of the bandpass, flux, and gain are applied onto the corrected raw data. Those calibrators can be found in Table \ref{tab:set}. The systematic uncertainty from the flux calibration is around 10\%, consistent with similar observations \citep[e.g.,][]{Sokolov2017}.

As most of the clumps in the selected sample are at the early evolutionary stage, the signal of the emission lines can be weak. Considering the reliability of the data, we applied two different settings of parameters in \textsc{tclean}, a CASA task which uses the multi-scale CLEAN algorithm \citep{2008ISTSP...2..793C} in the u-v space. The specific parameter settings used in multi-scale CLEAN are  1, 3, 9, and 27 pixels size to simultaneously recover both point sources and extended structures. In the primary setting, we use the natural weighting to optimize for searching the low signal-to-noise ratio (S/N) lines. Observed results are listed in Table \ref{tab:ob}. The aim of this setting is to detect the weakest signal and generate a complete detection result. Then we applied the second setting onto the clumps that contain the detected signal (S/N>3) from the primary setting, with a robust parameter of 0.5 under the Briggs weighting. Some of the clumps (e.g., G15.07, CFG64-B) that are detected in the first setting with the peak-S/N less than 5 are not detected in the second setting. 

We adopt the result which is based on the second setting as the final input data for the analysis. For example, after second \textsc{tclean} task, the mean synthesized beam of most detected data cubes is $2.8''\times4.2''$ with the mean position angle at 65$^ \circ $. Different observation configurations resulted in a slightly different angular resolution and the position angle, but the range of this difference is less than 20\% of the corresponding mean value. The physical resolution is determined both by the distance and the angular resolution. As presented in Table \ref{tab:sample}, the range of the distance is from less than 1 kpc to more than 5 kpc, the physical resolution also covers large range which may lead to a biased conclusion because of the selection effect. However, this effect is not seen in this study, as we discuss in Section \ref{results} and Section \ref{diss}. Due to the different integration times and the weather, the resultant rms noise range is from about 4 mJy beam$^{-1}$ to 10mJy beam$^{-1}$ with the mean noise value at 4.2~mJy $\rm{beam}^{-1}$. The following fitting is done on second \textsc{tclean} results, and each data cube has been smoothed to the same synthesized beam shape to NH$_3$\,(1,1). 

In Table \ref{tab:ob}, we present the detection result of each lines in this observation. Most clumps are detected in NH$_3$ (1,1) and (2,2) lines, and 10 of 32 are detected in NH$_3$ (3,3), which is helpful to revealing the ortho to para ratio (OPR). The high-excitation transitions of NH$_3$ of (4,4) or higher are also undetected.

Among those clumps, CGF49\_s1 has been detected in NH$_2$D , with a few pixels. CGF49\_s1 is optically dark but IR-bright at 4.5 $\mu$m. As the fitted gas properties in Table \ref{tab:re}, CGF49\_s1 is over 30 K with the lowest column density of this sample. This special gas properties may be resulted by the nearby HII region \citep{1958BAN....14..215W}. 

G111\_P8 and G14.99 have been detected in H$_2$O masers, a well known signpost of star formation \citep[e.g.][]{2011MNRAS.418.1689U}. Those two clumps harbor IR bright sources. But during the fitting process, we found that the noise of G111\_P8 is relatively high because of the bad weather conditions during observations. Only few pixels (less than 5) can be fitted to derive the gas properties, so we refrain from analyzing this source. The SNR of the H$_2$O maser (locates at RA=274.571\degree, Dec=-15.969\degree with a peak flux 28.1 mJy beam$^{-1}$ in Table \ref{tab:ob}) in G14.99 is much higher than G111\_P8 and the fitted result shows that G14.99 has a typical hot and dense core with supersonic turbulence. This detection result of G14.99 is consistent with the maser line and in an earlier study \citep{2012A&A...544A.146W}. The lack of other H$_2$O maser sources \citep{2012A&A...544A.146W} in G14.99 may due to the S/N or the variability of the H$_2$O maser \citep{2009ARep...53..420L,2019ARep...63.1022A}.

In this observation, G14.2\_P1 has a methanol maser (located at RA=274.552\degree, Dec=-16.825\degree, with a peak flux 105.3 mJy beam$^{-1}$ in Table \ref{tab:ob}), another tracer of the massive star formation. Although G14.2\_P1 is IR-dark at 8 $\mu$m, the methanol maser indicates that the protostar is formed in the center of the G14.2\_P1. This observation have detected 2 H$_2$O maser lines, 1 CH$_3$OH maser line, and 1 NH$_2$D line among 32 selected clumps. Considering the detection result, the following work is mainly based on NH$_3$ lines from (1,1) to (3,3). 

\begin{table*}[htp]
    \centering
    \setlength{\tabcolsep}{3pt}
    \caption{Observation results}
    \label{tab:ob}
    \begin{tabular}{lrrrrrrrrrr}
         \toprule
        Name&R.A.&DEC&NH$_3(1,1)$&NH$_3(2,2)$&NH$_3(3,3)$&NH$_3(4,4)$&NH$_3(5,5)$&NH$_3(6,6)$&NH$_3(7,7)$&NH$_2$D\\
         \hline
         G48.65&290.446\degree&13.822\degree&$\bigcirc$&$\bigcirc$&$\bigcirc$&1.3&1.1&1.6&$\times$&1.5\\
         G79.3-C12&307.993\degree&40.311\degree&$\bigcirc$&$\bigcirc$&1.9&1.9&2.1&1.7&$\times$&1.8\\
         G79.3-C19&308.087\degree&40.330\degree&$\bigcirc$&$\bigcirc$&2.0&2.1&2.2&1.9&$\times$&1.8\\
         G79.3-C25&308.122\degree&40.269\degree&$\bigcirc$&$\bigcirc$&$\bigcirc$&2.8&3.1&3.4&$\times$&2.5\\
         I18223-P1&276.296\degree&-12.702\degree&$\bigcirc$&$\bigcirc$&$\bigcirc$&3.8&3.7&3.7&3.4&3.9\\
         I18223-P2&276.292\degree&-12.719\degree&$\bigcirc$&$\bigcirc$&$\bigcirc$&3.6&3.8&3.9&3.6&3.3\\
         I18223-P3&276.292\degree&-12.735\degree&$\bigcirc$&$\bigcirc$&3.9&3.8&3.7&4.1&4.0&3.6\\
         I18223-P4&276.287\degree&-12.751\degree&$\bigcirc$&$\bigcirc$&$\bigcirc$&3.2&3.6&3.7&3.2&3.5\\
         IRAS18114&273.600\degree&-18.410\degree&$\bigcirc$&$\bigcirc$&3.2&3.5&4.6&3.1&2.8&2.7\\
         CFG47-P2&289.179\degree&12.656\degree&9.3&8.3&6.4&6.7&5.6&5.1&5.9&5.1\\
         CFG49-S1&290.616\degree&14.109\degree&$\bigcirc$&$\bigcirc$&$\bigcirc$&4.9&6.3&5.8&5.2&$\bigcirc$\\
         G111-P8&349.052\degree&61.373\degree&$\bigcirc$&$\bigcirc$&2.1&1.6&2.5&$\times$&$\times$&$\times$\\
         G14.2-P1&274.552\degree&-16.825\degree&$\bigcirc$&$\bigcirc$&3.9&3.7&3.2&3.4&3.4&3.7\\
         G14.2-P2&274.546\degree&-16.877\degree&$\bigcirc$&$\bigcirc$&3.6&3.5&3.2&3.5&3.7&3.6\\
         G14.2-P3&274.516\degree&-16.903\degree&$\bigcirc$&$\bigcirc$&3.5&3.6&3.4&3.6&3.5&3.4\\
         G14.2-P4&274.557\degree&-16.954\degree&$\bigcirc$&$\bigcirc$&$\bigcirc$&3.5&3.5&3.1&3.5&3.4\\
         G14.99&274.571\degree&-15.970\degree&$\bigcirc$&$\bigcirc$&7.4&4.3&5.6&$\times$&$\times$&$\times$\\
         G15.07&274.615\degree&-15.896\degree&$\bigcirc$&6.3&5.7&5.9&4.4&$\times$&$\times$&$\times$\\
         CFG64-A&298.743\degree&27.203\degree&6.8&6.4&4.5&4.6&4.8&3.7&3.7&4.5\\
         CFG64-B&298.745\degree&27.221\degree&$\bigcirc$&$\bigcirc$&5.5&5.5&6.1&3.1&3.5&4.9\\
         CFG64-C&298.786\degree&27.353\degree&$\bigcirc$&$\bigcirc$&5.2&5.2&4.9&3.1&3.2&4.6\\
         CFG64-D&298.796\degree&27.480\degree&11.3&11.6&7.3&5.5&4.3&4.2&4.1&5.2\\
         CFG64-E&298.800\degree&27.574\degree&12.4&15.2&8.3&5.3&4.9&4.3&4.7&5.4\\
         IRDC28.23-P1&280.882\degree&-4.224\degree&$\bigcirc$&$\bigcirc$&6.6&6.6&7.2&$\times$&$\times$&$\times$\\
         IRDC28.23-P2&280.873\degree&-4.213\degree&$\bigcirc$&$\bigcirc$&7.9&7.3&8.1&$\times$&$\times$&$\times$\\
         G11.11\_s5&272.577\degree&-19.410\degree&$\bigcirc$&$\bigcirc$&$\bigcirc$&3.9&4.5&$\times$&$\times$&$\times$\\
         G11.11\_s6&272.559\degree&-19.411\degree&$\bigcirc$&$\bigcirc$&4.9&4.7&5.4&$\times$&$\times$&$\times$\\
         G11.11\_s7&272.619\degree&-19.381\degree&$\bigcirc$&$\bigcirc$&$\bigcirc$&4.6&4.9&$\times$&$\times$&$\times$\\
         G11.11\_s8&272.637\degree&-19.375\degree&$\bigcirc$&$\bigcirc$&$\bigcirc$&4.8&4.7&$\times$&$\times$&$\times$\\
         G11.11\_s10&272.607\degree&-19.392\degree&$\bigcirc$&$\bigcirc$&5.9&4.8&5.3&$\times$&$\times$&$\times$\\
         G11.11\_s11&272.594\degree&-19.402\degree&$\bigcirc$&$\bigcirc$&5.7&4.2&4.8&$\times$&$\times$&$\times$\\
         G11.11\_s12&272.541\degree&-19.414\degree&$\bigcirc$&$\bigcirc$&5.6&4.8&4.9&$\times$&$\times$&$\times$\\
         \hline
    \end{tabular}
    \begin{tablenotes}
    \footnotesize
    \item Observation results of each line in 32 VLA pointings. $\bigcirc$ means that the flux peak of the emission line is larger than 3$\sigma$ (detected); $\times$ means that the range of the correlator setting not contain this line. For the undetected line, we mark its the upper limit at the 1$\sigma$ in mJy~beam$^{-1}$. The correlator is configured to cover the frequency range of 18-26.5~GHz, and the spectral resolution is 15.625 kHz (about 0.2\kms in this band). The rest frequencies of those lines are (in GHz): NH$_3(1,1)$ in 23.694, NH$_3(2,2)$ in 23.723, NH$_3(3,3)$ in 23.870, NH$_3(4,4)$ in 24.139, NH$_3(5,5)$ in 24.533, NH$_3(6,6)$ in 25.056, NH$_3(7,7)$ in 25.715, NH$_2$D$(4,4)$ in 25.024, NHD$_2$$(6,7)$ in 23.023, CCS$(2_1,1_2)$ in 22.344, CCS$(5_6,5_5)$ in 24.506, HC$^9$N-1$(40,39)$ in 23.241, HC$^9$N-2$(41,40)$ in 23.822, H64$\alpha$ in 24.509, CH$_3$OH$(6_2,6_1)$ in 25.018, H$_2$O$(6_16,5_23)$ in 22.235, NH$_3$-sp$(2,2)$ in 23.099, CC$^{34}$S(2,1) in 25.344, D$_2$O$(3,4)$ in 20.460, C$^{13}$CS(2,1) in 25.783. 
    \end{tablenotes}
\end{table*}
\addtocounter{table}{-1}

\begin{table*}[htp]
    \centering
    \setlength{\tabcolsep}{3pt}
    \caption{Observation results (continued).}
    \begin{tabular}{rrrrrrrrrrrr}
         \toprule
         NHD$_2$&CCS$(2_1,1_2)$&CCS$(5_6,5_5)$&HC$^9$N-1&HC$^9$N-2&H64$\alpha$&CH$_3$OH&H$_2$O&NH$_3$-sp&CC$^{34}$S(2,1)&D$_2$O&C$^{13}$CS(2,1)\\
         \hline
         1.2&$\times$&1.3&$\times$&$\times$&1.4&$\times$&$\times$&1.1&$\times$&$\times$&$\times$\\
         1.8&$\times$&1.9&$\times$&$\times$&1.6&$\times$&$\times$&1.7&$\times$&$\times$&$\times$\\
         1.7&$\times$&1.9&$\times$&$\times$&1.9&$\times$&$\times$&1.6&$\times$&$\times$&$\times$\\
         2.7&$\times$&2.8&$\times$&$\times$&3.6&$\times$&$\times$&2.7&$\times$&$\times$&$\times$\\
         3.5&$\times$&4.2&3.6&3.5&4.4&3.9&$\times$&$\times$&$\times$&$\times$&$\times$\\
         4.0&$\times$&3.6&3.7&3.2&3.7&3.2&$\times$&$\times$&$\times$&$\times$&$\times$\\
         4.2&$\times$&3.4&4.0&3.8&4.6&3.7&$\times$&$\times$&$\times$&$\times$&$\times$\\
         3.9&$\times$&3.5&3.5&3.1&3.9&4.1&$\times$&$\times$&$\times$&$\times$&$\times$\\
         3.1&$\times$&3.4&3.6&2.9&3.6&2.3&$\times$&$\times$&$\times$&$\times$&$\times$\\
         5.2&$\times$&4.5&4.1&3.9&4.7&5.1&$\times$&$\times$&$\times$&$\times$&$\times$\\
         5.1&$\times$&4.5&4.2&3.6&4.3&5.7&$\times$&$\times$&$\times$&$\times$&$\times$\\
         $\times$&2.6&2.0&$\times$&$\times$&2.1&$\times$&$\bigcirc$&$\times$&$\times$&$\times$&$\times$\\
         4.1&$\times$&4.3&3.6&3.6&3.2&$\bigcirc$&$\times$&$\times$&$\times$&$\times$&$\times$\\
         3.2&$\times$&3.6&3.2&3.6&3.1&3.1&$\times$&$\times$&$\times$&$\times$&$\times$\\
         3.5&$\times$&3.7&3.3&3.7&3.5&3.9&$\times$&$\times$&$\times$&$\times$&$\times$\\
         3.6&$\times$&3.2&3.4&3.5&3.6&3.6&$\times$&$\times$&$\times$&$\times$&$\times$\\
         $\times$&7.8&$\times$&$\times$&$\times$&$\times$&$\times$&$\bigcirc$&$\times$&$\times$&$\times$&$\times$\\
         $\times$&5.4&$\times$&$\times$&$\times$&$\times$&$\times$&9.7&$\times$&$\times$&$\times$&$\times$\\
         $\times$&7.9&$\times$&$\times$&$\times$&$\times$&$\times$&4.6&$\times$&$\times$&$\times$&$\times$\\
         4.3&5.4&3.5&2.8&2.3&3.1&4.5&8.1&$\times$&$\times$&6.1&$\times$\\
         4.6&6.7&4.2&3.4&2.9&3.8&5.4&8.8&$\times$&$\times$&7.1&$\times$\\
         4.8&6.4&4.4&2.9&2.5&3.4&4.4&9.2&$\times$&$\times$&6.4&$\times$\\
         5.3&6.4&3.3&3.1&3.2&3.4&5.3&8.5&$\times$&$\times$&6.6&$\times$\\
         5.7&6.4&4.1&3.4&3.1&3.7&6.2&8.9&$\times$&$\times$&7.1&$\times$\\
         $\times$&9.2&6.8&$\times$&$\times$&5.8&$\times$&9.5&$\times$&9.7&9.5&9.6\\
         $\times$&9.9&5.6&$\times$&$\times$&6.8&$\times$&9.4&$\times$&9.3&8.9&9.4\\
         $\times$&6.1&4.1&$\times$&$\times$&4.8&$\times$&8.7&$\times$&7.3&6.5&7.1\\
         $\times$&7.2&4.8&$\times$&$\times$&4.8&$\times$&9.4&$\times$&6.5&7.1&6.7\\
         $\times$&7.0&4.6&$\times$&$\times$&5.1&$\times$&9.3&$\times$&6.8&6.7&7.1\\
         $\times$&7.1&5.2&$\times$&$\times$&4.6&$\times$&10.5&$\times$&7.8&8.1&8.2\\
         $\times$&6.5&4.8&$\times$&$\times$&5.0&$\times$&9.6&$\times$&7.7&6.1&6.7\\
         $\times$&7.5&4.6&$\times$&$\times$&4.9&$\times$&9.8&$\times$&6.9&6.7&6.4\\
         $\times$&7.2&4.9&$\times$&$\times$&4.8&$\times$&9.7&$\times$&7.2&6.4&6.7\\
         \hline
    \end{tabular}
\end{table*}

\section{Identification and fitting}\label{analysis}

\subsection{Identification of cores}
As we introduced before, a core with more than 100 Jeans masses may fragment into lots of smaller cores. This fragmentation occurs in part of this sample. For example, Figure \ref{nh} shows that there are two dense cores in G14.2-P3. Thus, we integrated the intensity maps of the NH$_3$ (1,1) line for each detected clump. With the maps in Figure \ref{nh}, we find that there are multiple cores in some observed clumps. 


NH$_3$ has a good association with cores in dust continue map \citep[e.g.,][]{2018A&A...611L...3S,Lu_2014} which can be used to trace the dense gas in massive star-forming regions \citep{Lu_2018, 2022ApJ...926..165L}. Although NH$_3$ (1,1) may be optically thick in the most dense part, its hyperfine structures still can relatively well trace the gas temperature and the column density \citep{HoInterstellar} in this sample. Thus, the identification of the core and the analysis of its gas properties are mainly based on the fitted results from the NH$_3$ emission lines. In the rest of this work, we use "cores" to refer to the NH$_3$ cores.

As 21 VLA pointings detected in NH$_3$ (1,1) lines with more than five pixels per pointing, the criteria for the definition of a core for the following study are: (1) contains at least 10 pixels (larger than the beam size which means the core is resolved); (2) the fitting result must be continuous and the uncertainty of parameters should be less than 10\%; and (3) contains only one column density peak in the center of the core. The last requirement is based on the error study of \cite{Lu_2018}: the blending effect which may affects the following a study of the gas evolution \citep{2018RNAAS...2...52W}. We fitted radial profiles of the recognized cores of the temperature and the column density with one Gaussian component. We checked their residuals and ensure that there is little possibility for the existence of the second component with the current data.

We used the \textsc{imfit} task of the CASA to recognize the basic parameter of the possible core in the integrated intensity map of the NH$_3$ (1,1) line and found 32 cores in 21 VLA pointings. We labeled these cores ``parent molecular cloud+core number'' in Table \ref{tab:re}. Most pointings (14 out of 21) have 1 core but there are 7 pointings which are multi-cores systems: four clumps (G14.2-P1, G14.2-P3, G11.11\_s5 and G11.11\_s11) have two cores; two clumps (G79-C19 and G48.64) have three cores and IRAS18114 has four cores. More of the analysis are discussed in Section \ref{results}.

\subsection{Ammonia spectral line fitting}

We used the Python package PySpecKit \citep{2022AJ....163..291G} to fit the NH$_3$ (1,1) to (3.3) lines. The S/N threshold is 3$\sigma$ and the model \citep{HoInterstellar, 2008ApJS..175..509R} allows us to fit six parameters (excitation temperature ($T_{\mathrm{ex}}$), kinetic temperature ($T_{\mathrm{k}}$), column density ($N$(NH$_3$)), ortho-to-para ratio (OPR), centroid velocity (V$_{LSR}$), and velocity dispersion ($\sigma_v$)) together. For the cores detected in NH$_3$ (3,3) line, we try to fit their OPR. This parameter may trace the evolution of the massive star-forming regions \citep{2014MNRAS.439.3275W} but rarely be detected in previous NH$_3$ studies \cite[e.g.,][]{2018A&A...611L...3S}. Since NH$_3$ (3,3) line is rarely detected in massive star formation regions and can be observed with a non-local thermodynamic equilibrium (LTE) \cite{1983A&A...122..164W}, the fitted OPR may have a large uncertainties and only give the lower limit value. This work obtains the OPR values that should be taken as the lower limits for reference only, instead of strong constraints.

\begin{figure*}[htb!]
    \centering
    \includegraphics[width=0.9\linewidth]{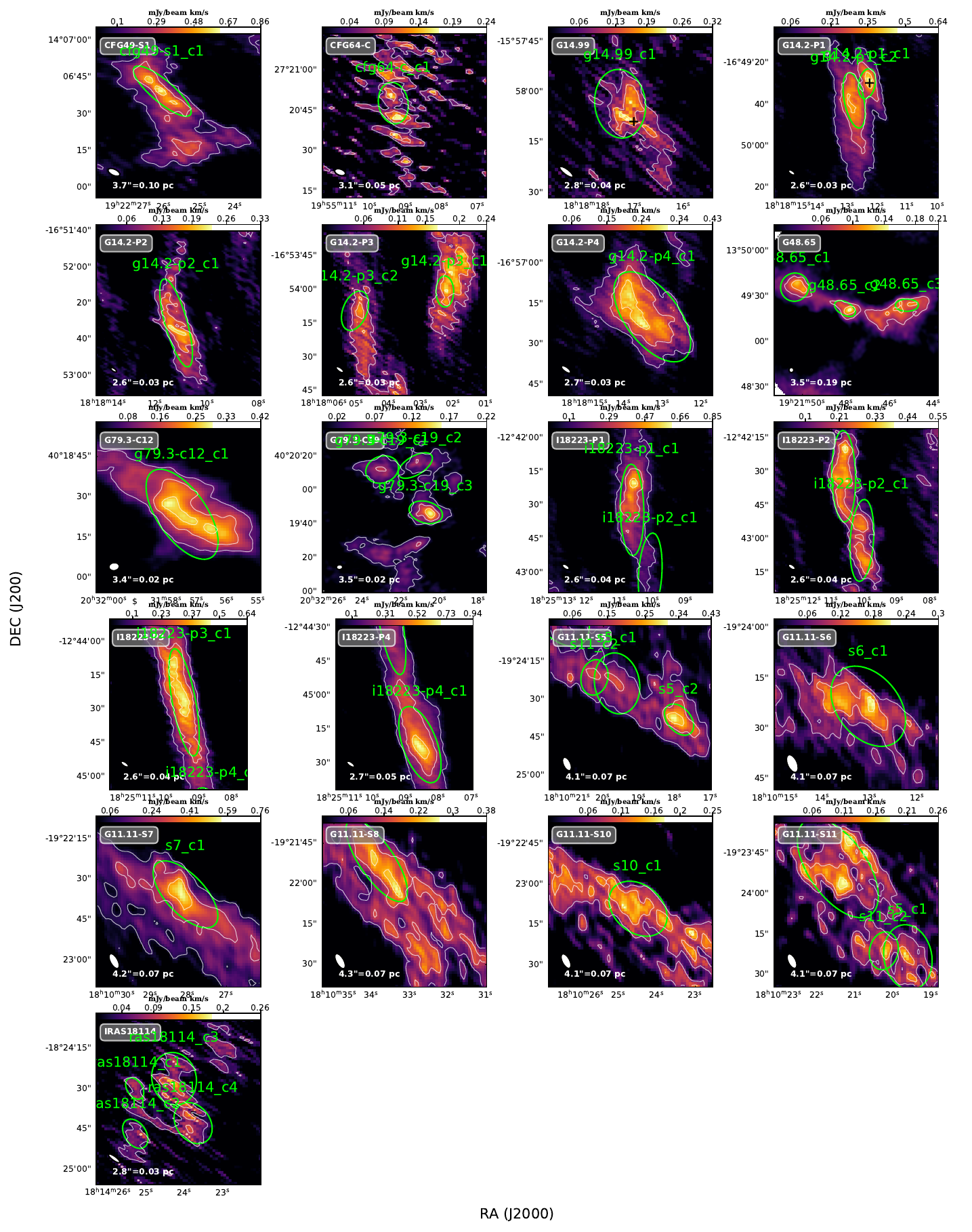}
    \caption{Flux-integrated maps of VLA NH$_3$ inversion emission line (1,1) from cores with a successful detection (detected in NH$_3$ (1,1) line with more than five pixels per pointing) and line fitting. Contours in each panel starts from 2$\sigma$ to the peak value in a linear scale. The green ellipse indicate the outlines of defined cores. Black crosses indicates the detected masers in this observation. White ellipse in bottom-left corner shows synthesized beam, with a corresponding physical scale. The color bar is shown at the top of each panel.}
    \label{nh}
\end{figure*}

In the fitting process, several cores do not converge so we have therefore excluded them from further statistical analysis. Figure \ref{nh} shows the integrated intensity map of the NH$_3$ (1,1) of all the successful fitted cores and the Table \ref{tab:re} presents the fitted parameters of each core.

We use parallel computing to speed up the fitting process of more than $10^9$ data points. Each fitted line is checked both by its residuals and manual inspection. 
The relative uncertainties for all fitted parameters in each line are required to be less than 10\%. To avoid local minima in the fit, we tried multiple sets of initial guesses evenly distributed throughout the parameter space, and compared their residuals to obtain the best fitting. The uncertainty of the fitting is mainly from the low S/N pixels. We tested the setting of the threshold higher to 5 and find that fitted data points are not enough to obtain statistics. The current threshold optimizes the weak signal and the uncertainty. We have also tested one- and multi-velocity components models in the fitting process to avoid the missing velocity components. By visually checking, we found that all the cores are dominated by one velocity component. In addition, we have also provided figure. \ref{nh3fit} as an example of spectral line fitting in the appendix for reference.

\begin{table*}[htp]
    \centering
    \caption{Fitted parameters.}
    \label{tab:re}
    \resizebox{\textwidth}{70mm}{
    \begin{tabular}{llllllllll}
         \toprule
         Core name&Temperature&Column density &Velocity&$\sigma_V$&OPR&M$_n$&size&M$_{esti}$&M$_{vir}$\\
         &K&$log_{10}$ cm$^{-2}$&km/s&km/s&$10^{-2}$& & pc&${M_\odot }$&${M_\odot }$\\
         \hline
         CFG49-S1-c1&35.9$\pm$8.2/33.2&12.5$\pm$0.1/12.5&67.7$\pm$0.7/68.0&1.6$\pm$0.6/1.5&$\times$&2.6$\pm$0.1/2.6&0.4&0.1&127.4\\
         CFG64-c1&10.0$\pm$4.0/10.8&14.2$\pm$0.3/14.3&22.0$\pm$0.1/22.0&0.2$\pm$0.1/0.2&$\times$&0.4$\pm$0.1/0.5&0.2&8.6&2.4\\
         G14.2-P1-c1&16.5$\pm$0.8/16.7&14.7$\pm$0.1/14.7&18.8$\pm$0.1/18.8&0.7$\pm$0.1/0.7&$\times$&2.4$\pm$0.2/2.4&0.1&1.7&8.1\\
         G14.2-P1-c2&14.7$\pm$1.0/14.6&14.8$\pm$0.1/14.8&19.7$\pm$0.4/19.7&0.8$\pm$0.1/0.9&$\times$&2.3$\pm$0.2/2.3&0.2&4.0&14.3\\
         G14.2-P2-c1&13.0$\pm$1.0/12.9&14.6$\pm$0.1/14.6&21.6$\pm$0.2/21.6&0.5$\pm$0.1/0.5&$\times$&1.3$\pm$0.1/1.3&0.2&7.4&9.7\\
         G14.2-P3-c1&12.0$\pm$1.7/11.9&14.7$\pm$0.3/14.7&21.3$\pm$0.1/21.3&0.3$\pm$0.1/0.3&$\times$&0.8$\pm$0.2/0.9&0.1&2.1&1.4\\
         G14.2-P3-c2&12.1$\pm$2.1/12.3&14.7$\pm$0.5/14.6&21.2$\pm$0.2/21.2&0.3$\pm$0.1/0.3&$\times$&1.0$\pm$0.1/1.0&0.1&3.3&2.0\\
         G14.2-P4-c1&16.7$\pm$1.9/16.5&14.7$\pm$0.2/14.7&19.9$\pm$0.5/19.9&0.7$\pm$0.3/0.8&2.1$\pm$1.3/2.0&0.9$\pm$0.1/0.9&0.3&9.1&19.0\\
         G14.99-c1&127.7$\pm$128.5/47.2&15.0$\pm$1.2/15.0&48.8$\pm$1.3/48.5&0.9$\pm$1.0/0.7&$\times$&0.3$\pm$0.1/0.3&0.2&196.7&94.3\\
         G48.65-c1&12.3$\pm$0.8/12.4&14.4$\pm$0.1/14.4&33.9$\pm$0.2/33.9&0.4$\pm$0.1/0.4&1.4$\pm$1.2/1.3&1.3$\pm$0.2/1.4&1.4&4.2&5.4\\
         G48.65-c2&12.7$\pm$0.4/12.7&14.4$\pm$0.1/14.4&33.9$\pm$0.2/33.9&0.3$\pm$0.0/0.3&2.0$\pm$1.9/1.9&1.0$\pm$0.1/1.0&0.8&0.5&1.1\\
         G48.65-c3&12.4$\pm$1.0/12.4&14.4$\pm$0.2/14.4&33.4$\pm$0.2/33.4&0.4$\pm$0.1/0.4&1.1$\pm$0.9/1.0&1.2$\pm$0.1/1.2&0.8&1.6&3.6\\
         G79.3-C12-c1&11.0$\pm$1.1/11.2&14.6$\pm$0.3/14.6&1.6$\pm$0.6/1.7&0.3$\pm$0.1/0.3&$\times$&0.4$\pm$0.0/0.4&0.2&1.8&1.4\\
         G79.3-C19-c1&17.3$\pm$17.9/15.9&14.4$\pm$0.4/14.4&0.4$\pm$0.1/0.4&0.4$\pm$0.3/0.4&$\times$&0.4$\pm$0.2/0.4&0.9&0.3&1.8\\
         G79.3-C19-c2&18.9$\pm$5.9/18.5&14.3$\pm$0.2/14.2&0.4$\pm$0.1/0.4&0.2$\pm$0.1/0.2&$\times$&0.3$\pm$0.1/0.3&0.9&0.2&0.5\\
         G79.3-C19-c3&10.1$\pm$3.4/10.6&15.8$\pm$0.7/15.9&0.4$\pm$0.1/0.4&0.8$\pm$0.2/0.7&$\times$&3.2$\pm$0.2/3.3&0.9&8.1&7.2\\
         I18223-P1-c1&16.0$\pm$2.2/15.5&14.8$\pm$0.1/14.8&45.0$\pm$0.3/44.9&0.5$\pm$0.2/0.4&2.1$\pm$1.0/2.0&1.1$\pm$0.1/1.1&0.2&20.1&12.6\\
         I18223-P2-c1&22.5$\pm$20.2/19.7&14.8$\pm$0.3/14.8&45.0$\pm$0.3/44.9&0.5$\pm$0.2/0.5&3.2$\pm$2.1/3.2&0.9$\pm$0.1/1.0&0.2&16.2&11.2\\
         I18223-P3-c1&14.8$\pm$0.9/14.9&14.8$\pm$0.2/14.8&45.7$\pm$0.6/45.8&0.7$\pm$0.2/0.7&$\times$&1.9$\pm$0.2/1.9&0.3&32.2&30.2\\
         I18223-P4-c1&14.3$\pm$1.4/14.2&14.8$\pm$0.1/14.8&45.9$\pm$0.3/45.8&0.7$\pm$0.1/0.7&4.0$\pm$2.4/3.4&2.1$\pm$0.1/2.1&0.3&34.4&30.2\\
         IRAS18114-c1&42.4$\pm$6.2/43.9&14.7$\pm$0.0/14.7&46.1$\pm$0.0/46.1&1.1$\pm$0.0/1.1&$\times$&2.6$\pm$0.1/2.6&0.1&2.1&20.2\\
         IRAS18114-c2&46.0$\pm$5.5/46.5&14.7$\pm$0.0/14.7&46.0$\pm$0.0/46.1&1.0$\pm$0.0/1.1&$\times$&2.4$\pm$0.1/2.4&0.1&3.9&22.4\\
         IRAS18114-c3&46.2$\pm$6.8/46.1&14.7$\pm$0.1/14.7&46.0$\pm$0.0/46.0&1.1$\pm$0.0/1.1&$\times$&2.3$\pm$0.1/2.3&0.1&13.0&54.2\\
         IRAS18114-c4&47.0$\pm$6.1/47.1&14.7$\pm$0.1/14.7&46.0$\pm$0.0/46.0&1.0$\pm$0.0/1.0&$\times$&2.3$\pm$0.0/2.3&0.1&7.8&35.6\\
         G11.11\_S5-c1&14.2$\pm$4.4/13.4&14.7$\pm$0.3/14.7&29.5$\pm$0.3/29.6&0.3$\pm$0.2/0.3&8.2$\pm$1.4/3.3&0.4$\pm$0.1/0.4&0.1&15.2&4.0\\
         G11.11\_S5-c2&6.4$\pm$0.8/6.4&16.1$\pm$0.3/16.2&30.1$\pm$0.1/30.1&0.4$\pm$0.1/0.4&$\times$&2.3$\pm$0.4/2.4&0.2&120.1&3.6\\
         G11.11\_S6-c1&12.4$\pm$3.1/11.8&14.8$\pm$0.6/14.7&30.1$\pm$0.3/30.0&0.3$\pm$0.2/0.2&$\times$&0.3$\pm$0.1/0.4&0.3&15.9&4.0\\
         G11.11\_S7-c1&16.3$\pm$6.7/15.2&14.8$\pm$0.2/14.9&29.5$\pm$0.4/29.5&0.5$\pm$0.2/0.5&3.2$\pm$3.0/2.4&0.7$\pm$0.0/0.7&0.2&30.3&14.0\\
         G11.11\_S8-c1&12.6$\pm$2.2/12.9&14.7$\pm$0.2/14.8&30.6$\pm$0.6/30.9&0.4$\pm$0.3/0.3&3.0$\pm$3.0/1.4&0.6$\pm$0.1/0.6&0.2&17.9&8.1\\
         G11.11\_S10-c1&11.8$\pm$2.2/11.6&14.6$\pm$0.3/14.7&28.8$\pm$0.2/28.8&0.3$\pm$0.1/0.3&$\times$&0.5$\pm$0.1/0.5&0.3&10.4&3.5\\
         G11.11\_S11-c1&13.0$\pm$3.6/12.1&14.5$\pm$0.5/14.6&28.9$\pm$0.3/28.8&0.3$\pm$0.2/0.3&$\times$&0.2$\pm$0.1/0.2&0.6&15.6&4.5\\
         G11.11\_S11-c2&13.1$\pm$3.7/12.8&14.8$\pm$0.5/14.7&29.7$\pm$0.4/29.7&0.4$\pm$0.3/0.3&$\times$&0.3$\pm$0.1/0.3&0.4&20.1&8.9\\
         \hline
    \end{tabular}}
    \begin{tablenotes}
    \footnotesize
    \item Fitted parameters of the 32 cores identified in NH$_3$(1,1) integrated emission. We adopt the fashion of ``mean value$\pm$error/median value". For three parameters (size, M$_{esti}$ and M$_{vir}$), only mean value is shown. The statistical parameters of M$_n$ are calculated from the red circles in Figure \ref{result}, Figure \ref{restresult1} and Figure \ref{restresult2}. The cross means that the core has not been detected in NH$_3$ (3,3) so we did not fit its OPR as an independent parameter. M$_{esti}$ is estimated from the $N_{\mathrm{NH_3}}/N_{\mathrm{H_2}}$ value ($4.6\times10^{-8}$) from \cite{2014ApJ...786..116B}.   
    \end{tablenotes}
\end{table*}

\begin{figure*}[htb!]
    \centering
    \includegraphics[width=1.0\textwidth]{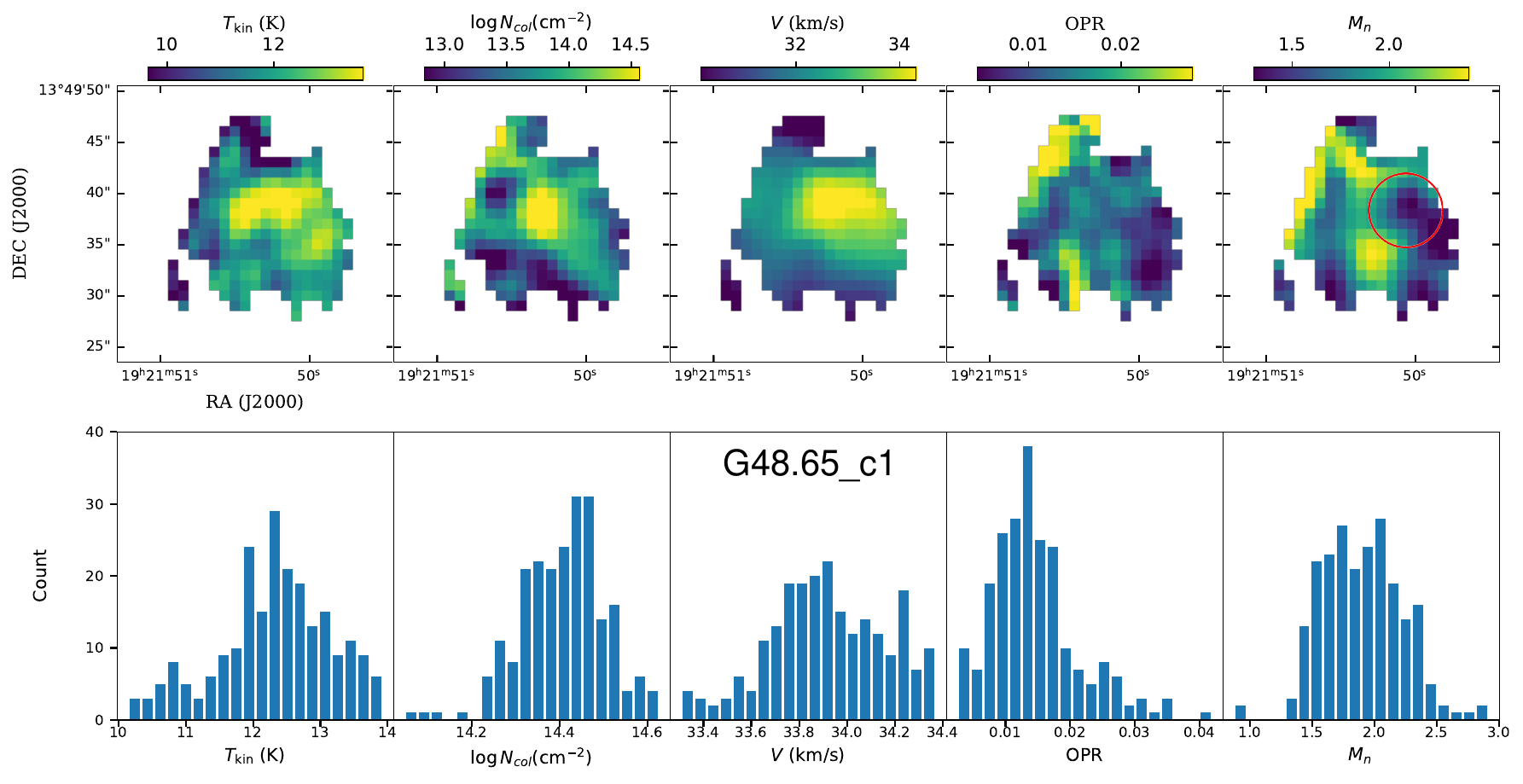}
    \caption{Fitting results of G48.65\_c1, as an example. \textbf{First row:} Temperature (K), column density (in log scale), centroid velocity (km ${\rm{s}^{-1}}$), OPR, and derived Mach number. The red circle is the sub-region we selected to study the core-averaged Mach number. \textbf{Second row:} Corresponding histograms of the parameter distribution among the emission regions.}
    \label{result}
\end{figure*}

Figure \ref{result} presents the map of fitted parameters of G48.65\_c1. The maps of the rest cores can be found in Figure \ref{restresult1} and Figure \ref{restresult2} in the appendix. 

\subsection{Mach number calculation} 

The Mach number is defined as ${\sigma _{{V_{non - th}}}}/{c_s}$ and the sound speed is calculated from ${c_s} = \sqrt {\frac{{k_B}{T_{kin}}}{{m_{\rm{H}}}{\mu _p}}}$, which is same to our previous work in G35.20-0.74 N \citep{2023A&A...674A..46W}.
We used the velocity dispersion along the line of sight, which is similar to \cite{2018A&A...611L...3S} and \cite{Lu_2014} to derive the non-thermal velocity dispersion: 

\begin{equation}
    {\sigma _{{V_{obs}}}}^2 ={\sigma _{{V_{th}}}}^2 +{\sigma _{{V_{channel}}}}^2+{\sigma _{{V_{grad}}}}^2+{\sigma _{{V_{non-th}}}}^2\;,
\end{equation}

where $\sigma _{{V_{th}}}$ is the thermal velocity dispersion,  defined as ${k_B}{T_{kin}}/{m_{\rm{NH}_3}}$ \citep{1983ApJ...270..105M}; $\sigma _{{V_{obs}}}$ and $T_{kin}$ are fitted parameters; and $m_{N{H_3}}$ is the ammonia mass.
The
$\sigma _{{V_{channel}}}$ measurement is due to the effect from the channel width (0.23 km~${\rm{s}^{-1}}$ in this work): 0.23 $\rm{km} \cdot {s^{ - 1}}/2\sqrt {2\ln 2}$. In most most previous works, the much larger channel width (0.6 km ${\rm{s}^{-1}}$) of the VLA observation \citep[e.g.,][]{2013MNRAS.432.3288S, Lu_2014} may be the reason of the supersonic turbulence  detection. 

The $\sigma _{{V_{grad}}}$ is the unresolved velocity gradient within the synthesized beam. We fitted a uniform velocity gradient at the large scale of each core and estimated the difference of  the velocity at two opposite edges of the synthesized beam, which may enlarge the measured velocity dispersion. For most filaments, their flat gradients are about 0.5 km~${\rm{s}^{-1}}$~${\rm{pc}^{-1}}$ \citep{wangke2016FL, 2018RNAAS...2...52W, GeYF2022FL, GeYF2023}, which contributes less than 2\% at the scale of the synthesized beam (the typical dispersion is larger than 0.1 km~${\rm{s}^{-1}}$) to the observed velocity dispersion, which can be ignored in most cases.

Since the conversion of the fitted NH$_3$ column density to H$_2$ column density may have large uncertainty, as noted in \cite{2014ApJ...786..116B}, and several cores have quite flat H$_2$ column density distributions, so the potential caveats exist that the region defined from the NH$_3$ column density peak may not trace dense and cold gas.

In the following analyses, we chose the sub-region (the red circles in Figure \ref{result}, Figure \ref{restresult1} and Figure \ref{restresult2}) with the lowest Mach number which covers more than three-beam size (more than 20 data points) for each core to study the core-averaged Mach number, which is listed as "$M_n$" in Table \ref{tab:re}. The reasons of this data selection are: (1) avoiding the influence from the external environment on the intrinsic turbulence and (2) these regions with the lowest Mach number are usually associated with the highest column density and can reveal the properties of the gas where the massive star may form.

\section{Results}\label{results}

\begin{figure*}[htb!]
    \centering
    \includegraphics[width=1.0\textwidth]{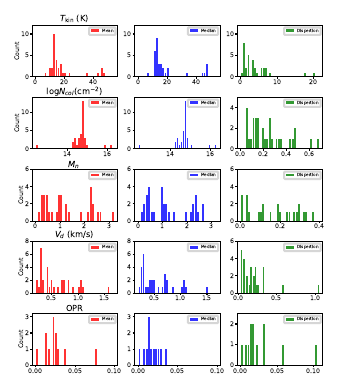}
    \caption{Histograms of fitted parameters of the 32 cores. From the top to the bottom: the parameters are temperature, column density, derived Mach number, velocity dispersion, and the OPR (for ten detected cores). From the left to the right: the mean value, the median value and the dispersion of each core. G14.99 is not included in the histogram of the temperature because of its extremely high value which is biased by the bright point source in the center of the core.}
    \label{sta-p}
\end{figure*}

Figure \ref{result} and Table \ref{tab:re} present the distribution map and the statistical results of the main parameters of G48.65\_c1. We add the median value as an auxiliary parameter because of the asymmetrical profile in part of the parameters' histograms. We calculated the mean value, the median value and the dispersion of each core's map and plotted the histograms of those statistical parameters in Figure \ref{sta-p}.

\subsection{Cores' Parameters and their Pearson correlation coefficient}

\subsubsection{Temperature}

As the temperature histograms presented as the first row in Figure \ref{sta-p}, 32 cores are mainly located in two ranges: 10-20 K and 40-50 K. Those two ranges in temperature histograms are consistent with two typical star-forming evolutionary stages: the prestellar (10-20 K) and the protostellar (40-50 K). Both the distributions of the histogram of the mean value and that of the median value are similar. G11.11\_S5-c2 is the coldest core (about 6 K) of the sample. CFG49\_S1-c1 (about 35 K) may have a protostar \citep{2015MNRAS.450.4043W}. Based on their temperatures, we divide cores into the prestellar group and the protostellar group, and represent them with different colors (blue and red) in the following statistical figures. We also check their IR images and previous studies, ensuring that the classification is consistent with previous studies.

The cores in the protostellar group are IRAS18114 (4 cores: c1 to c4), CFG49\_S1-c1 (HII region) and G14.99-c1 (maser) which contribute about about 18\% of the sample. The temperature of CFG49\_S1-c1 is relatively low in the protostellar group. On the contrary, the outer region of G14.99-c1 has the highest temperature in the protostellar group, but the temperature in the center of G14.99-c1 is much lower. The trend of the temperature of G14.99-c1 toward the center is decreasing. The four cores in IRAS18114 are typical protostellar with the warm and dense gas and form a multi-core system in Figure \ref{nh}. In the protostellar group, the gas motion may be highly affected by the environment, the fragmentation or the embedded protostellar outflows, which enhance the turbulence of the gas. 

The histogram of the temperature in prestellar group has a Gaussian profile. The peak is located at 14 K with the dispersion at 3 K. Those values are similar to previous studies of prestellar cores \citep[e.g.]{Lu_2014} which can reveal the initial condition of the turbulence. 


\subsubsection{Column density}

The range of the NH$_3$ column density is mainly from $10^{14.2}$ cm$^{-2}$ to $10^{15.1}$ cm$^{-2}$ with the peak at $10^{14.7}$ cm$^{-2}$. This peak value is lower than the mean column density of the massive cores in \cite{Lu_2018}, but higher than that of \cite{1999ApJS..125..161J}. Three data points that are out of this range are belong to CFG49\_S1-c1 ($10^{12.5}$ cm$^{-2}$), G79.3\_C19-C3 ($10^{15.8}$ cm$^{-2}$), and G11.11\_S5-c2 ($10^{16.1}$ cm$^{-2}$). The column density values of CFG49\_S1-c1, G79.3\_C19-C3, and G11.11\_S5-c2 are different to other cores. Similar situations also occurred in their temperatures. This may indicate their different evolutionary stages or environments. Although other cores belong to different temperature groups, they have similar column density values which means that the column density does not change a lot during the evolutionary stage: the evolution from the prestellar to the protostellar core may not enhance the cores' column density.  

Based on those fitted column density values and the sizes, we estimated cores' mass based on the [$N_{\mathrm{NH_3}}/N_{\mathrm{H_2}}$] value ($4.6\times10^{-8}$) from \cite{2014ApJ...786..116B}. The mean and median mass of all cores is 17.0/8.2${M_\odot }$ with uncertainties at about 10\%. This mean and median mass of cores is much larger than that of the low mass cores in \cite{1999ApJS..125..161J}. Otherwise, this mean/median mass is consistent with the results in \cite{Lu_2014} after the same mass conversion. Since the mean column density of our sample is slightly lower than that of \cite{Lu_2014} and cores in \cite{Lu_2014} are typical candidates of massive stars (\cite{Lu_2014} have used smaller convert factor as $3\times10^{-8}$), we deduce that cores in our sample may be at the earlier evolutionary stage, which would accrete the gas until growing into larger cores similar to the cores in \cite{Lu_2014}. As the following work of \cite{Lu_2014}, \cite{Lu_2018}  found transonic turbulence in similar cores. We expect to revealing the properties of the turbulence in our earlier cores.

\subsubsection{Velocity dispersion}

The histogram of the fitted (or observed) velocity dispersion in Figure \ref{sta-p} has a long tail up to about 1.6 km ${\rm{s}^{-1}}$ with the peak at 0.35 km ${\rm{s}^{-1}}$. Considering with the spectral resolution in this work (0.23 km ${\rm{s}^{-1}}$), the velocity dispersion in most cores is resolved. Even without the correction of the thermal motion and other effects, nearly one-third of the sources are sub- or transonic.

\subsubsection{Parameter correlations}

\begin{figure*}[htb!]
    \centering
    \includegraphics[width=1.0\textwidth]{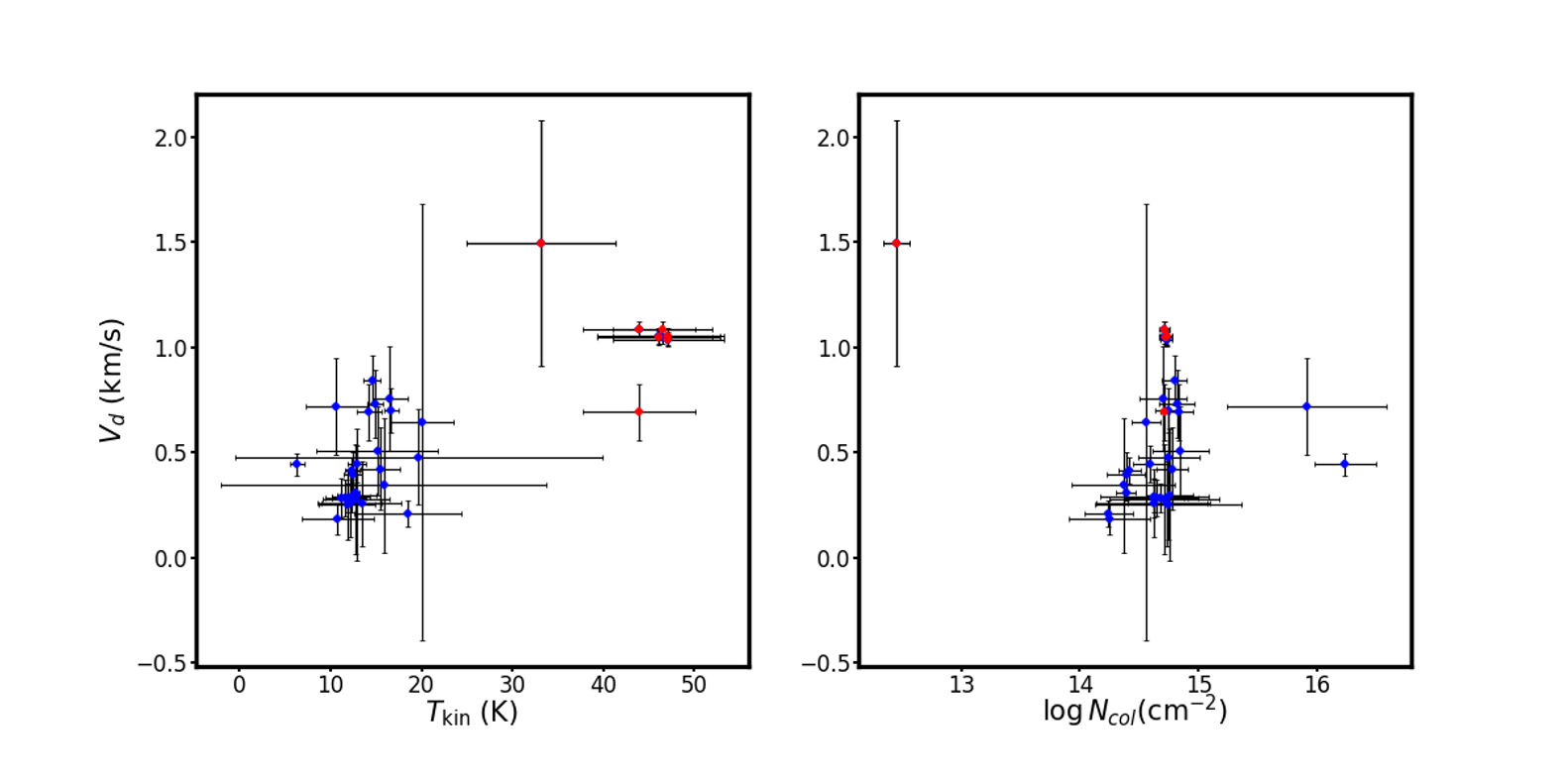}
    \includegraphics[width=1.0\textwidth]{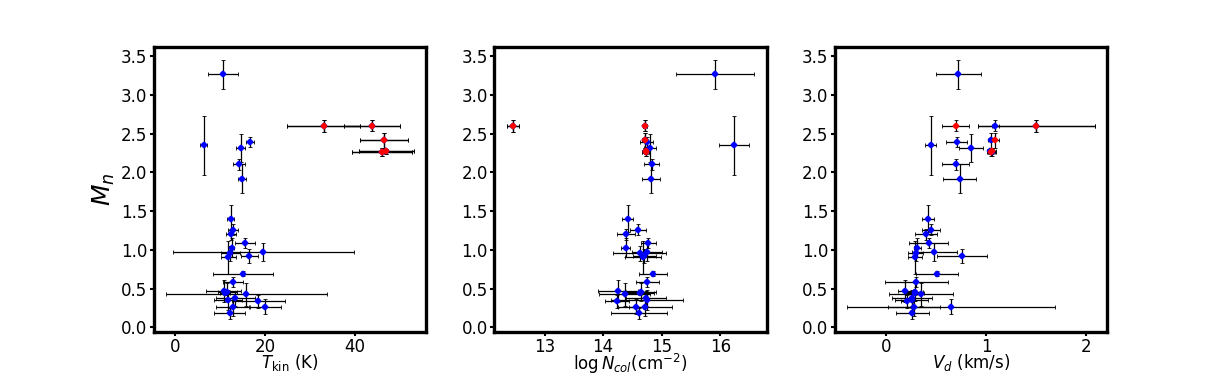}
    \caption{Relations among the main parameters. The different colors represent different temperature groups. \textbf{First row:}  Temperature versus the velocity dispersion, the column density versus the velocity dispersion, and the temperature versus the OPR (from left to right). The red points and black error bars are values of the fitted parameter and dispersion in Table \ref{tab:re}. G14.99 is not included in the figure because of its extremely high value which is biased by the bright point source in the center of the core. \textbf{Second row:} Relations between the Mach number and other parameters. Since the Mach number is calculated from both kinetic temperature and velocity dispersion,  relations between these parameters are expected.}
    \label{cor-p}
\end{figure*}

Figure \ref{cor-p} shows the relations of main parameters: temperature, column density, velocity dispersion, and the OPR. The temperature and the velocity dispersion have a positive correlation with the correlation parameter at 0.73. This correlation parameter is reasonable because the higher temperature means the stronger thermal motion of the gas, which results in the larger velocity dispersion. From the colder group to the warmer group, the mean velocity dispersion becomes larger as the mean temperature raises. On the other hand, the core with the higher temperature indicates the possible complex gas motion there which leads to the larger velocity dispersion. This enhancement may be more important in the protostellar group which could explain the weak correlation in the warmer group between the temperature and the velocity dispersion.
Since the NH$_3$ column density locates within a narrow range, the correlation between the column density and other parameters is rather weak. 
 

\subsection{Turbulence}\label{tur}

In a typical scenario of the massive star-forming, the turbulence of the gas gradually decays toward the center of the clump \citep{1981MNRAS.194..809L,2003ApJ...585..850M}. In the innermost region (the core scale which is about 0.1 pc), the turbulence could become sub- or transonic before the protostellar forming and giving its feedback \citep[e.g.,][]{WangKe2012,LuX2015IRDC,FengSY2016,LiuT2016outflow}. In order to avoid the interference of the surrounding gas in the turbulence study, we selected data points around the local minima in the Mach number map of each core, which is usually the center of each core. The selected sub-regions were resolved by three synthesized beams (covering more than 20 points).

\subsubsection{The distribution of the Mach number}

The histogram in the middle row of Figure \ref{sta-p} shows the statistical result of the Mach number. Its distribution is very different from that of the velocity dispersion in the fourth row of Figure \ref{sta-p}. The mean value of the Mach number histogram is 1.3, which means the turbulence of those region is mainly transonic, instead of the supersonic turbulence that has been predicted to be necessary in TCA. However, the multi-peak distribution  indicates that this mean value is not suitable enough to describe the whole properties of the turbulence in massive star-forming regions.

First, a multi-Gaussian model should be used to fit the Mach number distribution in order to more reliably determine the existence of different components.  We used the Gaussian mixture model (GMM) instead of the simple Gaussian fitting.  Based on the scikit-learn GaussianMixture, we estimated the number of the component with a GMM model onto the Mach number distribution of the 32 cores and we both used the Bayesian information criterion (BIC) and Akaike information criterion (AIC) to select the optimal model under different number of components. AIC suggests three to five Gaussian components as the most probable model and BIC suggests three components.
As fitting with the multi-Gaussian model with three conponents, we have found three main components, which are relatively independent to each other. Their peaks and dispersions are 0.4$\pm$0.1, 1.2$\pm$0.2, and 2.4$\pm$0.3. However, the sample size of each component is small (about 10) which is insufficient to definitively demonstrate the necessity of three components.  Since the Mach number distribution in this work exhibits continuity and this continuous trend is consistent with the evolutionary stages of massive stars which are not clearly delimited. Thus, we did not use the "component" to describe the Mach number distribution and  used velocity regimes such as "subsonic, transonic, and supersonic" to characterize different parts of the Mach number distribution.

The subsonic regime has 16 cores (50\%). Except for G14.99-c1, most cores in this part are the typical prestellar with cold (11-17 K) and dense ($10^{14-15}$ cm$^{-2}$) gas. The distribution maps of their column density and temperature are relatively flat.  

The transonic regime has seven cores (about 22\%). Cores in this regime is warmer (11-22 K) than those in the subsonic regime with the similar column density. The distribution maps of those cores' column density and temperature are not as flat as that of the cores in the subsonic regime. Those warmer cores may be at a later evolutionary stage.

The supersonic regime has nine cores (about 28\%). The gas properties of cores in this regime are very different to each other. CFG49\_S1-c1 has the warm (about 35 K) and thin ($10^{12.5}$ cm$^{-2}$) gas. Its high Mach number is more likely from the shock wave of the HII region rather than the feedback of the protostellar. Otherwise,  G11.11\_S5-c2 is very cold (6 K) and dense ($10^{16.2}$ cm$^{-2}$), which is similar to G79.3\_C19-C3 (10 K and $10^{15.8}$ cm$^{-2}$). Those two cores are not the typical prestellar or protostellar. The high Mach number may be due to the gas infall from the interaction of other cores in the same multi-core system. We discuss this further in Section \ref{diss}. The rest cores have large temperature dispersions. 

The histogram of the Mach number reveals a result: 
based on this sample, the sub- and transonic turbulence is prevalent (21 of 32, about 72\%) in massive star-forming regions and closely associated with the early evolutionary stage. Since the mean core-mass is relatively not significantly up to the high-mass stellar cores, we selected cores with more than 16\,\msun\ and find that about 78\% of them exhibit subsonic or transonic turbulence. This means that it is not the ubiquitously weak turbulence present in low-mass cores, as commonly assumed, that affected the conclusion. In the histograms of the Mach number, we found multiple regimes, suggesting that the intensity of turbulence varies in different evolutionary stages and tends to increase with evolution until becoming supersonic. This poses a challenge to the TCA \citep{2003ApJ...585..850M}: this model requires supersonic turbulence in the early evolutionary stage to slowdown the gravitational collapse so that massive stars can form. However, this is inconsistent with our results. Our study of the turbulence in massive star-forming regions indicates that sub- and transonic turbulence cannot provide enough pressure. Therefore, other pressure sources, such as the strong magnetic fields, may replace the role of the supersonic turbulence. The TAC model of the massive star formation need to be revised accordingly or be replaced by other models, for instance, GHC. We discuss this further in Section \ref{diss}.

\subsubsection{Correlation with other parameters}


The second row of Figure \ref{cor-p} presents the relations between the Mach number and other parameters. The Mach number has a weak relation with the temperature (correlation parameter at 0.36). Since the Mach number is calculated from both kinetic temperature and velocity dispersion, while the velocity dispersion displays a tight relation with the temperature, the weak  positive correlation is expected.

This is different from that of the velocity dispersion. As we discussed in Section \ref{results}, both the larger thermal motion and the possible more complex gas motion in warmer cores could enlarge the velocity dispersion. However, the  thermal motion part has been subtracted form the Mach number we used in the second row of Figure \ref{cor-p}. Thus, the relation between the Mach number and the temperature could reveals the change of the turbulence with the the raising temperature. In all of the cores, the turbulence raises with the temperature. However, this trend no longer exists in the protostellar group: the turbulence keeps supersonic, while the temperature changes from 40 K to 50 K. As we mentioned in Figure \ref{cor-p}, G14.99 is not included in the figure because of its extremely high value which is biased by the bright point source in the center of the core. The gas around the point source is still cold with the small velocity dispersion. 

The column density has a very weak-link with the Mach number. As what we discussed from Figure \ref{sta-p}, the column density keeps at $10^{14-15}$ cm$^{-2}$ and does not change a lot at different evolutionary stages. Thus, the change in the Mach number has little influence on the column density. 


The correlation parameter of the Mach number and the velocity dispersion is 0.78 from the second row of Figure \ref{cor-p}. But their profiles of the histograms are different. Beside the turbulence which is traced by the Mach number, the velocity dispersion contains the channel width, thermal motions, and other effects. In several early studies, the velocity dispersion are roughly used as the replacement of the Mach number to study the turbulence. The tightly linking between those two in this study supports this replacement. But our research reveals that the velocity dispersion (one-third are sub- and transonic) only inherits some properties of the Mach number (about 72\% are sub- and transonic) and their profiles could be different from each other.

\subsubsection{Turbulence and the temperature and column density distribution}

\begin{figure}[htb!]
    \centering
    \includegraphics[width=0.5\textwidth]{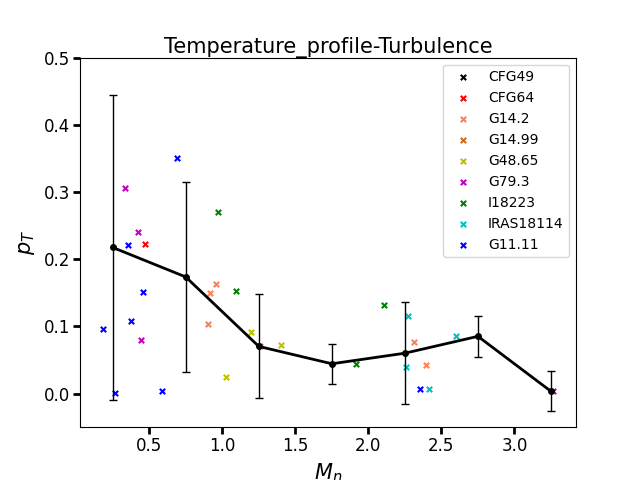}
    \caption{Fitted parameters of the temperature radial profile versus Mach numbers. Label colors represent different parental clumps. The black point and line indicate the mean and error of the data points in each beam of the Mach number, respectively.}
    \label{tp}
\end{figure}

\begin{figure}[htb!]
    \centering
    \includegraphics[width=0.5\textwidth]{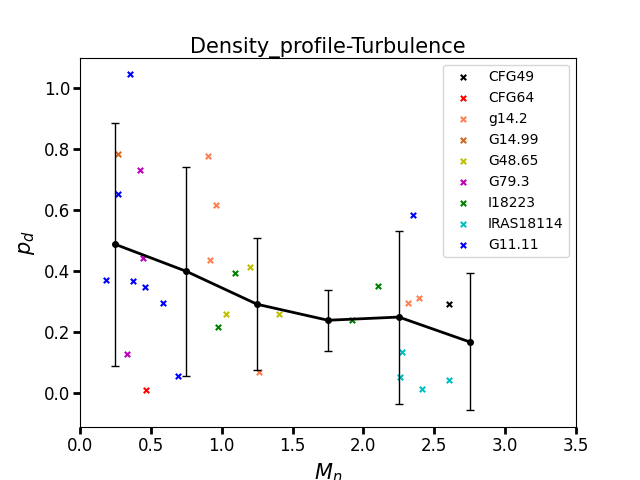}
    \caption{Fitted parameters of the column density radial profile versus Mach numbers. Label colors represent different parental clumps. The black point and line indicate  the mean and error of the data points in each beam of the Mach number, respectively. }
    \label{dp}
\end{figure}

Besides the effect from the turbulence on the whole core, the gas motion could also reshape the profile in the core. We fit the temperature and column density radial profile of each core with the power law model as $T \sim {r^{ - p_T}}$ and $N \sim {r^{ - p_d}}$, and we plotted those two parameters with the Mach number in Figs. \ref{tp} and \ref{dp}. We divide the total Mach number data into several bins with a step of 0.5. For each bin, we calculate the mean/dispersion value and plot them in Figs. \ref{tp} and \ref{dp}. Cores in the same clump are labeled with the same color. We present the sampling method and fitting model in Figure \ref{mfitting} of the appendix, with details introduced in the corresponding paragraphs of the appendix. 

The whole trends in those two figures are similar: as the Mach number increases, both the temperature and column density radial profile become flatter. In the high-value end of the Mach number, this trend becomes ambiguous. The reason of this trend may be the extra pressure of the turbulence. The stronger turbulence could disturb the original gas structure of the core and support more material which results in a larger core with a flatter profile. On the contrary, the sub- and transonic turbulence cannot support the growing gravity potential so the gas and dust will fall into the central region more easily, which makes a steeper profile. 

But this trend may not exist for a particular multi-core system. We first calculate the mean temperature and column density of seven multi-core systems. Besides IRAS18114 (about 45 K), all the other systems are cold (about 12 K) and dense. As IRAS18114 has several YSOs, its warm gas may be heated by those YSOs. Assuming that all the cores in seven multi-core systems are formed together from their parent molecular cloud, their initial conditions and environment should be similar. However, from Figs. \ref{tp} and \ref{dp}, their profiles indicate their complex evolutionary stages. For example, the range of the column density profile parameters of cores in the large filament:  ``Snake'' \citep{2014MNRAS.439.3275W,WangKe2015book,Pillai2019}, which contains clumps from G11.11\_S5 to G11.11\_S11 are from 0.06 to 1.04. Other multi-core systems have similar situations both in the column density and the temperature profiles. This large separation means that cores with similar conditions in the same clump could still have different evolutionary stages. 

\section{Discussion}\label{diss}

\subsection{Spatial distributions and core evolution}

We estimated the masses of each cores and found that most multi-core systems have equally shared the total mass of the clump with similar gas properties and profile slopes. However, the cores in G11.11\_S5 (dual system with 135.3 ${M_\odot }$) and G79.3\_C19 (triple system with 8.6 ${M_\odot }$) are very different. The masses of G11.11\_S5-c2 (120.1 ${M_\odot }$, 88.8\%) and G79.3\_C19-c3 (8.1 ${M_\odot }$, 94.2\%) \citep{2019ApJ...876...70L} dominate their whole system. Those two cores are very cold and dense: comparing with other cores in their systems, G11.11\_S5-c2 and G79.3\_C19-c3 are 6-10 K, with the column density being higher than an order of the magnitude. Besides, they are filled with the highly turbulent gas which has the Mach number at 2-3 (supersonic).  

As studies of hub-filament systems suggest, the mass distribution of the multi-cores systems could be similar \citep[e.g.,][]{2009ApJ...700.1609M, 2013A&A...555A.112P}. G11.11\_S5-c2 and G79.3\_C19-c3 are very different from those studies. Their existence indicates that the fragmentation in star-forming regions may be affected by some other factors that result in different masses and gas properties. For example, \citet{Xu2023SDC335} suggest the initial gas streams efficiently feed the central massive core in the SDC335 hub-filament system. The high efficiency can lead to a over-dense region in the center, supporting the different mass ratio among cores in such a system.

Similar to the SDC335 hub-filament system, G11.11\_S5-c2 and G79.3\_C19-c3 locate in the center of their multi-core systems, which can be well explained by the hub-filament system model: the gas falls into the central core along the filament structure and prompts this core into the evolutionary stage later than other outer cores; while the interaction of the gas enhances the line width and raises the Mach number. This scene is consistent with the simulation result of \cite{2018A&A...615A..94F}: under the low Mach number (e.g., Mach number at 3 which is similar to ours), the fragmentation is inhibited independent of the magnetic support and the filament structure appears. 

\cite{2018A&A...615A..94F} has studied the relationship between the turbulence, core number and geometrical morphology of the fragments. Their result supports that the subsonic turbulence helps form dense cores in the slender cloud under weak magnetic field. In Fig. \ref{nh}, G14-P1 and P3 are dual systems which only have the simple distribution. G11.11\_s5 and s11 are also dual systems but they are part of the larger filament G11.11 ( ``Snake''). This is a long S-shape filament with several dense cores \citep{2014MNRAS.439.3275W,WangKe2015book}. In Table \ref{tab:re} and Figure \ref{nh}, cores in G11.11 are all subsonic with the mean Mach number at 0.4, except for G11-s5-c2, which is totally supersonic. The triple system G79-C19 is similar to G11.11. The \nh3 cores in G79-C9 have a C-shape distribution, and the Mach number of other cores is at 0.4 except for G79-C19-c3: a supersonic dense core in the center of the threadlike filament. Another triple system G48.65 is a straight filament with three critical transonic cores. All those multi-core systems are slender and most cores are sub- or transonic. However, IRAS18114 is different from them: four supersonic cores form a clump with the irregular spatial distribution. Our work shows that the sub- and transonic turbulence core prefers to be formed in the slender filament but the supersonic core prefers the irregular clump. This trend is same as the result in \cite{2018A&A...615A..94F}: the spatial distributions could affect the gas properties and cores' evolution.

Based on those multi-core systems, we deduce that most multi-core systems may have similar cores with the same evolutionary stage at first. However, the evolution could be affected by the spatial distributions (both the shape and the relative location) and leads to the different evolutionary stages of cores. The dense and cold turbulent gases in G79.3\_C19-c3 and G11.11\_S5-c2 are probably due to the spatial distribution.

\subsection{Turbulence and the core's evolution}

As we mention in the introduction, in several recent studies of  massive star-forming regions, turbulence has been resolved as transonic or even subsonic under sufficient high spectral and spatial resolutions. However, most of these studies are case studies and it is difficult to demonstrate whether sub- and transonic turbulence is common in massive star-forming regions. Yet, there are a few statistical studies with large samples, among which the more representative ones are \cite{1999ApJS..125..161J} for low-mass stars and \cite{Lu_2014,LuX2015IRDC,Lu_2018} for high-mass stars.

\cite{1999ApJS..125..161J}  studied the gas properties and dynamics of 264 cores using NH$_3$ (1,1) and (2,2) lines. Similarly to us, they found that the non-thermal line width decreases with decreasing temperature. They also deduced that the core's environment plays an important role in turbulence and core's evolution, which is similar to our previous subsection. Limited by the spectral resolution, their average line width (0.74 km/s) is larger than ours (0.54 km/s). However, their study still implied the possibility of subsonic turbulence. As we presented in Section \ref{result}, the core mass and the density of \cite{1999ApJS..125..161J} indicate that their study focuses on low-mass stars. The general properties of turbulence in massive star-forming regions may be different.

\cite{Lu_2014} have studied the properties and dynamics of the gas in 62 high-mass star-forming regions with NH$_3$ (1,1) and (2,2) lines. They identified 174 cores and derived their line width (1.1 km/s), temperature (18 K), NH$_3$ column density ($10^{15}$ cm$^{-2}$), and mass (67 \msun). Their following work \citep{Lu_2018} further found that transonic turbulence exists in massive star-forming regions and the fragmentation of cores cannot be explained solely by the support of thermal or turbulent pressure. 

With sufficiently high spectral and spatial resolutions, we have found sub- and transonic turbulence in massive star-forming regions, further confirming that such weak turbulence is common (about 72\%). Since the cores in our sample are slightly less massive, colder, and more tenuous than that of \cite{Lu_2018}, our cores are more likely to at the earlier stages. This explains the narrower line width and weaker turbulence we measured. 

Besides, \cite{1999ApJS..125..161J} have pointed the influence from the associated YSOs or clumps onto the turbulence and \cite{Lu_2018} also deduced that the non-thermal motion could be enhanced in the filaments by the feedback or accretion. They also fitted the radial temperature distribution of the cores by power-law and found the range of the slope is from -0.18 to -0.35 which is similar to ours.
Combined with our results, their inferences lead us to prefer that the role of turbulence in massive star formation as follows: the turbulence is weak (sub- and transonic) at the early stage. Then it intensifies with the feedback and accretion/cores' interaction until becoming supersonic which in turn affects the evolution of the host core (e.g., supporting more accreted material and form a high-mass star).

Under this situation, other mechanisms, such as magnetic fields, are needed to provide enough support in the early stages when subsonic or transonic turbulence dominates the gas and the TAC \citep{2003ApJ...585..850M} needs revision by including these factors to account for the formation of massive stars: a combination of turbulence and other mechanisms drives the evolution from cores to massive stars. 

\subsection{Effects of the distance}

\begin{figure*}[htb!]
    \centering
    \includegraphics[width=1.0\textwidth]{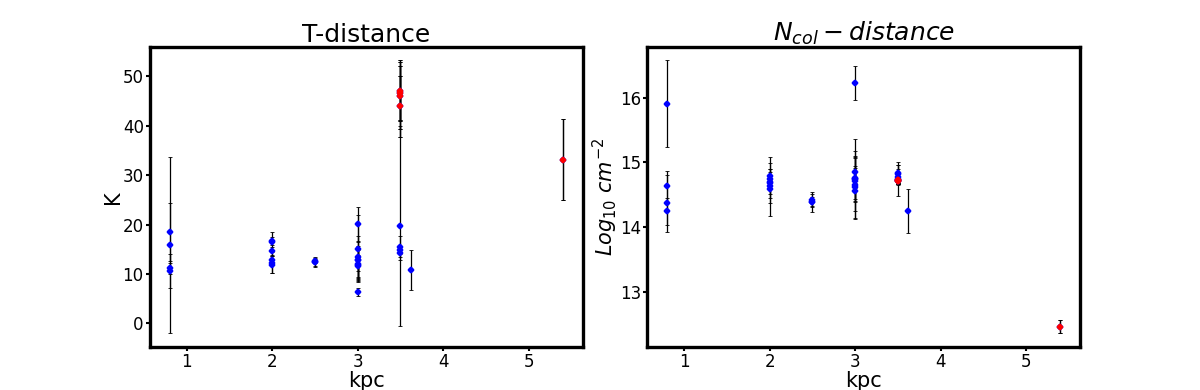}
    \caption{Distances versus other parameters. The color is same as in Fig. \ref{cor-p}}
    \label{dis}
\end{figure*}

As the broad distance range (0.9-5.4 pc) of the sample we used, we discuss the potential selection effect in this study. Thus we plot the distances versus other parameters of each core in Figure \ref{dis}. If the selection effect really exists, the  core further on would have higher temperature and column density, which would be more easily detected. In this case, those warmer cores are more likely at the later evolutionary stage, with higher Mach numbers. 
 
In Figure \ref{dis}, both the temperature and the column density of cores show the weak relation of the distance. Although some of the further cores have relatively higher temperature and column density, their Mach numbers are rarely affected by the distance. This indicates that the selection effect is not important.

Another effect is the corresponding pixel scale under the similar resolution (3") at different distances. As we calculated in Section \ref{analysis}, the pixel size will contribute the extra velocity dispersion into the line width within the velocity gradient at a large scale. As a typical large velocity gradient in dense cores (for example, about 1 km/s pc$^{-1}$ in Orion \citep{2021RAA....21...24Y}) which locate at 5 kpc with the resolution as 1", it will contribute less than 2\% of the line width. Although  this effect is subtracted before the calculation of the Mach number, it still enlarges the uncertainty and makes the Mach number slightly larger in a more complex environment. In this study, we carefully check the velocity gradient fitting in all the cores especially for the further cores. All of them have smooth velocity distribution maps, which means this influence of the far distance is not important.

\subsection{Peaks' separation}

\begin{figure*}[htb!]
    \centering
    \includegraphics[width=1.0\textwidth]{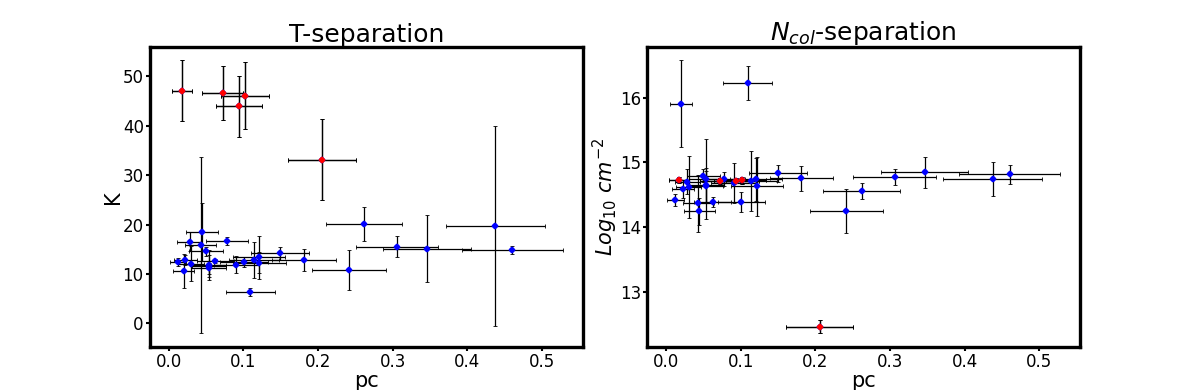}
    \caption{Separation of the temperature and the column density.}
    \label{sep}
\end{figure*}

During the identification of the NH$_3$ cores, we have found that the peak of the temperature map is slightly different from that of the column density map. First, this difference may be the result of the optically thick NH$_3$ lines. Thus, we checked the fitted lines especially for the lines with high column density values ($10^{15}$ cm$^{-2}$) and most of them are optically thin ($\tau  \sim {\rm{0}}{\rm{.1}}$). 

Then we compare the separation and the beam size. As the mean value of the beam size is about 3", most separations of the cores are larger than the corresponding spatial resolution. Thus, this separation is true and generally exists in star-forming regions. 

Another guess is the NH$_3$ depletion. Similarly to the depletion of the CO in massive star-forming regions \citep{2007A&A...467..207P,2021ApJ...923..147M,2022ApJ...936...80S}, the NH$_3$ could deplete at the highest column density region which leads to the peak shifting. 

In Figure \ref{sep}, we plot the peak separation with other parameters (temperature and column density) for further studying. If the separation is related to the NH$_3$ depletion, cold and dense cores will have larger shift values. However, in Figure \ref{sep}, the separation do not have obvious relation with either the temperature or the column density. Besides, we check the spatial distribution maps of all the cores and have not found any ring-like or ark-like structure within them. Thus, we deduce that the separation is not mainly because of the NH$_3$ depletion.

\section{Summary}\label{sum}

We use the Very Large Array (VLA) to observe 20 emission lines with the high spectral and spatial resolution (0.23 km ${\rm{s}^{-1}}$ and 3") in a sample of 13 massive star-forming regions. With such a high spectral resolution, we resolve the intrinsic turbulence excluding the thermal motion and other effects. We find that the sub- and transonic turbulence is prevalent found in dense cores. The finding challenges the important role of turbulence to support the gravitational collapse in massive star formation and suggests that other internal pressure candidates or massive-star formation theories are needed. Here, we summarise our work:

\begin{enumerate}
    \item In 32 selected clumps, 21 have been detected in NH$_3$ emission lines, and 2 of them exhibit a H$_2$O maser, 1 exhibits a CH$_3$OH maser, and 1 exhibits a NH$_2$D line. The NH$_3$ are usually detected in only (1,1) and (2,2) lines. The lack of higher excitation lines confirms that the selected sample is mainly at the early evolutionary stage.
    \item Based on NH$_3$ lines, we fit gas properties (excitation temperature (T$_{ex}$), kinetic temperature (T$_k$), column density, centroid velocity (V$_{LSR}$) and velocity dispersion ($\sigma_v$)) of 32 recognized cores in 21 VLA pointings. Besides, we fit the ortho-to-para ratio (OPR) for cores with the NH$_3$ (3,3) detection.
    \item The histograms of the Mach number of 32 cores are distributed in three regimes. The sub- and transonic turbulence is ubiquitously found (72\%) in the early evolutionary stages and this fraction is higher (78\%) among cores more massive than 16\,\msun\. This fraction may challenge the important role of turbulence to support the gravitational collapse in massive star formation and suggests that other internal pressure candidates (e.g., magnetic field) or theories (e.g., GHC) may be needed.
    \item The 32 cores are classified into two groups based on their temperature histogram, which are thought to trace two evolutionary stages. Combined with the column density histogram, the temperature raises during the evolution but change of the column density is not obvious. 
    \item The turbulence may affect the radial profile of cores. Cores with higher Mach number have flatter distribution profile both of temperature and column density. 
    \item There are seven multi-core systems in this sample, and within each system, most cores equally share the clump mass with similar gas properties. We have found two cases that the system is dominated by a highly turbulent core which locates at the center of the system. Those systems support the prediction from the hub-filament model: the spatial distribution can affect the evolution of cores.
\end{enumerate}

\begin{acknowledgements}

We thank Siju Zhang and Wenyu Jiao for valuable discussion, and an anonymous referee for constructive comments that helped improve this paper. We acknowledge support from the National Science Foundation of China (11973013, 12033005, 12041305,11721303), the China Manned Space Project (CMS-CSST-2021-A09), the National Key Research and Development Program of China (2022YFA1603102, 2019YFA0405100), and the High-Performance Computing Platform of Peking University.
PS was partially supported by a Grant-in-Aid for Scientific Research (KAKENHI Number JP22H01271 and JP23H01221) of JSPS. 
The National Radio Astronomy Observatory is a facility of the National Science Foundation operated under cooperative agreement by Associated Universities, Inc.
D.L. is supported by NSFC grant No. 11988101. F.W.X. acknowledges the support by NSFC through grant No. 12033005. H.B.L. is supported by the National Science and Technology Council (NSTC) of Taiwan (Grant Nos. 111-2112-M-110-022-MY3).
GB acknowledges support from the PID2020-117710GB-I00 grant funded by MCIN/ AEI /10.13039/501100011033.

\end{acknowledgements}


\bibliographystyle{plainnat}
\bibliography{ref}

\begin{thebibliography}{87}
\providecommand{\natexlab}[1]{#1}
\providecommand{\url}[1]{\texttt{#1}}
\expandafter\ifx\csname urlstyle\endcsname\relax
  \providecommand{\doi}[1]{doi: #1}\else
  \providecommand{\doi}{doi: \begingroup \urlstyle{rm}\Url}\fi

\bibitem[{A{\~n}ez-L{\'o}pez} et~al.(2020){A{\~n}ez-L{\'o}pez}, {Busquet}, {Koch}, {Girart}, {Liu}, {Santos}, {Chapman}, {Novak}, {Palau}, {Ho}, and {Zhang}]{2020A&A...644A..52A}
N.~{A{\~n}ez-L{\'o}pez}, G.~{Busquet}, P.~M. {Koch}, J.~M. {Girart}, H.~B. {Liu}, F.~{Santos}, N.~L. {Chapman}, G.~{Novak}, A.~{Palau}, P.~T.~P. {Ho}, and Q.~{Zhang}.
\newblock {Role of the magnetic field in the fragmentation process: the case of G14.225-0.506}.
\newblock \emph{\aap}, 644:\penalty0 A52, December 2020.
\newblock \doi{10.1051/0004-6361/202039152}.

\bibitem[{Ashimbaeva} et~al.(2019){Ashimbaeva}, {Colom}, {Lekht}, {Pashchenko}, {Rudnitskii}, and {Tolmachev}]{2019ARep...63.1022A}
N.~T. {Ashimbaeva}, P.~{Colom}, E.~E. {Lekht}, M.~I. {Pashchenko}, G.~M. {Rudnitskii}, and A.~M. {Tolmachev}.
\newblock {Evolution of the H$^{2}$O Maser Emission in the Star-Forming Region S252A}.
\newblock \emph{Astronomy Reports}, 63\penalty0 (12):\penalty0 1022--1034, December 2019.
\newblock \doi{10.1134/S1063772919120011}.

\bibitem[{Bally} and {Zinnecker}(2005)]{2005AJ....129.2281B}
John {Bally} and Hans {Zinnecker}.
\newblock {The Birth of High-Mass Stars: Accretion and/or Mergers?}
\newblock \emph{\aj}, 129\penalty0 (5):\penalty0 2281--2293, May 2005.
\newblock \doi{10.1086/429098}.

\bibitem[{Battersby} et~al.(2014){Battersby}, {Bally}, {Dunham}, {Ginsburg}, {Longmore}, and {Darling}]{2014ApJ...786..116B}
Cara {Battersby}, John {Bally}, Miranda {Dunham}, Adam {Ginsburg}, Steve {Longmore}, and Jeremy {Darling}.
\newblock {The Comparison of Physical Properties Derived from Gas and Dust in a Massive Star-forming Region}.
\newblock \emph{\apj}, 786\penalty0 (2):\penalty0 116, May 2014.
\newblock \doi{10.1088/0004-637X/786/2/116}.

\bibitem[{Beuther} et~al.(2018){Beuther}, {Mottram}, {Ahmadi}, {Bosco}, {Linz}, {Henning}, {Klaassen}, {Winters}, {Maud}, {Kuiper}, {Semenov}, {Gieser}, {Peters}, {Urquhart}, {Pudritz}, {Ragan}, {Feng}, {Keto}, {Leurini}, {Cesaroni}, {Beltran}, {Palau}, {S{\'a}nchez-Monge}, {Galvan-Madrid}, {Zhang}, {Schilke}, {Wyrowski}, {Johnston}, {Longmore}, {Lumsden}, {Hoare}, {Menten}, and {Csengeri}]{2018A&A...617A.100B}
H.~{Beuther}, J.~C. {Mottram}, A.~{Ahmadi}, F.~{Bosco}, H.~{Linz}, Th. {Henning}, P.~{Klaassen}, J.~M. {Winters}, L.~T. {Maud}, R.~{Kuiper}, D.~{Semenov}, C.~{Gieser}, T.~{Peters}, J.~S. {Urquhart}, R.~{Pudritz}, S.~E. {Ragan}, S.~{Feng}, E.~{Keto}, S.~{Leurini}, R.~{Cesaroni}, M.~{Beltran}, A.~{Palau}, {\'A}.~{S{\'a}nchez-Monge}, R.~{Galvan-Madrid}, Q.~{Zhang}, P.~{Schilke}, F.~{Wyrowski}, K.~G. {Johnston}, S.~N. {Longmore}, S.~{Lumsden}, M.~{Hoare}, K.~M. {Menten}, and T.~{Csengeri}.
\newblock {Fragmentation and disk formation during high-mass star formation. IRAM NOEMA (Northern Extended Millimeter Array) large program CORE}.
\newblock \emph{\aap}, 617:\penalty0 A100, September 2018.
\newblock \doi{10.1051/0004-6361/201833021}.

\bibitem[{Bonnell} et~al.(2001){Bonnell}, {Bate}, {Clarke}, and {Pringle}]{2001MNRAS.323..785B}
I.~A. {Bonnell}, M.~R. {Bate}, C.~J. {Clarke}, and J.~E. {Pringle}.
\newblock {Competitive accretion in embedded stellar clusters}.
\newblock \emph{\mnras}, 323\penalty0 (4):\penalty0 785--794, May 2001.
\newblock \doi{10.1046/j.1365-8711.2001.04270.x}.

\bibitem[{Bonnell} et~al.(2004){Bonnell}, {Vine}, and {Bate}]{2004MNRAS.349..735B}
Ian~A. {Bonnell}, Stephen~G. {Vine}, and Matthew~R. {Bate}.
\newblock {Massive star formation: nurture, not nature}.
\newblock \emph{\mnras}, 349\penalty0 (2):\penalty0 735--741, April 2004.
\newblock \doi{10.1111/j.1365-2966.2004.07543.x}.

\bibitem[{Busquet} et~al.(2013){Busquet}, {Zhang}, {Palau}, {Liu}, {S{\'a}nchez-Monge}, {Estalella}, {Ho}, {de Gregorio-Monsalvo}, {Pillai}, {Wyrowski}, {Girart}, {Santos}, and {Franco}]{2013ApJ...764L..26B}
Gemma {Busquet}, Qizhou {Zhang}, Aina {Palau}, Hauyu~Baobab {Liu}, {\'A}lvaro {S{\'a}nchez-Monge}, Robert {Estalella}, Paul T.~P. {Ho}, Itziar {de Gregorio-Monsalvo}, Thushara {Pillai}, Friedrich {Wyrowski}, Josep~M. {Girart}, F{\'a}bio~P. {Santos}, and Gabriel A.~P. {Franco}.
\newblock {Unveiling a Network of Parallel Filaments in the Infrared Dark Cloud G14.225-0.506}.
\newblock \emph{\apjl}, 764\penalty0 (2):\penalty0 L26, February 2013.
\newblock \doi{10.1088/2041-8205/764/2/L26}.

\bibitem[{Busquet} et~al.(2016){Busquet}, {Estalella}, {Palau}, {Liu}, {Zhang}, {Girart}, {de Gregorio-Monsalvo}, {Pillai}, {Anglada}, and {Ho}]{2016ApJ...819..139B}
Gemma {Busquet}, Robert {Estalella}, Aina {Palau}, Hauyu~Baobab {Liu}, Qizhou {Zhang}, Josep~Miquel {Girart}, Itziar {de Gregorio-Monsalvo}, Thushara {Pillai}, Guillem {Anglada}, and Paul T.~P. {Ho}.
\newblock {What Is Controlling the Fragmentation in the Infrared Dark Cloud G14.225-0.506?: Different Levels of Fragmentation in Twin Hubs.}
\newblock \emph{\apj}, 819\penalty0 (2):\penalty0 139, March 2016.
\newblock \doi{10.3847/0004-637X/819/2/139}.

\bibitem[{Contreras} et~al.(2013){Contreras}, {Schuller}, {Urquhart}, {Csengeri}, {Wyrowski}, {Beuther}, {Bontemps}, {Bronfman}, {Henning}, {Menten}, {Schilke}, {Walmsley}, {Wienen}, {Tackenberg}, and {Linz}]{2013A&A...549A..45C}
Y.~{Contreras}, F.~{Schuller}, J.~S. {Urquhart}, T.~{Csengeri}, F.~{Wyrowski}, H.~{Beuther}, S.~{Bontemps}, L.~{Bronfman}, T.~{Henning}, K.~M. {Menten}, P.~{Schilke}, C.~M. {Walmsley}, M.~{Wienen}, J.~{Tackenberg}, and H.~{Linz}.
\newblock {ATLASGAL - compact source catalogue: 330{\textdegree} < {\ensuremath{\ell}} < 21{\textdegree}}.
\newblock \emph{\aap}, 549:\penalty0 A45, January 2013.
\newblock \doi{10.1051/0004-6361/201220155}.

\bibitem[{Cornwell}(2008)]{2008ISTSP...2..793C}
T.~J. {Cornwell}.
\newblock {Multiscale CLEAN Deconvolution of Radio Synthesis Images}.
\newblock \emph{IEEE Journal of Selected Topics in Signal Processing}, 2\penalty0 (5):\penalty0 793--801, Nov 2008.
\newblock \doi{10.1109/JSTSP.2008.2006388}.

\bibitem[{Eden} et~al.(2019){Eden}, {Liu}, {Kim}, {Juvela}, {Liu}, {Tatematsu}, {Francesco}, {Wang}, {Wu}, {Thompson}, {Fuller}, {Li}, {Ristorcelli}, {Kang}, {Hirano}, {Johnstone}, {Lin}, {He}, {Koch}, {Sanhueza}, {Qin}, {Zhang}, {Goldsmith}, {Evans}, {Yuan}, {Zhang}, {White}, {Choi}, {Lee}, {Toth}, {Mairs}, {Yi}, {Tang}, {Soam}, {Peretto}, {Samal}, {Fich}, {Parsons}, {Malinen}, {Bendo}, {Rivera-Ingraham}, {Liu}, {Wouterloot}, {Li}, {Qian}, {Rawlings}, {Rawlings}, {Feng}, {Wang}, {Li}, {Liu}, {Luo}, {Marston}, {Pattle}, {Pelkonen}, {Rigby}, {Zahorecz}, {Zhang}, {B{\H{o}}gner}, {Aikawa}, {Akhter}, {Alina}, {Bell}, {Bernard}, {Blain}, {Bronfman}, {Byun}, {Chapman}, {Chen}, {Chen}, {Chen}, {Chen}, {Chen}, {Chrysostomou}, {Chu}, {Chung}, {Cornu}, {Cosentino}, {Cunningham}, {Demyk}, {Drabek-Maunder}, {Doi}, {Eswaraiah}, {Falgarone}, {Feh{\'e}r}, {Fraser}, {Friberg}, {Garay}, {Ge}, {Gear}, {Greaves}, {Guan}, {Harvey-Smith}, {Hasegawa}, {He}, {Henkel}, {Hirota}, {Holland}, {Hughes}, {Jarken}, {Ji}, {Jimenez-Serra},
  {Kang}, {Kawabata}, {Kim}, {Kim}, {Kim}, {Kim}, {Koo}, {Kwon}, {Kuan}, {Lacaille}, {Lai}, {Lee}, {Lee}, {Lee}, {Li}, {Lo}, {Lopez}, {Lu}, {Lyo}, {Mardones}, {McGehee}, {Meng}, {Montier}, {Montillaud}, {Moore}, {Morata}, {Moriarty-Schieven}, {Ohashi}, {Pak}, {Park}, {Paladini}, {Pech}, {Qiu}, {Ren}, {Richer}, {Sakai}, {Shang}, {Shinnaga}, {Stamatellos}, {Tang}, {Traficante}, {Vastel}, {Viti}, {Walsh}, {Wang}, {Wang}, {Ward-Thompson}, {Whitworth}, {Wilson}, {Xu}, {Yang}, {Yuan}, {Yuan}, {Zavagno}, {Zhang}, {Zhang}, {Zhang}, {Zhou}, {Zhou}, {Zhu}, and {Zuo}]{2019MNRAS.485.2895E}
D.~J. {Eden}, Tie {Liu}, Kee-Tae {Kim}, M.~{Juvela}, S.~Y. {Liu}, K.~{Tatematsu}, J.~Di {Francesco}, K.~{Wang}, Y.~{Wu}, M.~A. {Thompson}, G.~A. {Fuller}, Di~{Li}, I.~{Ristorcelli}, Sung-ju {Kang}, N.~{Hirano}, D.~{Johnstone}, Y.~{Lin}, J.~H. {He}, P.~M. {Koch}, Patricio {Sanhueza}, S.~L. {Qin}, Q.~{Zhang}, P.~F. {Goldsmith}, N.~J. {Evans}, J.~{Yuan}, C.~P. {Zhang}, G.~J. {White}, Minho {Choi}, Chang~Won {Lee}, L.~V. {Toth}, S.~{Mairs}, H.~W. {Yi}, M.~{Tang}, A.~{Soam}, N.~{Peretto}, M.~R. {Samal}, M.~{Fich}, H.~{Parsons}, J.~{Malinen}, G.~J. {Bendo}, A.~{Rivera-Ingraham}, H.~L. {Liu}, J.~{Wouterloot}, P.~S. {Li}, L.~{Qian}, J.~{Rawlings}, M.~G. {Rawlings}, S.~{Feng}, B.~{Wang}, Dalei {Li}, M.~{Liu}, G.~{Luo}, A.~P. {Marston}, K.~M. {Pattle}, V.~M. {Pelkonen}, A.~J. {Rigby}, S.~{Zahorecz}, G.~{Zhang}, R.~{B{\H{o}}gner}, Y.~{Aikawa}, S.~{Akhter}, D.~{Alina}, G.~{Bell}, J.~P. {Bernard}, A.~{Blain}, L.~{Bronfman}, D.~Y. {Byun}, S.~{Chapman}, H.~R. {Chen}, M.~{Chen}, W.~P. {Chen}, X.~{Chen}, Xuepeng {Chen},
  A.~{Chrysostomou}, Y.~H. {Chu}, E.~J. {Chung}, D.~{Cornu}, G.~{Cosentino}, M.~R. {Cunningham}, K.~{Demyk}, E.~{Drabek-Maunder}, Y.~{Doi}, C.~{Eswaraiah}, E.~{Falgarone}, O.~{Feh{\'e}r}, H.~{Fraser}, P.~{Friberg}, G.~{Garay}, J.~X. {Ge}, W.~K. {Gear}, J.~{Greaves}, X.~{Guan}, L.~{Harvey-Smith}, T.~{Hasegawa}, Y.~{He}, C.~{Henkel}, T.~{Hirota}, W.~{Holland}, A.~{Hughes}, E.~{Jarken}, T.~G. {Ji}, I.~{Jimenez-Serra}, M.~{Kang}, K.~S. {Kawabata}, Gwanjeong {Kim}, Jungha {Kim}, Jongsoo {Kim}, S.~{Kim}, B.~C. {Koo}, Woojin {Kwon}, Y.~J. {Kuan}, K.~M. {Lacaille}, S.~P. {Lai}, C.~F. {Lee}, J.~E. {Lee}, Y.~U. {Lee}, H.~{Li}, N.~{Lo}, J.~A.~P. {Lopez}, X.~{Lu}, A.~R. {Lyo}, D.~{Mardones}, P.~{McGehee}, F.~{Meng}, L.~{Montier}, J.~{Montillaud}, T.~J.~T. {Moore}, O.~{Morata}, G.~H. {Moriarty-Schieven}, S.~{Ohashi}, S.~{Pak}, Geumsook {Park}, R.~{Paladini}, G.~{Pech}, K.~{Qiu}, Z.~Y. {Ren}, J.~{Richer}, T.~{Sakai}, H.~{Shang}, H.~{Shinnaga}, D.~{Stamatellos}, Y.~W. {Tang}, A.~{Traficante}, C.~{Vastel}, S.~{Viti},
  A.~{Walsh}, H.~{Wang}, J.~{Wang}, D.~{Ward-Thompson}, A.~{Whitworth}, C.~D. {Wilson}, Y.~{Xu}, J.~{Yang}, Y.~L. {Yuan}, L.~{Yuan}, A.~{Zavagno}, C.~{Zhang}, G.~{Zhang}, H.~W. {Zhang}, C.~{Zhou}, J.~{Zhou}, L.~{Zhu}, and P.~{Zuo}.
\newblock {SCOPE: SCUBA-2 Continuum Observations of Pre-protostellar Evolution - survey description and compact source catalogue}.
\newblock \emph{\mnras}, 485\penalty0 (2):\penalty0 2895--2908, May 2019.
\newblock \doi{10.1093/mnras/stz574}.

\bibitem[{Ellsworth-Bowers} et~al.(2015{\natexlab{a}}){Ellsworth-Bowers}, {Glenn}, {Riley}, {Rosolowsky}, {Ginsburg}, {Evans}, {Bally}, {Battersby}, {Shirley}, and {Merello}]{EB2015BGPSPhy}
Timothy~P. {Ellsworth-Bowers}, Jason {Glenn}, Allyssa {Riley}, Erik {Rosolowsky}, Adam {Ginsburg}, II~{Evans}, Neal~J., John {Bally}, Cara {Battersby}, Yancy~L. {Shirley}, and Manuel {Merello}.
\newblock {The Bolocam Galactic Plane Survey. XIII. Physical Properties and Mass Functions of Dense Molecular Cloud Structures}.
\newblock \emph{\apj}, 805\penalty0 (2):\penalty0 157, June 2015{\natexlab{a}}.
\newblock \doi{10.1088/0004-637X/805/2/157}.

\bibitem[{Ellsworth-Bowers} et~al.(2015{\natexlab{b}}){Ellsworth-Bowers}, {Rosolowsky}, {Glenn}, {Ginsburg}, {Evans}, {Battersby}, {Shirley}, and {Svoboda}]{EB2015BGPSdistance}
Timothy~P. {Ellsworth-Bowers}, Erik {Rosolowsky}, Jason {Glenn}, Adam {Ginsburg}, II~{Evans}, Neal~J., Cara {Battersby}, Yancy~L. {Shirley}, and Brian {Svoboda}.
\newblock {The Bolocam Galactic Plane Survey. XII. Distance Catalog Expansion Using Kinematic Isolation of Dense Molecular Cloud Structures with $^{13}$CO(1-0)}.
\newblock \emph{\apj}, 799\penalty0 (1):\penalty0 29, January 2015{\natexlab{b}}.
\newblock \doi{10.1088/0004-637X/799/1/29}.

\bibitem[{Feng} et~al.(2016){Feng}, {Beuther}, {Zhang}, {Liu}, {Zhang}, {Wang}, and {Qiu}]{FengSY2016}
Siyi {Feng}, Henrik {Beuther}, Qizhou {Zhang}, Hauyu~Baobab {Liu}, Zhiyu {Zhang}, Ke~{Wang}, and Keping {Qiu}.
\newblock {Outflow Detection in a 70 {\ensuremath{\mu}}m Dark High-Mass Core}.
\newblock \emph{\apj}, 828\penalty0 (2):\penalty0 100, September 2016.
\newblock \doi{10.3847/0004-637X/828/2/100}.

\bibitem[{Fontani} et~al.(2018){Fontani}, {Commer{\c{c}}on}, {Giannetti}, {Beltr{\'a}n}, {S{\'a}nchez-Monge}, {Testi}, {Brand}, and {Tan}]{2018A&A...615A..94F}
F.~{Fontani}, B.~{Commer{\c{c}}on}, A.~{Giannetti}, M.~T. {Beltr{\'a}n}, {\'A}.~{S{\'a}nchez-Monge}, L.~{Testi}, J.~{Brand}, and J.~C. {Tan}.
\newblock {Fragmentation properties of massive protocluster gas clumps: an ALMA study}.
\newblock \emph{\aap}, 615:\penalty0 A94, July 2018.
\newblock \doi{10.1051/0004-6361/201832672}.

\bibitem[{Ge} et~al.(2023){Ge}, {Wang}, {Duarte-Cabral}, {Pettitt}, {Dobbs}, {S{\'a}nchez-Monge}, {Neralwar}, {Urquhart}, {Colombo}, {Dur{\'a}n-Camacho}, {Beuther}, {Bronfman}, {Rigby}, {Eden}, {Neupane}, {Barnes}, {Henning}, and {Yang}]{GeYF2023}
Y.~{Ge}, K.~{Wang}, A.~{Duarte-Cabral}, A.~R. {Pettitt}, C.~L. {Dobbs}, A.~{S{\'a}nchez-Monge}, K.~R. {Neralwar}, J.~S. {Urquhart}, D.~{Colombo}, E.~{Dur{\'a}n-Camacho}, H.~{Beuther}, L.~{Bronfman}, A.~J. {Rigby}, D.~{Eden}, S.~{Neupane}, P.~{Barnes}, T.~{Henning}, and A.~Y. {Yang}.
\newblock {Large-scale velocity-coherent filaments in the SEDIGISM survey: Association with spiral arms and the fraction of dense gas}.
\newblock \emph{\aap}, 675:\penalty0 A119, July 2023.
\newblock \doi{10.1051/0004-6361/202245784}.

\bibitem[{Ge} and {Wang}(2022)]{GeYF2022FL}
Yifei {Ge} and Ke~{Wang}.
\newblock {A Census of 163 Large-scale ({\ensuremath{\geq}}10 pc), Velocity-coherent Filaments in the Inner Galactic Plane: Physical Properties, Dense-gas Fraction, and Association with Spiral Arms}.
\newblock \emph{\apjs}, 259\penalty0 (2):\penalty0 36, April 2022.
\newblock \doi{10.3847/1538-4365/ac4a76}.

\bibitem[{Ginsburg} et~al.(2022){Ginsburg}, {Sokolov}, {de Val-Borro}, {Rosolowsky}, {Pineda}, {Sip{\H{o}}cz}, and {Henshaw}]{2022AJ....163..291G}
Adam {Ginsburg}, Vlas {Sokolov}, Miguel {de Val-Borro}, Erik {Rosolowsky}, Jaime~E. {Pineda}, Brigitta~M. {Sip{\H{o}}cz}, and Jonathan~D. {Henshaw}.
\newblock {Pyspeckit: A Spectroscopic Analysis and Plotting Package}.
\newblock \emph{\aj}, 163\penalty0 (6):\penalty0 291, June 2022.
\newblock \doi{10.3847/1538-3881/ac695a}.

\bibitem[{Henshaw} et~al.(2014){Henshaw}, {Caselli}, {Fontani}, {Jim{\'e}nez-Serra}, and {Tan}]{2014MNRAS.440.2860H}
J.~D. {Henshaw}, P.~{Caselli}, F.~{Fontani}, I.~{Jim{\'e}nez-Serra}, and J.~C. {Tan}.
\newblock {The dynamical properties of dense filaments in the infrared dark cloud G035.39-00.33}.
\newblock \emph{\mnras}, 440\penalty0 (3):\penalty0 2860--2881, May 2014.
\newblock \doi{10.1093/mnras/stu446}.

\bibitem[{Ho} and {Townes}(1983)]{HoInterstellar}
P.~T.~P. {Ho} and C.~H. {Townes}.
\newblock {Interstellar ammonia.}
\newblock \emph{\araa}, 21:\penalty0 239--270, January 1983.
\newblock \doi{10.1146/annurev.aa.21.090183.001323}.

\bibitem[{Jijina} et~al.(1999){Jijina}, {Myers}, and {Adams}]{1999ApJS..125..161J}
J.~{Jijina}, P.~C. {Myers}, and Fred~C. {Adams}.
\newblock {Dense Cores Mapped in Ammonia: A Database}.
\newblock \emph{\apjs}, 125\penalty0 (1):\penalty0 161--236, November 1999.
\newblock \doi{10.1086/313268}.

\bibitem[{Kong} et~al.(2018){Kong}, {Tan}, {Caselli}, {Fontani}, {Wang}, and {Butler}]{2018ApJ...867...94K}
Shuo {Kong}, Jonathan~C. {Tan}, Paola {Caselli}, Francesco {Fontani}, Ke~{Wang}, and Michael~J. {Butler}.
\newblock {Zooming in to Massive Star Birth}.
\newblock \emph{\apj}, 867\penalty0 (2):\penalty0 94, November 2018.
\newblock \doi{10.3847/1538-4357/aae1b2}.

\bibitem[{Kong} et~al.(2021){Kong}, {Arce}, {Shirley}, and {Glasgow}]{2021ApJ...912..156K}
Shuo {Kong}, H{\'e}ctor~G. {Arce}, Yancy {Shirley}, and Colton {Glasgow}.
\newblock {Evidence of Core Growth in the Dragon Infrared Dark Cloud: A Path for Massive Star Formation}.
\newblock \emph{\apj}, 912\penalty0 (2):\penalty0 156, May 2021.
\newblock \doi{10.3847/1538-4357/abefe7}.

\bibitem[{Larson}(1981)]{1981MNRAS.194..809L}
R.~B. {Larson}.
\newblock {Turbulence and star formation in molecular clouds.}
\newblock \emph{\mnras}, 194:\penalty0 809--826, March 1981.
\newblock \doi{10.1093/mnras/194.4.809}.

\bibitem[{Laws} et~al.(2019){Laws}, {Hora}, and {Zhang}]{2019ApJ...876...70L}
Anna S.~E. {Laws}, Joseph~L. {Hora}, and Qizhou {Zhang}.
\newblock {Massive Young Stellar Objects and Outflow in the Infrared Dark Cloud G79.3+0.3}.
\newblock \emph{\apj}, 876\penalty0 (1):\penalty0 70, May 2019.
\newblock \doi{10.3847/1538-4357/ab1423}.

\bibitem[{Lekht} et~al.(2009){Lekht}, {Slysh}, and {Krasnov}]{2009ARep...53..420L}
E.~E. {Lekht}, V.~I. {Slysh}, and V.~V. {Krasnov}.
\newblock {Variability of the spectrum and spatial structure of the H$_{2}$O maser in W75N-VLA 1}.
\newblock \emph{Astronomy Reports}, 53\penalty0 (5):\penalty0 420--427, May 2009.
\newblock \doi{10.1134/S1063772909050059}.

\bibitem[{Li} et~al.(2013){Li}, {Kauffmann}, {Zhang}, and {Chen}]{2013ApJ...768L...5L}
D.~{Li}, J.~{Kauffmann}, Q.~{Zhang}, and W.~{Chen}.
\newblock {Massive Quiescent Cores in Orion: Dynamical State Revealed by High-resolution Ammonia Maps}.
\newblock \emph{\apjl}, 768\penalty0 (1):\penalty0 L5, May 2013.
\newblock \doi{10.1088/2041-8205/768/1/L5}.

\bibitem[{Li} et~al.(2020){Li}, {Zhang}, {Liu}, {Beuther}, {Palau}, {Girart}, {Smith}, {Hora}, {Lin}, {Qiu}, {Strom}, {Wang}, {Li}, and {Yue}]{2020ApJ...896..110L}
Shanghuo {Li}, Qizhou {Zhang}, Hauyu~Baobab {Liu}, Henrik {Beuther}, Aina {Palau}, Josep~Miquel {Girart}, Howard {Smith}, Joseph~L. {Hora}, Yuxing {Lin}, Keping {Qiu}, Shaye {Strom}, Junzhi {Wang}, Fei {Li}, and Nannan {Yue}.
\newblock {ALMA Observations of NGC 6334S. I. Forming Massive Stars and Clusters in Subsonic and Transonic Filamentary Clouds}.
\newblock \emph{\apj}, 896\penalty0 (2):\penalty0 110, June 2020.
\newblock \doi{10.3847/1538-4357/ab84f1}.

\bibitem[{Li} et~al.(2022){Li}, {Sanhueza}, {Lee}, {Zhang}, {Beuther}, {Palau}, {Liu}, {Smith}, {Liu}, {Jim{\'e}nez-Serra}, {Kim}, {Feng}, {Liu}, {Wang}, {Li}, {Qiu}, {Lu}, {Girart}, {Wang}, {Li}, {Li}, {Cao}, {Kim}, and {Strom}]{2022ApJ...926..165L}
Shanghuo {Li}, Patricio {Sanhueza}, Chang~Won {Lee}, Qizhou {Zhang}, Henrik {Beuther}, Aina {Palau}, Hong-Li {Liu}, Howard~A. {Smith}, Hauyu~Baobab {Liu}, Izaskun {Jim{\'e}nez-Serra}, Kee-Tae {Kim}, Siyi {Feng}, Tie {Liu}, Junzhi {Wang}, Di~{Li}, Keping {Qiu}, Xing {Lu}, Josep~Miquel {Girart}, Ke~{Wang}, Fei {Li}, Juan {Li}, Yue {Cao}, Shinyoung {Kim}, and Shaye {Strom}.
\newblock {ALMA Observations of NGC 6334S. II. Subsonic and Transonic Narrow Filaments in a High-mass Star Formation Cloud}.
\newblock \emph{\apj}, 926\penalty0 (2):\penalty0 165, February 2022.
\newblock \doi{10.3847/1538-4357/ac3df8}.

\bibitem[{Li} et~al.(2023){Li}, {Sanhueza}, {Zhang}, {Guido}, {Sabatini}, {Morii}, {Lu}, {Tafoya}, {Nakamura}, {Izumi}, {Tatematsu}, and {Li}]{2023ApJ...949..109L}
Shanghuo {Li}, Patricio {Sanhueza}, Qizhou {Zhang}, Garay {Guido}, Giovanni {Sabatini}, Kaho {Morii}, Xing {Lu}, Daniel {Tafoya}, Fumitaka {Nakamura}, Natsuko {Izumi}, Ken'ichi {Tatematsu}, and Fei {Li}.
\newblock {The ALMA Survey of 70 {\ensuremath{\mu}}m Dark High-mass Clumps in Early Stages (ASHES). VIII. Dynamics of Embedded Dense Cores}.
\newblock \emph{\apj}, 949\penalty0 (2):\penalty0 109, June 2023.
\newblock \doi{10.3847/1538-4357/acc58f}.

\bibitem[{Liu} et~al.(2012){Liu}, {Quintana-Lacaci}, {Wang}, {Ho}, {Li}, {Zhang}, and {Zhang}]{2012ApJ...745...61L}
Hauyu~Baobab {Liu}, Guillermo {Quintana-Lacaci}, Ke~{Wang}, Paul T.~P. {Ho}, Zhi-Yun {Li}, Qizhou {Zhang}, and Zhi-Yu {Zhang}.
\newblock {The Origin of OB Clusters: From 10 pc to 0.1 pc}.
\newblock \emph{\apj}, 745\penalty0 (1):\penalty0 61, January 2012.
\newblock \doi{10.1088/0004-637X/745/1/61}.

\bibitem[{Liu} et~al.(2013){Liu}, {Ho}, {Wright}, {Su}, {Hsieh}, {Sun}, {Kim}, and {Minh}]{2013ApJ...770...44L}
Hauyu~Baobab {Liu}, Paul T.~P. {Ho}, Melvyn C.~H. {Wright}, Yu-Nung {Su}, Pei-Ying {Hsieh}, Ai-Lei {Sun}, Sungsoo~S. {Kim}, and Young~Chol {Minh}.
\newblock {Interstellar Medium Processing in the Inner 20 pc in Galactic Center}.
\newblock \emph{\apj}, 770\penalty0 (1):\penalty0 44, June 2013.
\newblock \doi{10.1088/0004-637X/770/1/44}.

\bibitem[{Liu} et~al.(2019){Liu}, {Chen}, {Rom{\'a}n-Z{\'u}{\~n}iga}, {Galv{\'a}n-Madrid}, {Ginsburg}, {Ho}, {Minh}, {Jim{\'e}nez-Serra}, {Testi}, and {Zhang}]{2019ApJ...871..185L}
Hauyu~Baobab {Liu}, Huei-Ru~Vivien {Chen}, Carlos~G. {Rom{\'a}n-Z{\'u}{\~n}iga}, Roberto {Galv{\'a}n-Madrid}, Adam {Ginsburg}, Paul T.~P. {Ho}, Young~Chol {Minh}, Izaskun {Jim{\'e}nez-Serra}, Leonardo {Testi}, and Qizhou {Zhang}.
\newblock {Investigating Fragmentation of Gas Structures in OB Cluster-forming Molecular Clump G33.92+0.11 with 1000 au Resolution Observations of ALMA}.
\newblock \emph{\apj}, 871\penalty0 (2):\penalty0 185, February 2019.
\newblock \doi{10.3847/1538-4357/aaf6b4}.

\bibitem[{Liu} et~al.(2016){Liu}, {Zhang}, {Kim}, {Wu}, {Lee}, {Goldsmith}, {Li}, {Liu}, {Chen}, {Tatematsu}, {Wang}, {Lee}, {Qin}, {Mardones}, and {Cho}]{LiuT2016outflow}
Tie {Liu}, Qizhou {Zhang}, Kee-Tae {Kim}, Yuefang {Wu}, Chang-Won {Lee}, Paul~F. {Goldsmith}, Di~{Li}, Sheng-Yuan {Liu}, Huei-Ru {Chen}, Ken'ichi {Tatematsu}, Ke~{Wang}, Jeong-Eun {Lee}, Sheng-Li {Qin}, Diego {Mardones}, and Se-Hyung {Cho}.
\newblock {Discovery of an Extremely Wide-angle Bipolar Outflow in AFGL 5142}.
\newblock \emph{\apj}, 824\penalty0 (1):\penalty0 31, June 2016.
\newblock \doi{10.3847/0004-637X/824/1/31}.

\bibitem[Lu et~al.(2014)Lu, Zhang, Liu, Wang, and Gu]{Lu_2014}
Xing Lu, Qizhou Zhang, Hauyu~Baobab Liu, Junzhi Wang, and Qiusheng Gu.
\newblock {VERY} {LARGE} {ARRAY} {OBSERVATIONS} {OF} {AMMONIA} {IN} {HIGH}-{MASS} {STAR} {FORMATION} {REGIONS}.
\newblock \emph{The Astrophysical Journal}, 790\penalty0 (2):\penalty0 84, jul 2014.
\newblock \doi{10.1088/0004-637x/790/2/84}.
\newblock URL \url{https://doi.org/10.1088\%2F0004-637x\%2F790\%2F2\%2F84}.

\bibitem[{Lu} et~al.(2015){Lu}, {Zhang}, {Wang}, and {Gu}]{LuX2015IRDC}
Xing {Lu}, Qizhou {Zhang}, Ke~{Wang}, and Qiusheng {Gu}.
\newblock {Initial Fragmentation in the Infrared Dark Cloud G28.53-0.25}.
\newblock \emph{\apj}, 805\penalty0 (2):\penalty0 171, June 2015.
\newblock \doi{10.1088/0004-637X/805/2/171}.

\bibitem[Lu et~al.(2018)Lu, Zhang, Liu, Sanhueza, Tatematsu, Feng, Smith, Myers, Sridharan, and Gu]{Lu_2018}
Xing Lu, Qizhou Zhang, Hauyu~Baobab Liu, Patricio Sanhueza, Ken'ichi Tatematsu, Siyi Feng, Howard~A. Smith, Philip~C. Myers, T.~K. Sridharan, and Qiusheng Gu.
\newblock Filamentary fragmentation and accretion in high-mass star-forming molecular clouds.
\newblock \emph{The Astrophysical Journal}, 855\penalty0 (1):\penalty0 9, feb 2018.
\newblock \doi{10.3847/1538-4357/aaad11}.
\newblock URL \url{https://doi.org/10.3847\%2F1538-4357\%2Faaad11}.

\bibitem[{McKee} and {Tan}(2003)]{2003ApJ...585..850M}
Christopher~F. {McKee} and Jonathan~C. {Tan}.
\newblock {The Formation of Massive Stars from Turbulent Cores}.
\newblock \emph{\apj}, 585\penalty0 (2):\penalty0 850--871, Mar 2003.
\newblock \doi{10.1086/346149}.

\bibitem[{McMullin} et~al.(2007){McMullin}, {Waters}, {Schiebel}, {Young}, and {Golap}]{2007ASPC..376..127M}
J.~P. {McMullin}, B.~{Waters}, D.~{Schiebel}, W.~{Young}, and K.~{Golap}.
\newblock {CASA Architecture and Applications}.
\newblock In R.~A. {Shaw}, F.~{Hill}, and D.~J. {Bell}, editors, \emph{Astronomical Data Analysis Software and Systems XVI}, volume 376 of \emph{Astronomical Society of the Pacific Conference Series}, page 127, October 2007.

\bibitem[{Monsch} et~al.(2018){Monsch}, {Pineda}, {Liu}, {Zucker}, {How-Huan Chen}, {Pattle}, {Offner}, {Di Francesco}, {Ginsburg}, {Ercolano}, {Arce}, {Friesen}, {Kirk}, {Caselli}, and {Goodman}]{2018ApJ...861...77M}
Kristina {Monsch}, Jaime~E. {Pineda}, Hauyu~Baobab {Liu}, Catherine {Zucker}, Hope {How-Huan Chen}, Kate {Pattle}, Stella S.~R. {Offner}, James {Di Francesco}, Adam {Ginsburg}, Barbara {Ercolano}, H{\'e}ctor~G. {Arce}, Rachel {Friesen}, Helen {Kirk}, Paola {Caselli}, and Alyssa~A. {Goodman}.
\newblock {Dense Gas Kinematics and a Narrow Filament in the Orion A OMC1 Region Using NH$_{3}$}.
\newblock \emph{\apj}, 861\penalty0 (2):\penalty0 77, July 2018.
\newblock \doi{10.3847/1538-4357/aac8da}.

\bibitem[{Morii} et~al.(2021){Morii}, {Sanhueza}, {Nakamura}, {Jackson}, {Li}, {Beuther}, {Zhang}, {Feng}, {Tafoya}, {Guzm{\'a}n}, {Izumi}, {Sakai}, {Lu}, {Tatematsu}, {Ohashi}, {Silva}, {Olguin}, and {Contreras}]{2021ApJ...923..147M}
Kaho {Morii}, Patricio {Sanhueza}, Fumitaka {Nakamura}, James~M. {Jackson}, Shanghuo {Li}, Henrik {Beuther}, Qizhou {Zhang}, Siyi {Feng}, Daniel {Tafoya}, Andr{\'e}s~E. {Guzm{\'a}n}, Natsuko {Izumi}, Takeshi {Sakai}, Xing {Lu}, Ken'ichi {Tatematsu}, Satoshi {Ohashi}, Andrea {Silva}, Fernando~A. {Olguin}, and Yanett {Contreras}.
\newblock {The ALMA Survey of 70 {\ensuremath{\mu}}m Dark High-mass Clumps in Early Stages (ASHES). IV. Star Formation Signatures in G023.477}.
\newblock \emph{\apj}, 923\penalty0 (2):\penalty0 147, December 2021.
\newblock \doi{10.3847/1538-4357/ac2365}.

\bibitem[{Morii} et~al.(2023){Morii}, {Sanhueza}, {Nakamura}, {Zhang}, {Sabatini}, {Beuther}, {Lu}, {Li}, {Garay}, {Jackson}, {Olguin}, {Tafoya}, {Tatematsu}, {Izumi}, {Sakai}, and {Silva}]{Morii2023ASHES}
Kaho {Morii}, Patricio {Sanhueza}, Fumitaka {Nakamura}, Qizhou {Zhang}, Giovanni {Sabatini}, Henrik {Beuther}, Xing {Lu}, Shanghuo {Li}, Guido {Garay}, James~M. {Jackson}, Fernando~A. {Olguin}, Daniel {Tafoya}, Ken'ichi {Tatematsu}, Natsuko {Izumi}, Takeshi {Sakai}, and Andrea {Silva}.
\newblock {The ALMA Survey of 70 $\mu$m Dark High-mass Clumps in Early Stages (ASHES). IX. Physical Properties and Spatial Distribution of Cores in IRDCs}.
\newblock \emph{arXiv e-prints}, art. arXiv:2304.01757, April 2023.
\newblock \doi{10.48550/arXiv.2304.01757}.

\bibitem[{Myers}(1983)]{1983ApJ...270..105M}
P.~C. {Myers}.
\newblock {Dense cores in dark clouds. III. Subsonic turbulence.}
\newblock \emph{\apj}, 270:\penalty0 105--118, July 1983.
\newblock \doi{10.1086/161101}.

\bibitem[{Myers}(2009)]{2009ApJ...700.1609M}
Philip~C. {Myers}.
\newblock {Filamentary Structure of Star-forming Complexes}.
\newblock \emph{\apj}, 700\penalty0 (2):\penalty0 1609--1625, August 2009.
\newblock \doi{10.1088/0004-637X/700/2/1609}.

\bibitem[{Ohashi} et~al.(2016){Ohashi}, {Sanhueza}, {Chen}, {Zhang}, {Busquet}, {Nakamura}, {Palau}, and {Tatematsu}]{2016ApJ...833..209O}
Satoshi {Ohashi}, Patricio {Sanhueza}, Huei-Ru~Vivien {Chen}, Qizhou {Zhang}, Gemma {Busquet}, Fumitaka {Nakamura}, Aina {Palau}, and Ken'ichi {Tatematsu}.
\newblock {Dense Core Properties in the Infrared Dark Cloud G14.225-0.506 Revealed by ALMA}.
\newblock \emph{\apj}, 833\penalty0 (2):\penalty0 209, December 2016.
\newblock \doi{10.3847/1538-4357/833/2/209}.

\bibitem[{Padoan} et~al.(2020){Padoan}, {Pan}, {Juvela}, {Haugb{\o}lle}, and {Nordlund}]{2020ApJ...900...82P}
Paolo {Padoan}, Liubin {Pan}, Mika {Juvela}, Troels {Haugb{\o}lle}, and {\r{A}}ke {Nordlund}.
\newblock {The Origin of Massive Stars: The Inertial-inflow Model}.
\newblock \emph{\apj}, 900\penalty0 (1):\penalty0 82, September 2020.
\newblock \doi{10.3847/1538-4357/abaa47}.

\bibitem[{Palau} et~al.(2015){Palau}, {Ballesteros-Paredes}, {V{\'a}zquez-Semadeni}, {S{\'a}nchez-Monge}, {Estalella}, {Fall}, {Zapata}, {Camacho}, {G{\'o}mez}, {Naranjo-Romero}, {Busquet}, and {Fontani}]{2015MNRAS.453.3785P}
Aina {Palau}, Javier {Ballesteros-Paredes}, Enrique {V{\'a}zquez-Semadeni}, {\'A}lvaro {S{\'a}nchez-Monge}, Robert {Estalella}, S.~Michael {Fall}, Luis~A. {Zapata}, Vianey {Camacho}, Laura {G{\'o}mez}, Ra{\'u}l {Naranjo-Romero}, Gemma {Busquet}, and Francesco {Fontani}.
\newblock {Gravity or turbulence? - III. Evidence of pure thermal Jeans fragmentation at {\ensuremath{\sim}}0.1 pc scale}.
\newblock \emph{\mnras}, 453\penalty0 (4):\penalty0 3785--3797, November 2015.
\newblock \doi{10.1093/mnras/stv1834}.

\bibitem[{Pelkonen} et~al.(2021){Pelkonen}, {Padoan}, {Haugb{\o}lle}, and {Nordlund}]{2021MNRAS.504.1219P}
V.~M. {Pelkonen}, P.~{Padoan}, T.~{Haugb{\o}lle}, and {\r{A}}.~{Nordlund}.
\newblock {From the CMF to the IMF: beyond the core-collapse model}.
\newblock \emph{\mnras}, 504\penalty0 (1):\penalty0 1219--1236, June 2021.
\newblock \doi{10.1093/mnras/stab844}.

\bibitem[{Peretto} et~al.(2013){Peretto}, {Fuller}, {Duarte-Cabral}, {Avison}, {Hennebelle}, {Pineda}, {Andr{\'e}}, {Bontemps}, {Motte}, {Schneider}, and {Molinari}]{2013A&A...555A.112P}
N.~{Peretto}, G.~A. {Fuller}, A.~{Duarte-Cabral}, A.~{Avison}, P.~{Hennebelle}, J.~E. {Pineda}, Ph. {Andr{\'e}}, S.~{Bontemps}, F.~{Motte}, N.~{Schneider}, and S.~{Molinari}.
\newblock {Global collapse of molecular clouds as a formation mechanism for the most massive stars}.
\newblock \emph{\aap}, 555:\penalty0 A112, July 2013.
\newblock \doi{10.1051/0004-6361/201321318}.

\bibitem[{Pillai} et~al.(2007){Pillai}, {Wyrowski}, {Hatchell}, {Gibb}, and {Thompson}]{2007A&A...467..207P}
T.~{Pillai}, F.~{Wyrowski}, J.~{Hatchell}, A.~G. {Gibb}, and M.~A. {Thompson}.
\newblock {Probing the initial conditions of high mass star formation. I. Deuteration and depletion in high mass pre/protocluster clumps}.
\newblock \emph{\aap}, 467\penalty0 (1):\penalty0 207--216, May 2007.
\newblock \doi{10.1051/0004-6361:20065682}.

\bibitem[{Pillai} et~al.(2019){Pillai}, {Kauffmann}, {Zhang}, {Sanhueza}, {Leurini}, {Wang}, {Sridharan}, and {K{\"o}nig}]{Pillai2019}
Thushara {Pillai}, Jens {Kauffmann}, Qizhou {Zhang}, Patricio {Sanhueza}, Silvia {Leurini}, Ke~{Wang}, T.~K. {Sridharan}, and Carsten {K{\"o}nig}.
\newblock {Massive and low-mass protostars in massive ``starless'' cores}.
\newblock \emph{\aap}, 622:\penalty0 A54, February 2019.
\newblock \doi{10.1051/0004-6361/201732570}.

\bibitem[{Redaelli} et~al.(2022){Redaelli}, {Bovino}, {Sanhueza}, {Morii}, {Sabatini}, {Caselli}, {Giannetti}, and {Li}]{2022ApJ...936..169R}
Elena {Redaelli}, Stefano {Bovino}, Patricio {Sanhueza}, Kaho {Morii}, Giovanni {Sabatini}, Paola {Caselli}, Andrea {Giannetti}, and Shanghuo {Li}.
\newblock {The Core Population and Kinematics of a Massive Clump at Early Stages: An Atacama Large Millimeter/submillimeter Array View}.
\newblock \emph{\apj}, 936\penalty0 (2):\penalty0 169, September 2022.
\newblock \doi{10.3847/1538-4357/ac85b4}.

\bibitem[{Rizzo} et~al.(2014){Rizzo}, {Palau}, {Jim{\'e}nez-Esteban}, and {Henkel}]{2014A&A...564A..21R}
J.~R. {Rizzo}, Aina {Palau}, F.~{Jim{\'e}nez-Esteban}, and C.~{Henkel}.
\newblock {Ammonia observations in the LBV nebula G79.29+0.46. Discovery of a cold ring and some warm spots}.
\newblock \emph{\aap}, 564:\penalty0 A21, April 2014.
\newblock \doi{10.1051/0004-6361/201323170}.

\bibitem[{Rosolowsky} et~al.(2008){Rosolowsky}, {Pineda}, {Foster}, {Borkin}, {Kauffmann}, {Caselli}, {Myers}, and {Goodman}]{2008ApJS..175..509R}
E.~W. {Rosolowsky}, J.~E. {Pineda}, J.~B. {Foster}, M.~A. {Borkin}, J.~{Kauffmann}, P.~{Caselli}, P.~C. {Myers}, and A.~A. {Goodman}.
\newblock {An Ammonia Spectral Atlas of Dense Cores in Perseus}.
\newblock \emph{\apjs}, 175\penalty0 (2):\penalty0 509--521, April 2008.
\newblock \doi{10.1086/524299}.

\bibitem[{Sabatini} et~al.(2022){Sabatini}, {Bovino}, {Sanhueza}, {Morii}, {Li}, {Redaelli}, {Zhang}, {Lu}, {Feng}, {Tafoya}, {Izumi}, {Sakai}, {Tatematsu}, and {Allingham}]{2022ApJ...936...80S}
Giovanni {Sabatini}, Stefano {Bovino}, Patricio {Sanhueza}, Kaho {Morii}, Shanghuo {Li}, Elena {Redaelli}, Qizhou {Zhang}, Xing {Lu}, Siyi {Feng}, Daniel {Tafoya}, Natsuko {Izumi}, Takeshi {Sakai}, Ken'ichi {Tatematsu}, and David {Allingham}.
\newblock {The ALMA Survey of 70 {\ensuremath{\mu}}m Dark High-mass Clumps in Early Stages (ASHES). VI. The Core-scale CO Depletion}.
\newblock \emph{\apj}, 936\penalty0 (1):\penalty0 80, September 2022.
\newblock \doi{10.3847/1538-4357/ac83aa}.

\bibitem[{S{\'a}nchez-Monge} et~al.(2013){S{\'a}nchez-Monge}, {Palau}, {Fontani}, {Busquet}, {Ju{\'a}rez}, {Estalella}, {Tan}, {Sep{\'u}lveda}, {Ho}, {Zhang}, and {Kurtz}]{2013MNRAS.432.3288S}
{\'A}lvaro {S{\'a}nchez-Monge}, Aina {Palau}, Francesco {Fontani}, Gemma {Busquet}, Carmen {Ju{\'a}rez}, Robert {Estalella}, Jonathan~C. {Tan}, Inma {Sep{\'u}lveda}, Paul T.~P. {Ho}, Qizhou {Zhang}, and Stan {Kurtz}.
\newblock {Properties of dense cores in clustered massive star-forming regions at high angular resolution}.
\newblock \emph{\mnras}, 432\penalty0 (4):\penalty0 3288--3319, July 2013.
\newblock \doi{10.1093/mnras/stt679}.

\bibitem[{Sanhueza} et~al.(2012){Sanhueza}, {Jackson}, {Foster}, {Garay}, {Silva}, and {Finn}]{2012ApJ...756...60S}
Patricio {Sanhueza}, James~M. {Jackson}, Jonathan~B. {Foster}, Guido {Garay}, Andrea {Silva}, and Susanna~C. {Finn}.
\newblock {Chemistry in Infrared Dark Cloud Clumps: A Molecular Line Survey at 3 mm}.
\newblock \emph{\apj}, 756\penalty0 (1):\penalty0 60, September 2012.
\newblock \doi{10.1088/0004-637X/756/1/60}.

\bibitem[{Sanhueza} et~al.(2013){Sanhueza}, {Jackson}, {Foster}, {Jimenez-Serra}, {Dirienzo}, and {Pillai}]{2013ApJ...773..123S}
Patricio {Sanhueza}, James~M. {Jackson}, Jonathan~B. {Foster}, Izaskun {Jimenez-Serra}, William~J. {Dirienzo}, and Thushara {Pillai}.
\newblock {Distinct Chemical Regions in the ``Prestellar'' Infrared Dark Cloud G028.23-00.19}.
\newblock \emph{\apj}, 773\penalty0 (2):\penalty0 123, August 2013.
\newblock \doi{10.1088/0004-637X/773/2/123}.

\bibitem[{Sanhueza} et~al.(2017){Sanhueza}, {Jackson}, {Zhang}, {Guzm{\'a}n}, {Lu}, {Stephens}, {Wang}, and {Tatematsu}]{2017ApJ...841...97S}
Patricio {Sanhueza}, James~M. {Jackson}, Qizhou {Zhang}, Andr{\'e}s~E. {Guzm{\'a}n}, Xing {Lu}, Ian~W. {Stephens}, Ke~{Wang}, and Ken'ichi {Tatematsu}.
\newblock {A Massive Prestellar Clump Hosting No High-mass Cores}.
\newblock \emph{\apj}, 841\penalty0 (2):\penalty0 97, June 2017.
\newblock \doi{10.3847/1538-4357/aa6ff8}.

\bibitem[{Sanhueza} et~al.(2019){Sanhueza}, {Contreras}, {Wu}, {Jackson}, {Guzm{\'a}n}, {Zhang}, {Li}, {Lu}, {Silva}, {Izumi}, {Liu}, {Miura}, {Tatematsu}, {Sakai}, {Beuther}, {Garay}, {Ohashi}, {Saito}, {Nakamura}, {Saigo}, {Veena}, {Nguyen-Luong}, and {Tafoya}]{2019ApJ...886..102S}
Patricio {Sanhueza}, Yanett {Contreras}, Benjamin {Wu}, James~M. {Jackson}, Andr{\'e}s~E. {Guzm{\'a}n}, Qizhou {Zhang}, Shanghuo {Li}, Xing {Lu}, Andrea {Silva}, Natsuko {Izumi}, Tie {Liu}, Rie~E. {Miura}, Ken'ichi {Tatematsu}, Takeshi {Sakai}, Henrik {Beuther}, Guido {Garay}, Satoshi {Ohashi}, Masao {Saito}, Fumitaka {Nakamura}, Kazuya {Saigo}, V.~S. {Veena}, Quang {Nguyen-Luong}, and Daniel {Tafoya}.
\newblock {The ALMA Survey of 70 {\ensuremath{\mu}}m Dark High-mass Clumps in Early Stages (ASHES). I. Pilot Survey: Clump Fragmentation}.
\newblock \emph{\apj}, 886\penalty0 (2):\penalty0 102, December 2019.
\newblock \doi{10.3847/1538-4357/ab45e9}.

\bibitem[{Sanhueza} et~al.(2021){Sanhueza}, {Girart}, {Padovani}, {Galli}, {Hull}, {Zhang}, {Cortes}, {Stephens}, {Fern{\'a}ndez-L{\'o}pez}, {Jackson}, {Frau}, {Kock}, {Wu}, {Zapata}, {Olguin}, {Lu}, {Silva}, {Tang}, {Sakai}, {Guzm{\'a}n}, {Tatematsu}, {Nakamura}, and {Chen}]{2021ApJ...915L..10S}
Patricio {Sanhueza}, Josep~Miquel {Girart}, Marco {Padovani}, Daniele {Galli}, Charles L.~H. {Hull}, Qizhou {Zhang}, Paulo {Cortes}, Ian~W. {Stephens}, Manuel {Fern{\'a}ndez-L{\'o}pez}, James~M. {Jackson}, Pau {Frau}, Patrick~M. {Kock}, Benjamin {Wu}, Luis~A. {Zapata}, Fernando {Olguin}, Xing {Lu}, Andrea {Silva}, Ya-Wen {Tang}, Takeshi {Sakai}, Andr{\'e}s~E. {Guzm{\'a}n}, Ken'ichi {Tatematsu}, Fumitaka {Nakamura}, and Huei-Ru~Vivien {Chen}.
\newblock {Gravity-driven Magnetic Field at 1000 au Scales in High-mass Star Formation}.
\newblock \emph{\apjl}, 915\penalty0 (1):\penalty0 L10, July 2021.
\newblock \doi{10.3847/2041-8213/ac081c}.

\bibitem[{Schuller} et~al.(2009){Schuller}, {Menten}, {Contreras}, {Wyrowski}, {Schilke}, {Bronfman}, {Henning}, {Walmsley}, {Beuther}, {Bontemps}, {Cesaroni}, {Deharveng}, {Garay}, {Herpin}, {Lefloch}, {Linz}, {Mardones}, {Minier}, {Molinari}, {Motte}, {Nyman}, {Reveret}, {Risacher}, {Russeil}, {Schneider}, {Testi}, {Troost}, {Vasyunina}, {Wienen}, {Zavagno}, {Kovacs}, {Kreysa}, {Siringo}, and {Wei{\ss}}]{2009A&A...504..415S}
F.~{Schuller}, K.~M. {Menten}, Y.~{Contreras}, F.~{Wyrowski}, P.~{Schilke}, L.~{Bronfman}, T.~{Henning}, C.~M. {Walmsley}, H.~{Beuther}, S.~{Bontemps}, R.~{Cesaroni}, L.~{Deharveng}, G.~{Garay}, F.~{Herpin}, B.~{Lefloch}, H.~{Linz}, D.~{Mardones}, V.~{Minier}, S.~{Molinari}, F.~{Motte}, L.~{\r{A}}. {Nyman}, V.~{Reveret}, C.~{Risacher}, D.~{Russeil}, N.~{Schneider}, L.~{Testi}, T.~{Troost}, T.~{Vasyunina}, M.~{Wienen}, A.~{Zavagno}, A.~{Kovacs}, E.~{Kreysa}, G.~{Siringo}, and A.~{Wei{\ss}}.
\newblock {ATLASGAL - The APEX telescope large area survey of the galaxy at 870 {\ensuremath{\mu}}m}.
\newblock \emph{\aap}, 504\penalty0 (2):\penalty0 415--427, September 2009.
\newblock \doi{10.1051/0004-6361/200811568}.

\bibitem[{Simon} et~al.(2006){Simon}, {Rathborne}, {Shah}, {Jackson}, and {Chambers}]{2006ApJ...653.1325S}
R.~{Simon}, J.~M. {Rathborne}, R.~Y. {Shah}, J.~M. {Jackson}, and E.~T. {Chambers}.
\newblock {The Characterization and Galactic Distribution of Infrared Dark Clouds}.
\newblock \emph{\apj}, 653\penalty0 (2):\penalty0 1325--1335, December 2006.
\newblock \doi{10.1086/508915}.

\bibitem[{Sokolov} et~al.(2017){Sokolov}, {Wang}, {Pineda}, {Caselli}, {Henshaw}, {Tan}, {Fontani}, {Jim{\'e}nez-Serra}, and {Lim}]{Sokolov2017}
Vlas {Sokolov}, Ke~{Wang}, Jaime~E. {Pineda}, Paola {Caselli}, Jonathan~D. {Henshaw}, Jonathan~C. {Tan}, Francesco {Fontani}, Izaskun {Jim{\'e}nez-Serra}, and Wanggi {Lim}.
\newblock {Temperature structure and kinematics of the IRDC G035.39-00.33}.
\newblock \emph{\aap}, 606:\penalty0 A133, October 2017.
\newblock \doi{10.1051/0004-6361/201630350}.

\bibitem[{Sokolov} et~al.(2018){Sokolov}, {Wang}, {Pineda}, {Caselli}, {Henshaw}, {Barnes}, {Tan}, {Fontani}, {Jim{\'e}nez-Serra}, and {Zhang}]{2018A&A...611L...3S}
Vlas {Sokolov}, Ke~{Wang}, Jaime~E. {Pineda}, Paola {Caselli}, Jonathan~D. {Henshaw}, Ashley~T. {Barnes}, Jonathan~C. {Tan}, Francesco {Fontani}, Izaskun {Jim{\'e}nez-Serra}, and Qizhou {Zhang}.
\newblock {Subsonic islands within a high-mass star-forming infrared dark cloud}.
\newblock \emph{\aap}, 611:\penalty0 L3, Mar 2018.
\newblock \doi{10.1051/0004-6361/201832746}.

\bibitem[{Spitzer Science Center (SSC)} and {Infrared Science Archive (IRSA)}(2021)]{2021yCat.2368....0S}
{Spitzer Science Center (SSC)} and {Infrared Science Archive (IRSA)}.
\newblock {VizieR Online Data Catalog: The Spitzer (SEIP) source list (SSTSL2) (Spitzer Science Center, 2021)}.
\newblock \emph{VizieR Online Data Catalog}, art. II/368, March 2021.

\bibitem[{Tackenberg} et~al.(2014){Tackenberg}, {Beuther}, {Henning}, {Linz}, {Sakai}, {Ragan}, {Krause}, {Nielbock}, {Hennemann}, {Pitann}, and {Schmiedeke}]{2014A&A...565A.101T}
J.~{Tackenberg}, H.~{Beuther}, Th. {Henning}, H.~{Linz}, T.~{Sakai}, S.~E. {Ragan}, O.~{Krause}, M.~{Nielbock}, M.~{Hennemann}, J.~{Pitann}, and A.~{Schmiedeke}.
\newblock {Kinematic structure of massive star-forming regions. I. Accretion along filaments}.
\newblock \emph{\aap}, 565:\penalty0 A101, May 2014.
\newblock \doi{10.1051/0004-6361/201321555}.

\bibitem[{Tan} et~al.(2013){Tan}, {Kong}, {Butler}, {Caselli}, and {Fontani}]{2013ApJ...779...96T}
Jonathan~C. {Tan}, Shuo {Kong}, Michael~J. {Butler}, Paola {Caselli}, and Francesco {Fontani}.
\newblock {The Dynamics of Massive Starless Cores with ALMA}.
\newblock \emph{\apj}, 779\penalty0 (2):\penalty0 96, December 2013.
\newblock \doi{10.1088/0004-637X/779/2/96}.

\bibitem[{Urquhart} et~al.(2011){Urquhart}, {Morgan}, {Figura}, {Moore}, {Lumsden}, {Hoare}, {Oudmaijer}, {Mottram}, {Davies}, and {Dunham}]{2011MNRAS.418.1689U}
J.~S. {Urquhart}, L.~K. {Morgan}, C.~C. {Figura}, T.~J.~T. {Moore}, S.~L. {Lumsden}, M.~G. {Hoare}, R.~D. {Oudmaijer}, J.~C. {Mottram}, B.~{Davies}, and M.~K. {Dunham}.
\newblock {The Red MSX Source survey: ammonia and water maser analysis of massive star-forming regions}.
\newblock \emph{\mnras}, 418\penalty0 (3):\penalty0 1689--1706, December 2011.
\newblock \doi{10.1111/j.1365-2966.2011.19594.x}.

\bibitem[{V{\'a}zquez-Semadeni} et~al.(2019){V{\'a}zquez-Semadeni}, {Palau}, {Ballesteros-Paredes}, {G{\'o}mez}, and {Zamora-Avil{\'e}s}]{2019MNRAS.490.3061V}
Enrique {V{\'a}zquez-Semadeni}, Aina {Palau}, Javier {Ballesteros-Paredes}, Gilberto~C. {G{\'o}mez}, and Manuel {Zamora-Avil{\'e}s}.
\newblock {Global hierarchical collapse in molecular clouds. Towards a comprehensive scenario}.
\newblock \emph{\mnras}, 490\penalty0 (3):\penalty0 3061--3097, December 2019.
\newblock \doi{10.1093/mnras/stz2736}.

\bibitem[{Walmsley} and {Ungerechts}(1983)]{1983A&A...122..164W}
C.~M. {Walmsley} and H.~{Ungerechts}.
\newblock {Ammonia as a molecular cloud thermometer.}
\newblock \emph{\aap}, 122:\penalty0 164--170, June 1983.

\bibitem[{Wang} and {Wang}(2023)]{2023A&A...674A..46W}
Chao {Wang} and Ke~{Wang}.
\newblock {Highly structured turbulence in high-mass star formation: An evolved infrared-dark cloud G35.20-0.74 N}.
\newblock \emph{\aap}, 674:\penalty0 A46, June 2023.
\newblock \doi{10.1051/0004-6361/202244525}.

\bibitem[{Wang}(2015)]{WangKe2015book}
Ke~{Wang}.
\newblock \emph{{The Earliest Stages of Massive Clustered Star Formation: Fragmentation of Infrared Dark Clouds}}.
\newblock 2015.
\newblock \doi{10.1007/978-3-662-44969-1}.

\bibitem[{Wang}(2018)]{2018RNAAS...2...52W}
Ke~{Wang}.
\newblock {Temperature, Turbulence, and Kinematics of the Giant Filamentary Infrared Dark Cloud G28.34+0.06 at {\ensuremath{\sim}}0.1{\,}pc Resolution}.
\newblock \emph{Research Notes of the American Astronomical Society}, 2\penalty0 (2):\penalty0 52, June 2018.
\newblock \doi{10.3847/2515-5172/aacb29}.

\bibitem[{Wang} et~al.(2011){Wang}, {Zhang}, {Wu}, and {Zhang}]{2011ApJ...735...64W}
Ke~{Wang}, Qizhou {Zhang}, Yuefang {Wu}, and Huawei {Zhang}.
\newblock {Hierarchical Fragmentation and Jet-like Outflows in IRDC G28.34+0.06: A Growing Massive Protostar Cluster}.
\newblock \emph{\apj}, 735\penalty0 (1):\penalty0 64, July 2011.
\newblock \doi{10.1088/0004-637X/735/1/64}.

\bibitem[{Wang} et~al.(2012){Wang}, {Zhang}, {Wu}, {Li}, and {Zhang}]{WangKe2012}
Ke~{Wang}, Qizhou {Zhang}, Yuefang {Wu}, Hua-bai {Li}, and Huawei {Zhang}.
\newblock {Protostellar Outflow Heating in a Growing Massive Protocluster}.
\newblock \emph{\apjl}, 745\penalty0 (2):\penalty0 L30, February 2012.
\newblock \doi{10.1088/2041-8205/745/2/L30}.

\bibitem[{Wang} et~al.(2014){Wang}, {Zhang}, {Testi}, {van der Tak}, {Wu}, {Zhang}, {Pillai}, {Wyrowski}, {Carey}, {Ragan}, and {Henning}]{2014MNRAS.439.3275W}
Ke~{Wang}, Qizhou {Zhang}, Leonardo {Testi}, Floris {van der Tak}, Yuefang {Wu}, Huawei {Zhang}, Thushara {Pillai}, Friedrich {Wyrowski}, Sean {Carey}, Sarah~E. {Ragan}, and Thomas {Henning}.
\newblock {Hierarchical fragmentation and differential star formation in the Galactic `Snake': infrared dark cloud G11.11-0.12}.
\newblock \emph{\mnras}, 439\penalty0 (4):\penalty0 3275--3293, April 2014.
\newblock \doi{10.1093/mnras/stu127}.

\bibitem[{Wang} et~al.(2015){Wang}, {Testi}, {Ginsburg}, {Walmsley}, {Molinari}, and {Schisano}]{2015MNRAS.450.4043W}
Ke~{Wang}, Leonardo {Testi}, Adam {Ginsburg}, C.~Malcolm {Walmsley}, Sergio {Molinari}, and Eugenio {Schisano}.
\newblock {Large-scale filaments associated with Milky Way spiral arms}.
\newblock \emph{\mnras}, 450\penalty0 (4):\penalty0 4043--4049, July 2015.
\newblock \doi{10.1093/mnras/stv735}.

\bibitem[{Wang} et~al.(2016){Wang}, {Testi}, {Burkert}, {Walmsley}, {Beuther}, and {Henning}]{wangke2016FL}
Ke~{Wang}, Leonardo {Testi}, Andreas {Burkert}, C.~Malcolm {Walmsley}, Henrik {Beuther}, and Thomas {Henning}.
\newblock {A Census of Large-scale ({\ensuremath{\geq}}10 PC), Velocity-coherent, Dense Filaments in the Northern Galactic Plane: Automated Identification Using Minimum Spanning Tree}.
\newblock \emph{\apjs}, 226\penalty0 (1):\penalty0 9, September 2016.
\newblock \doi{10.3847/0067-0049/226/1/9}.

\bibitem[{Wang} et~al.(2008){Wang}, {Zhang}, {Pillai}, {Wyrowski}, and {Wu}]{2008ApJ...672L..33W}
Y.~{Wang}, Q.~{Zhang}, T.~{Pillai}, F.~{Wyrowski}, and Y.~{Wu}.
\newblock {NH$_{3}$ Observations of the Infrared Dark Cloud G28.34+0.06}.
\newblock \emph{\apjl}, 672\penalty0 (1):\penalty0 L33, January 2008.
\newblock \doi{10.1086/524949}.

\bibitem[{Westerhout}(1958)]{1958BAN....14..215W}
G.~{Westerhout}.
\newblock {A survey of the continuous radiation from the Galactic System at a frequency of 1390 Mc/s}.
\newblock \emph{\bain}, 14:\penalty0 215, December 1958.

\bibitem[{Wienen} et~al.(2012){Wienen}, {Wyrowski}, {Schuller}, {Menten}, {Walmsley}, {Bronfman}, and {Motte}]{2012A&A...544A.146W}
M.~{Wienen}, F.~{Wyrowski}, F.~{Schuller}, K.~M. {Menten}, C.~M. {Walmsley}, L.~{Bronfman}, and F.~{Motte}.
\newblock {Ammonia from cold high-mass clumps discovered in the inner Galactic disk by the ATLASGAL survey}.
\newblock \emph{\aap}, 544:\penalty0 A146, August 2012.
\newblock \doi{10.1051/0004-6361/201118107}.

\bibitem[{Xu} et~al.(2023){Xu}, {Wang}, {Liu}, {Goldsmith}, {Zhang}, {Juvela}, {Liu}, {Qin}, {Li}, {Tej}, {Garay}, {Bronfman}, {Li}, {Wu}, {G{\'o}mez}, {V{\'a}zquez-Semadeni}, {Tatematsu}, {Ren}, {Zhang}, {Toth}, {Liu}, {Yue}, {Zhang}, {Baug}, {Issac}, {Stutz}, {Liu}, {Fuller}, {Tang}, {Zhang}, {Dewangan}, {Lee}, {Zhou}, {Xie}, {Jiao}, {Wang}, {Liu}, {Luo}, {Soam}, and {Eswaraiah}]{Xu2023SDC335}
Feng-Wei {Xu}, Ke~{Wang}, Tie {Liu}, Paul~F. {Goldsmith}, Qizhou {Zhang}, Mika {Juvela}, Hong-Li {Liu}, Sheng-Li {Qin}, Guang-Xing {Li}, Anandmayee {Tej}, Guido {Garay}, Leonardo {Bronfman}, Shanghuo {Li}, Yue-Fang {Wu}, Gilberto~C. {G{\'o}mez}, Enrique {V{\'a}zquez-Semadeni}, Ken'ichi {Tatematsu}, Zhiyuan {Ren}, Yong {Zhang}, L.~Viktor {Toth}, Xunchuan {Liu}, Nannan {Yue}, Siju {Zhang}, Tapas {Baug}, Namitha {Issac}, Amelia~M. {Stutz}, Meizhu {Liu}, Gary~A. {Fuller}, Mengyao {Tang}, Chao {Zhang}, Lokesh {Dewangan}, Chang~Won {Lee}, Jianwen {Zhou}, Jinjin {Xie}, Wenyu {Jiao}, Chao {Wang}, Rong {Liu}, Qiuyi {Luo}, Archana {Soam}, and Chakali {Eswaraiah}.
\newblock {ATOMS: ALMA Three-millimeter Observations of Massive Star-forming regions - XV. Steady accretion from global collapse to core feeding in massive hub-filament system SDC335}.
\newblock \emph{\mnras}, 520\penalty0 (3):\penalty0 3259--3285, April 2023.
\newblock \doi{10.1093/mnras/stad012}.

\bibitem[{Xu} et~al.(2011){Xu}, {Moscadelli}, {Reid}, {Menten}, {Zhang}, {Zheng}, and {Brunthaler}]{2011ApJ...733...25X}
Y.~{Xu}, L.~{Moscadelli}, M.~J. {Reid}, K.~M. {Menten}, B.~{Zhang}, X.~W. {Zheng}, and A.~{Brunthaler}.
\newblock {Trigonometric Parallaxes of Massive Star-forming Regions. VIII. G12.89+0.49, G15.03-0.68 (M17), and G27.36-0.16}.
\newblock \emph{\apj}, 733\penalty0 (1):\penalty0 25, May 2011.
\newblock \doi{10.1088/0004-637X/733/1/25}.

\bibitem[{Yuan} et~al.(2012){Yuan}, {Li}, {Huang}, {Hsia}, and {Miao}]{2012A&A...540A..95Y}
Jing-Hua {Yuan}, Jin~Zeng {Li}, Ya~Fang {Huang}, Chih-Hao {Hsia}, and Jingqi {Miao}.
\newblock {The discovery based on GLIMPSE data of a protostar driving a bipolar outflow}.
\newblock \emph{\aap}, 540:\penalty0 A95, April 2012.
\newblock \doi{10.1051/0004-6361/201117858}.

\bibitem[{Yue} et~al.(2021){Yue}, {Li}, {Zhang}, {Zhu}, {Henshaw}, {Mardones}, and {Ren}]{2021RAA....21...24Y}
Nan-Nan {Yue}, Di~{Li}, Qi-Zhou {Zhang}, Lei {Zhu}, Jonathan {Henshaw}, Diego {Mardones}, and Zhi-Yuan {Ren}.
\newblock {Resolution-dependent subsonic non-thermal line dispersion revealed by ALMA}.
\newblock \emph{Research in Astronomy and Astrophysics}, 21\penalty0 (1):\penalty0 024, January 2021.
\newblock \doi{10.1088/1674-4527/21/1/24}.

\end{thebibliography}


\begin{thebibliography}{}

  \bibitem[1966]{baker} Baker, N. 1966,
      in Stellar Evolution,
      ed.\ R. F. Stein,\& A. G. W. Cameron
      (Plenum, New York) 333

   \bibitem[1988]{balluch} Balluch, M. 1988,
      A\&A, 200, 58

   \bibitem[1980]{cox} Cox, J. P. 1980,
      Theory of Stellar Pulsation
      (Princeton University Press, Princeton) 165

   \bibitem[1969]{cox69} Cox, A. N.,\& Stewart, J. N. 1969,
      Academia Nauk, Scientific Information 15, 1

   \bibitem[1980]{mizuno} Mizuno H. 1980,
      Prog. Theor. Phys., 64, 544
   
   \bibitem[1987]{tscharnuter} Tscharnuter W. M. 1987,
      A\&A, 188, 55
  
   \bibitem[1992]{terlevich} Terlevich, R. 1992, in ASP Conf. Ser. 31, 
      Relationships between Active Galactic Nuclei and Starburst Galaxies, 
      ed. A. V. Filippenko, 13

   \bibitem[1980a]{yorke80a} Yorke, H. W. 1980a,
      A\&A, 86, 286

   \bibitem[1997]{zheng} Zheng, W., Davidsen, A. F., Tytler, D. \& Kriss, G. A.
      1997, preprint
\end{thebibliography}

\begin{appendix}

\section{Fitting of ammonia spectral lines}
\begin{figure*}[htb!]
    \centering
    \includegraphics[width=0.9\textwidth]{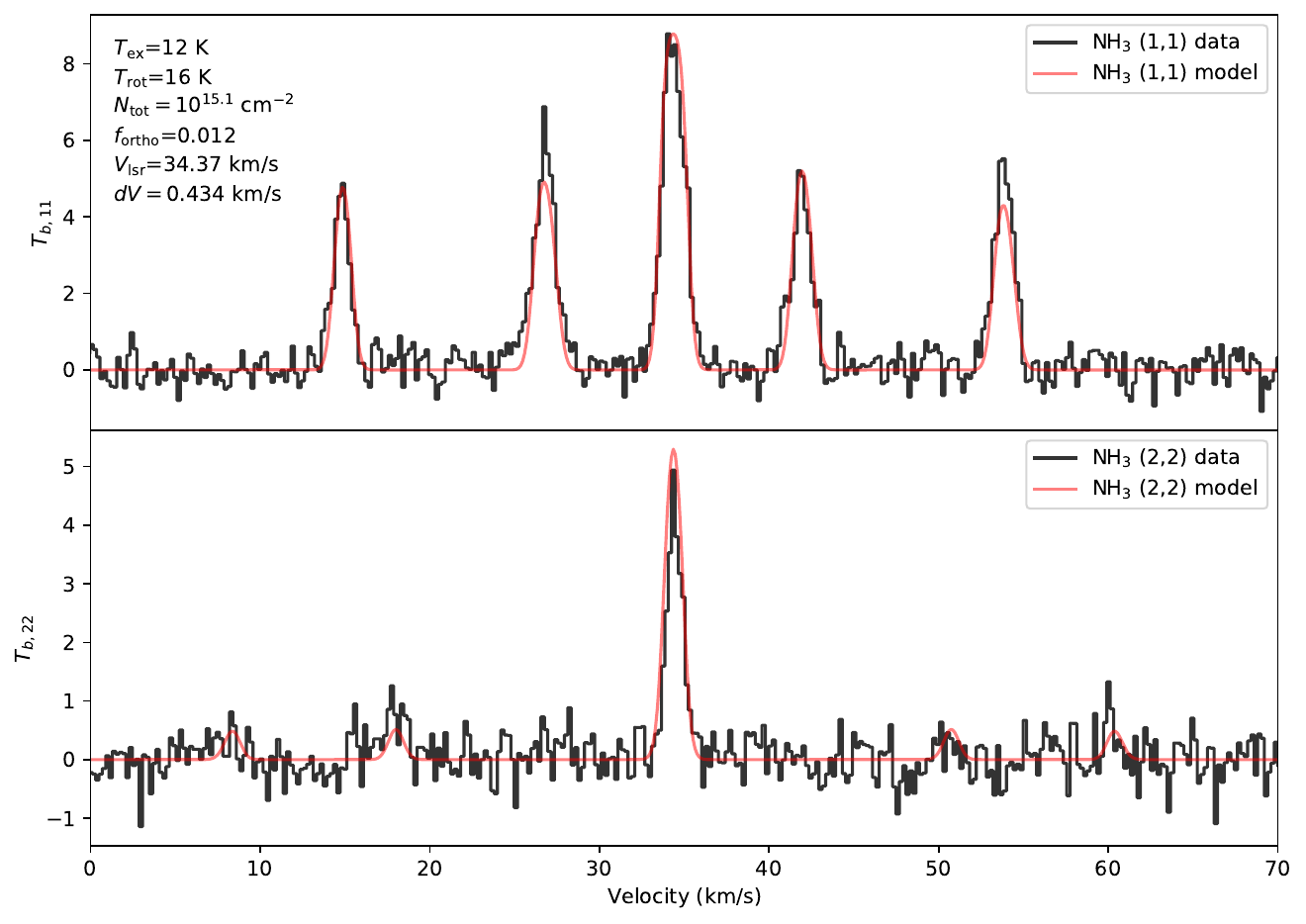}
    \caption{Raw data (black) and fitted model (red) of NH$_3$ (1,1) at the top and (2,2) at the bottom lines from the center of G48.64\_c1. Fitting results are noted in the upper left corner. }
    \label{nh3fit}
\end{figure*}

We use the PySpecKit \citep{2022AJ....163..291G} to fit the NH$_3$ (1,1) to (3.3) lines and here we present part of the result as the example. Figure. \ref{nh3fit} is the NH$_3$ (1,1) and (2,2) lines from the center of G48.64\_c1. The model well meets the raw data. From the integrated flux maps, we found that there exists elongated structures and negative absorption, but after the SNR checking, the data in those regions were not fitted due to insufficient S/N (less than 3), and did not impact the results further. 

\section{Fitting of the power law model}
\begin{figure*}[htb!]
    \centering
    \includegraphics[width=0.9\textwidth]{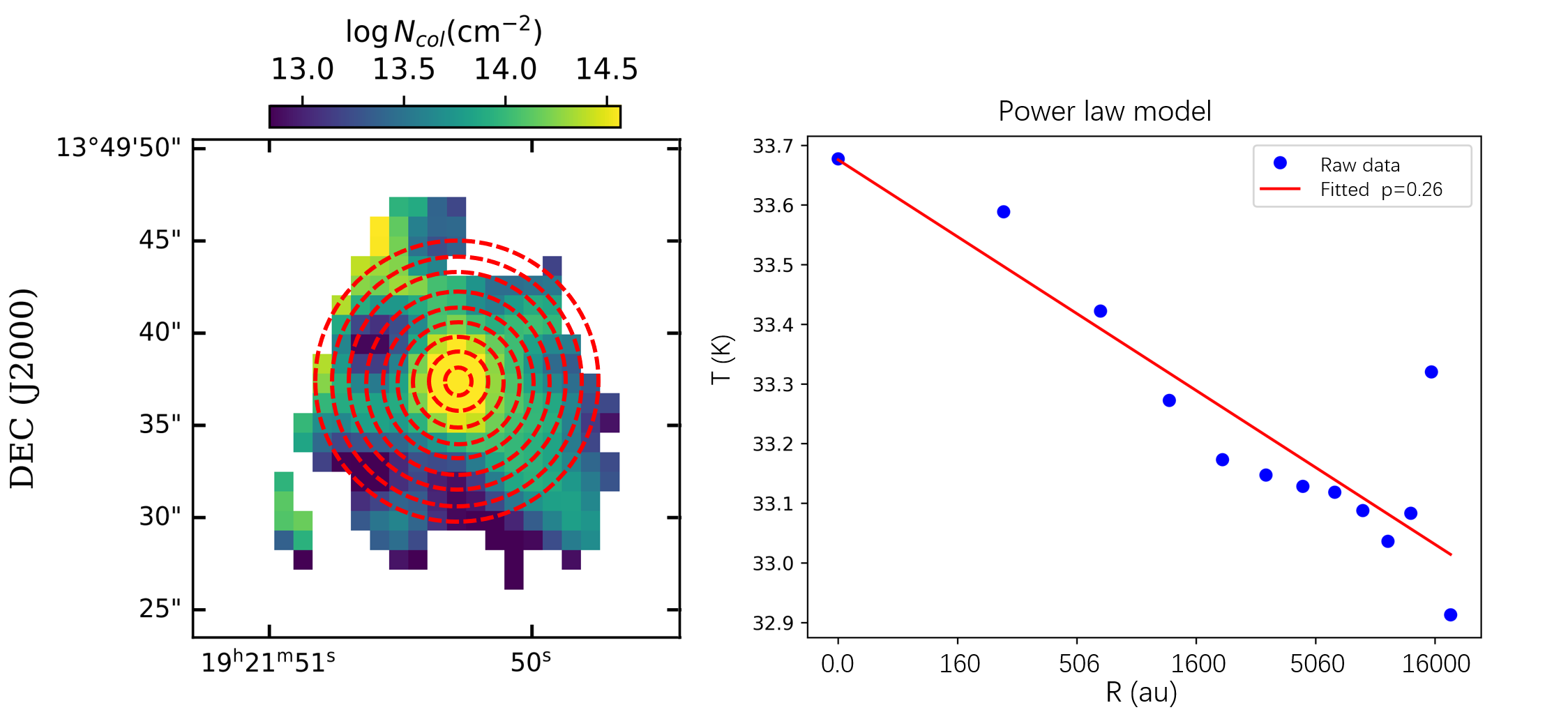}
    \caption{Sampling method of the fitted power-law model and fitting results (G48.65\_c1 as an example). \textbf{Left:} Column density distribution of G48\_c1, with the red dashed lines representing concentric circles centered on the peak value. \textbf{Right:} Raw data (blue dots) and fitted result (red line) when fitting the power-law model: ${N_{col}} \propto {r^{ - p}}$. }
    \label{mfitting}
\end{figure*}

We present the sampling method and one of the fitted result in Fig. \ref{mfitting}. First we obtain the peak position of temperature and column density of the core by fitting a Gaussian model. Then with this position as the center, and a step of half the beam size, we average the data at equal distances as the mean value at that radius. We then fit a power-law model (${N_{col}} \propto {r^{ - p}}$) to obtain the power-law index $p$ which is the parameter of the temperature and column density.

\section{Maps of all fitted parameters}
\begin{figure*}[htb!]
    \begin{minipage}[c]{1.0\linewidth}
        \centerline{\includegraphics[width=\textwidth]{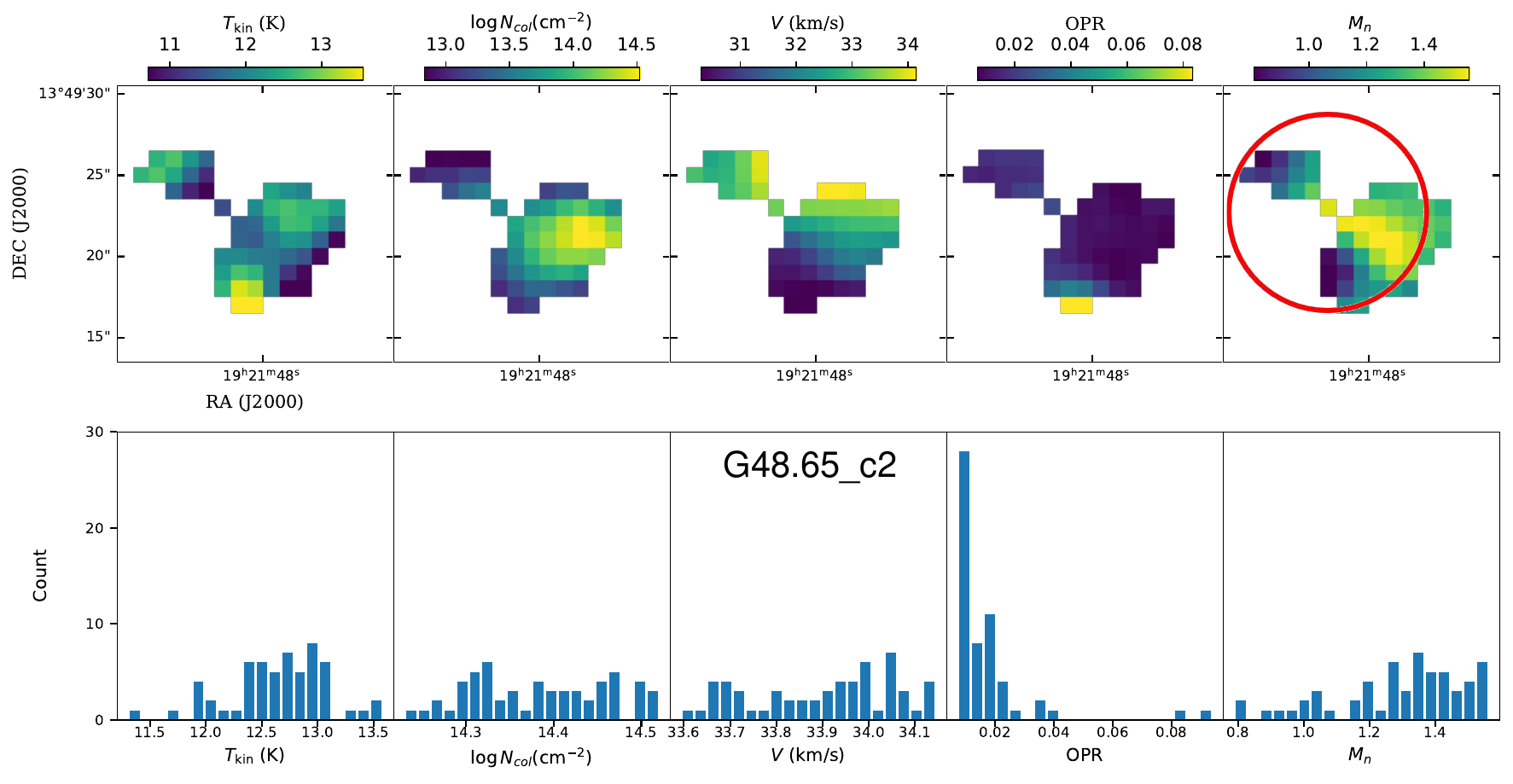}}
        \centerline{\includegraphics[width=\textwidth]{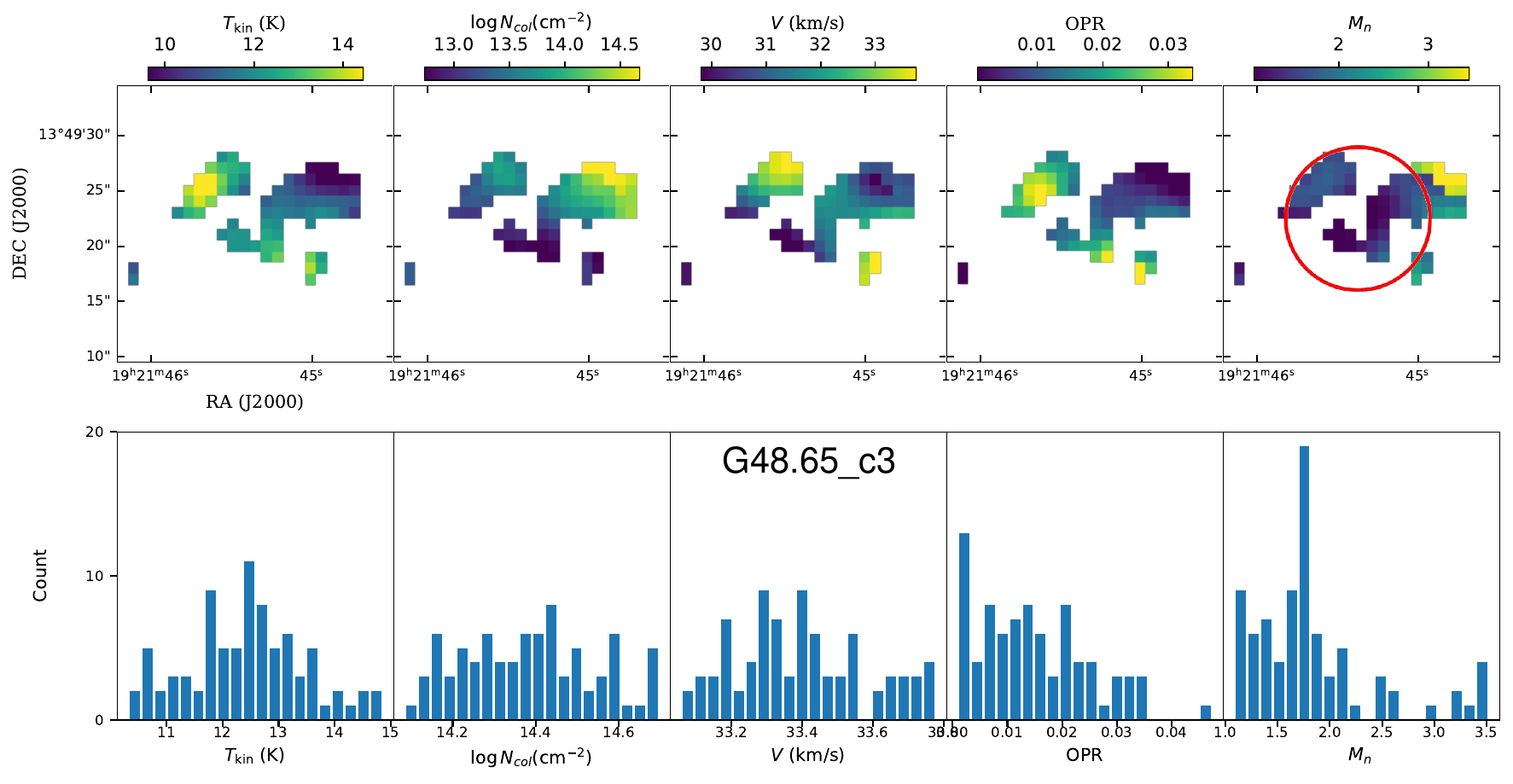}}
    \end{minipage}
    \caption{Maps and histograms of the fitted cores. {First row:}  Temperature (K), column density (in log scale), centroid velocity (km ${\rm{s}^{-1}}$), OPR,  and Mach number (from left
to right). The red circle is the sub-region we selected to study the core-averaged Mach number. \textbf{Second row:} Corresponding histograms of the parameter distribution among the emission regions.}
    \label{restresult1}
\end{figure*}

 Here, we present the maps and their histograms of fitted parameters of the rest of the cores. The setting of the Figure \ref{restresult1}  is the same sa it is to the Figure \ref{result} which shows the results of cores with the detection of the  NH$_3$ (3,3). Otherwise, the Figure \ref{restresult2} shows the results of cores without the detection of the NH$_3$ (3,3).


\addtocounter{figure}{-1}

\begin{figure*}[htb!]
    \begin{minipage}[c]{1.0\linewidth}
        \centerline{\includegraphics[width=\textwidth]{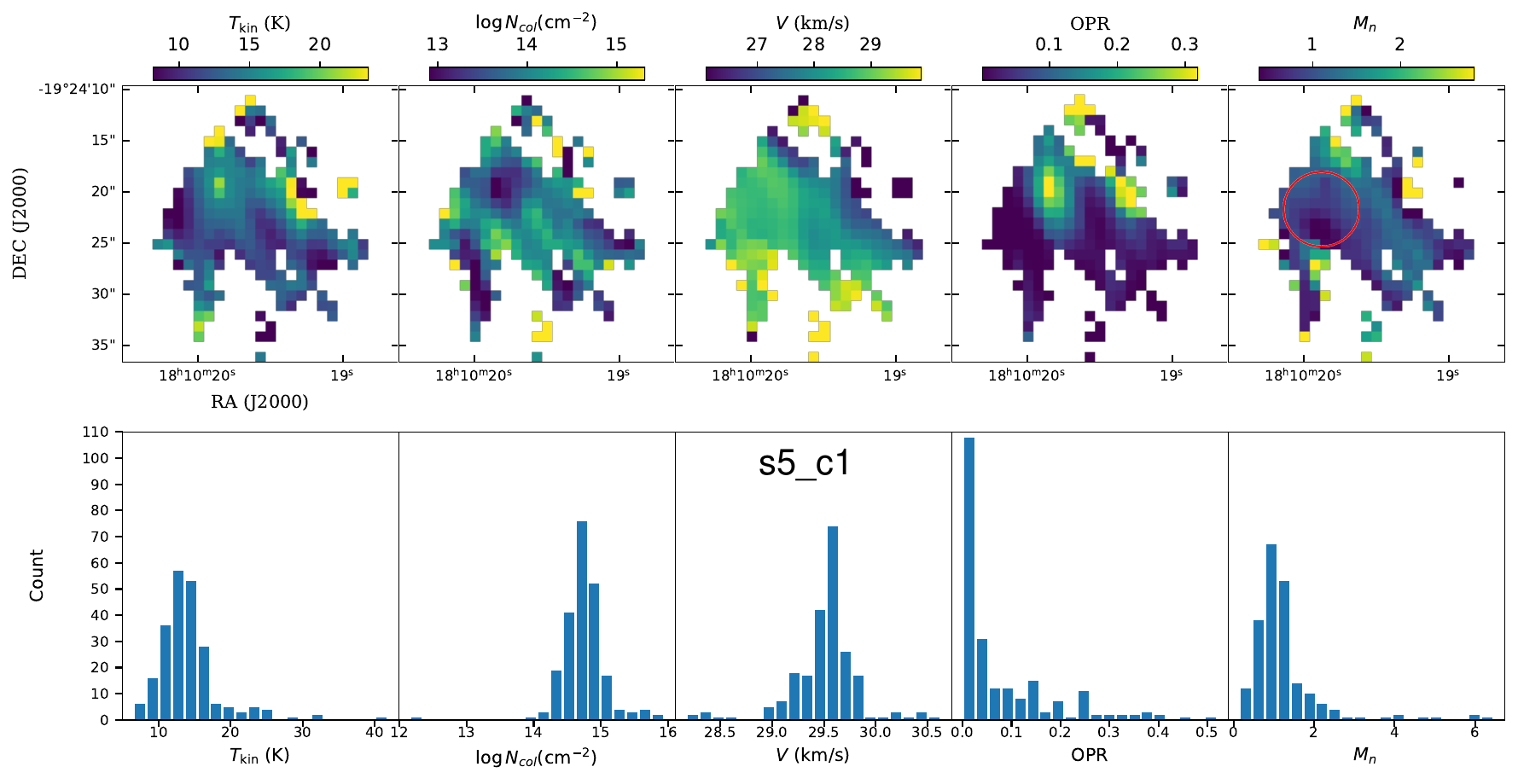}}
        \centerline{\includegraphics[width=\textwidth]{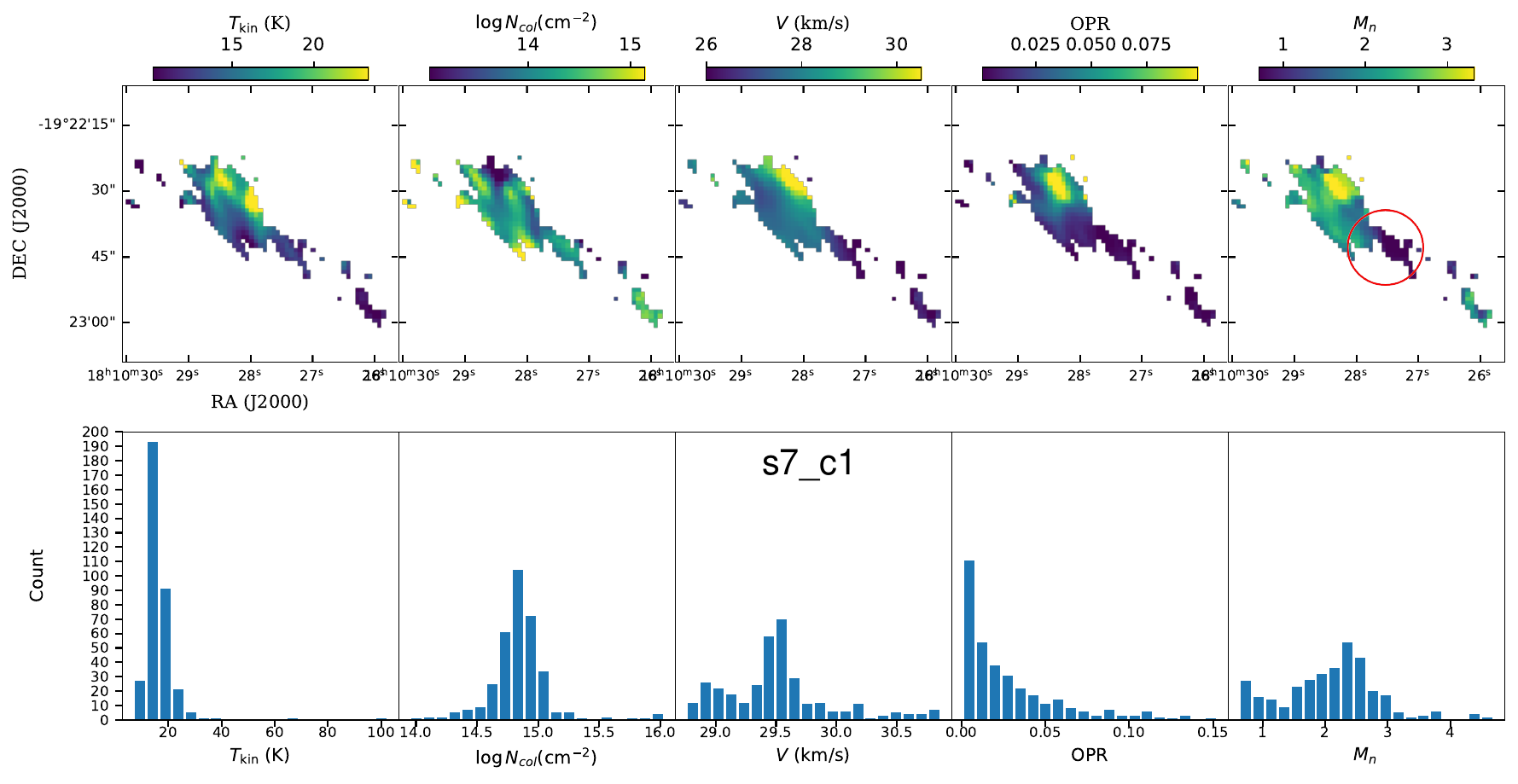}}
    \end{minipage}
    \caption{Maps and histograms of the fitted cores (continued).}
\end{figure*}
\addtocounter{figure}{-1}

\begin{figure*}[htb!]
    \begin{minipage}[c]{1.0\linewidth}
        \centerline{\includegraphics[width=\textwidth]{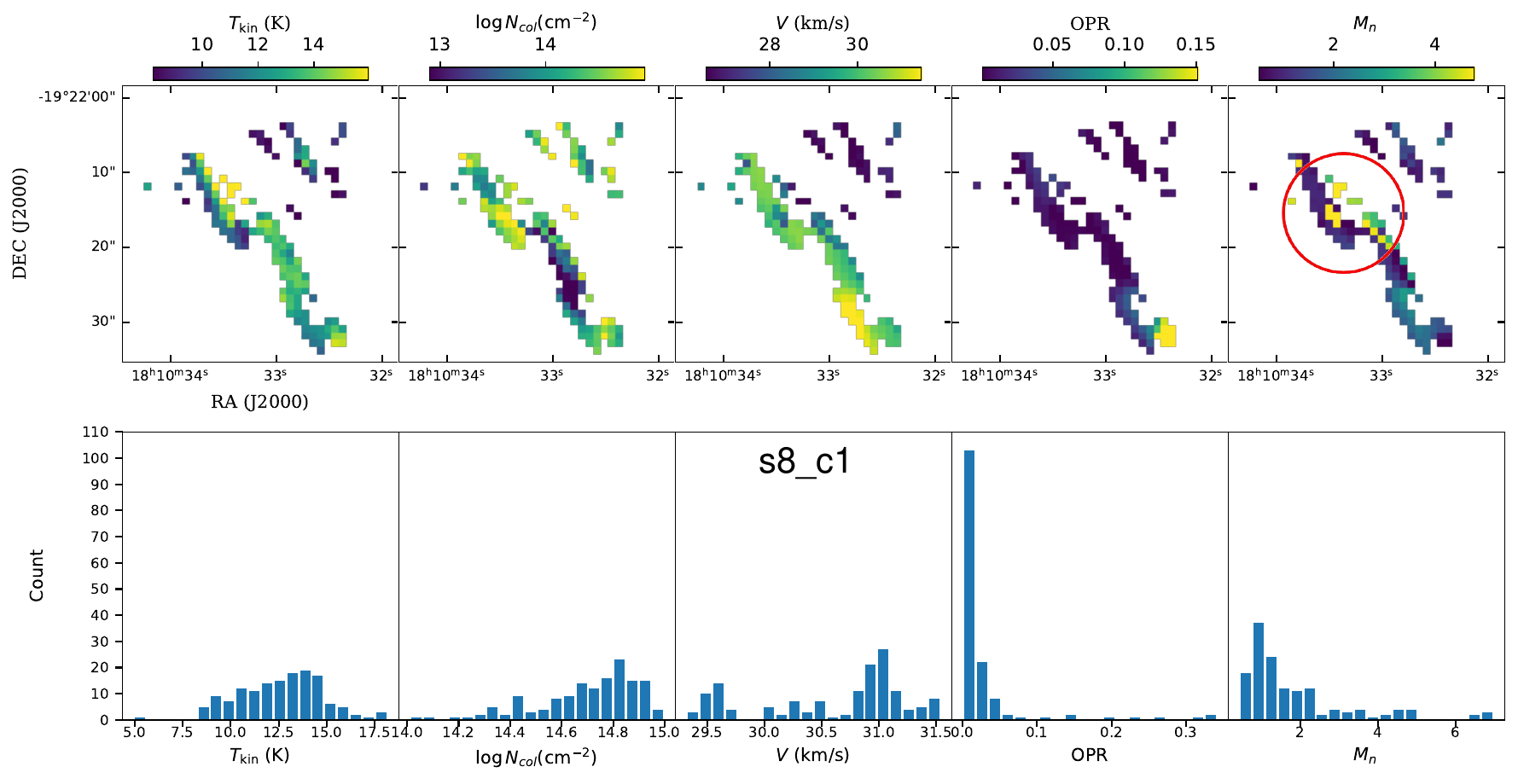}}
        \centerline{\includegraphics[width=\textwidth]{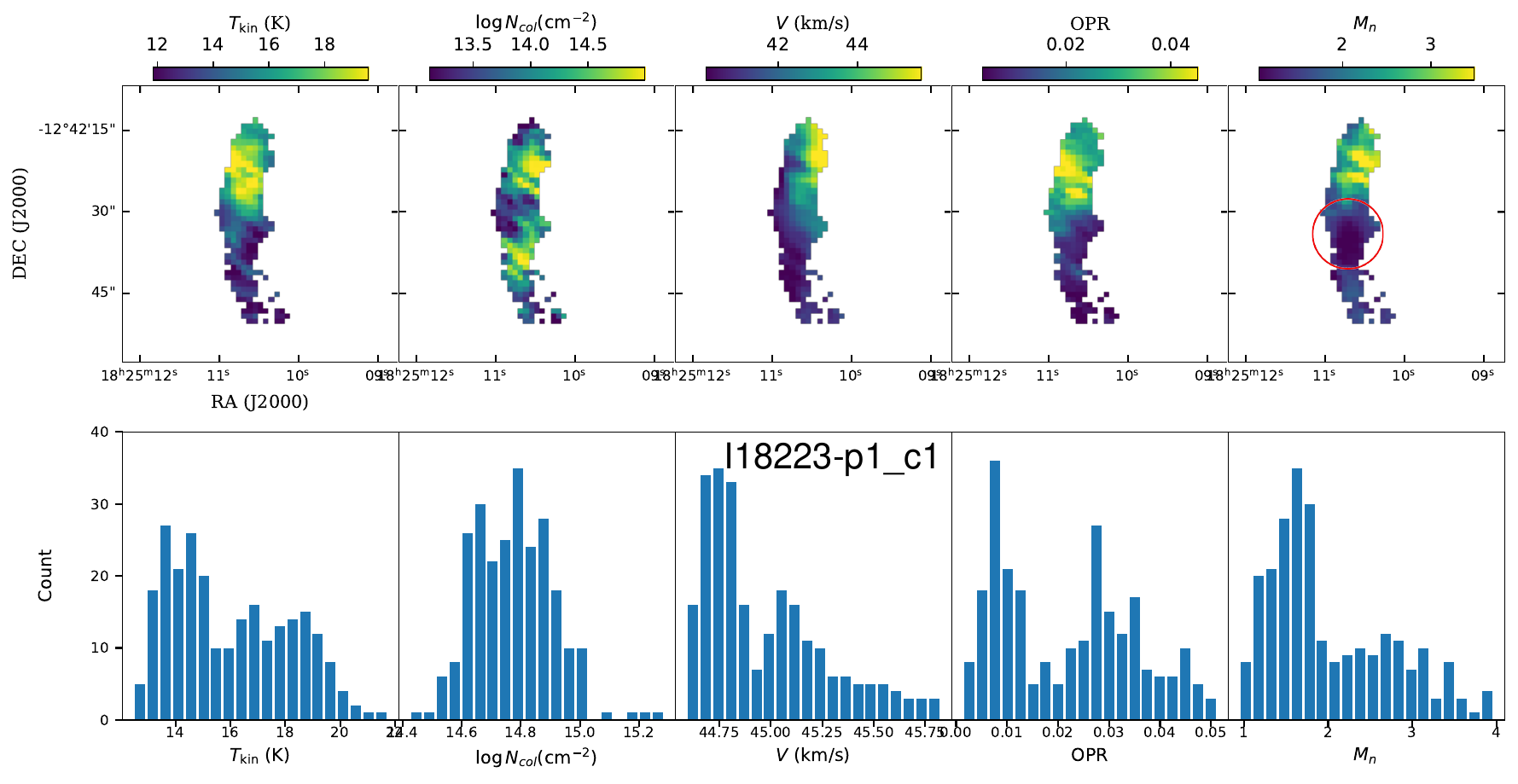}}
    \end{minipage}
    \caption{Maps and histograms of the fitted cores (continued).}
\end{figure*}
\addtocounter{figure}{-1}

\begin{figure*}[htb!]
    \begin{minipage}[c]{1.0\linewidth}
        \centerline{\includegraphics[width=\textwidth]{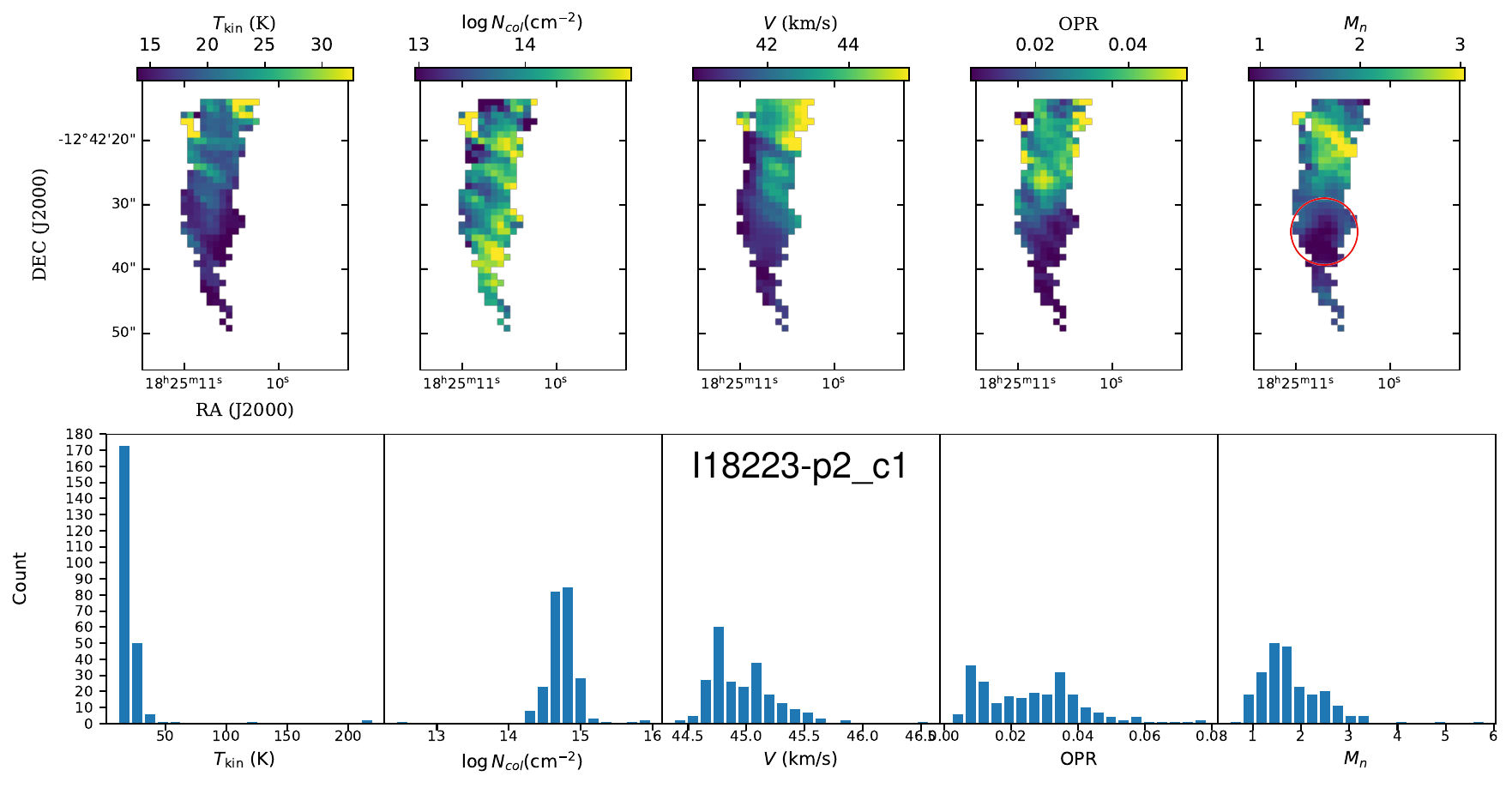}}
        \centerline{\includegraphics[width=\textwidth]{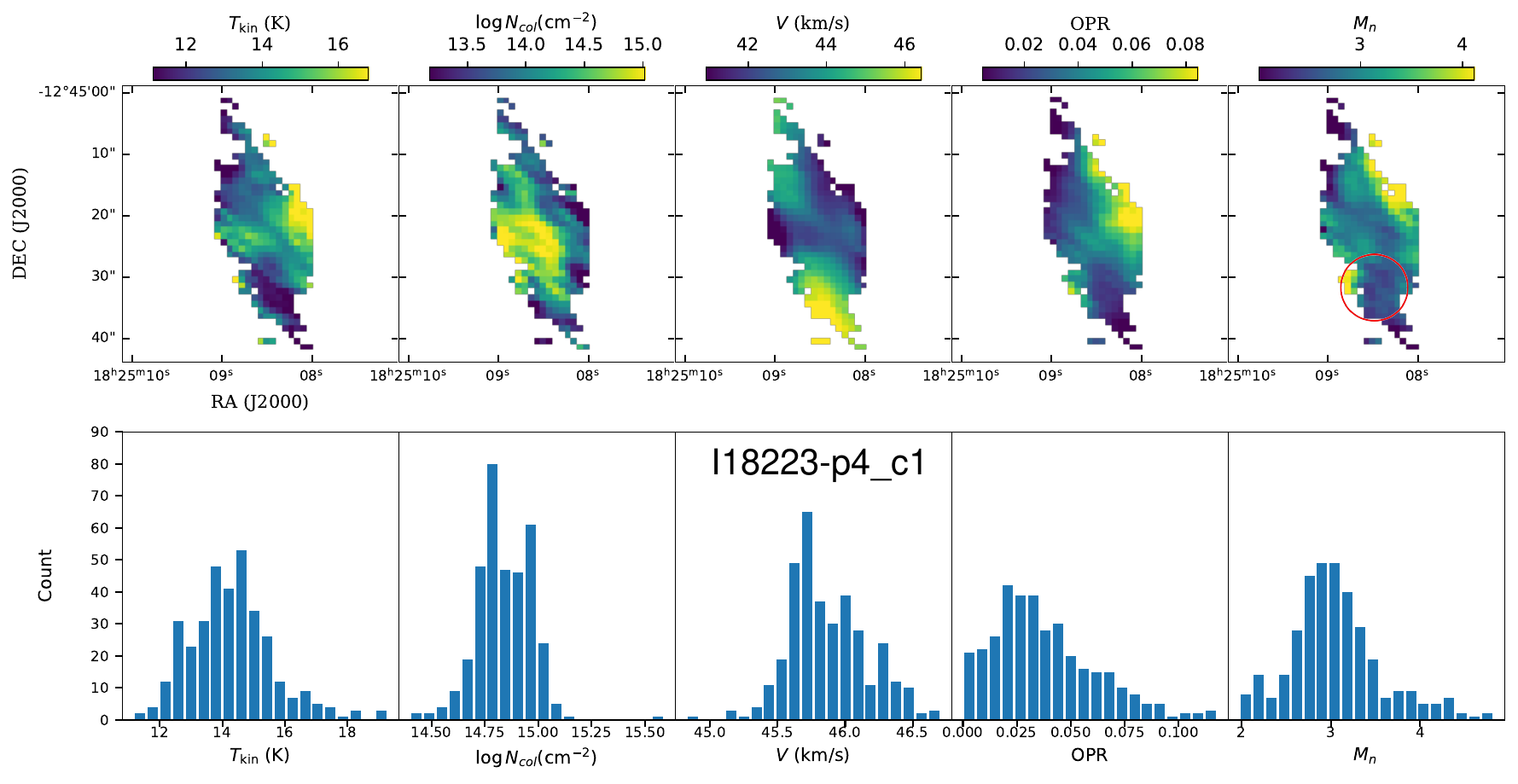}}
    \end{minipage}
    \caption{Maps and histograms of the fitted cores (continued).}
\end{figure*}
\addtocounter{figure}{-1}

\begin{figure*}[htb!]
    \begin{minipage}[c]{1.0\linewidth}
        \centerline{\includegraphics[width=\textwidth]{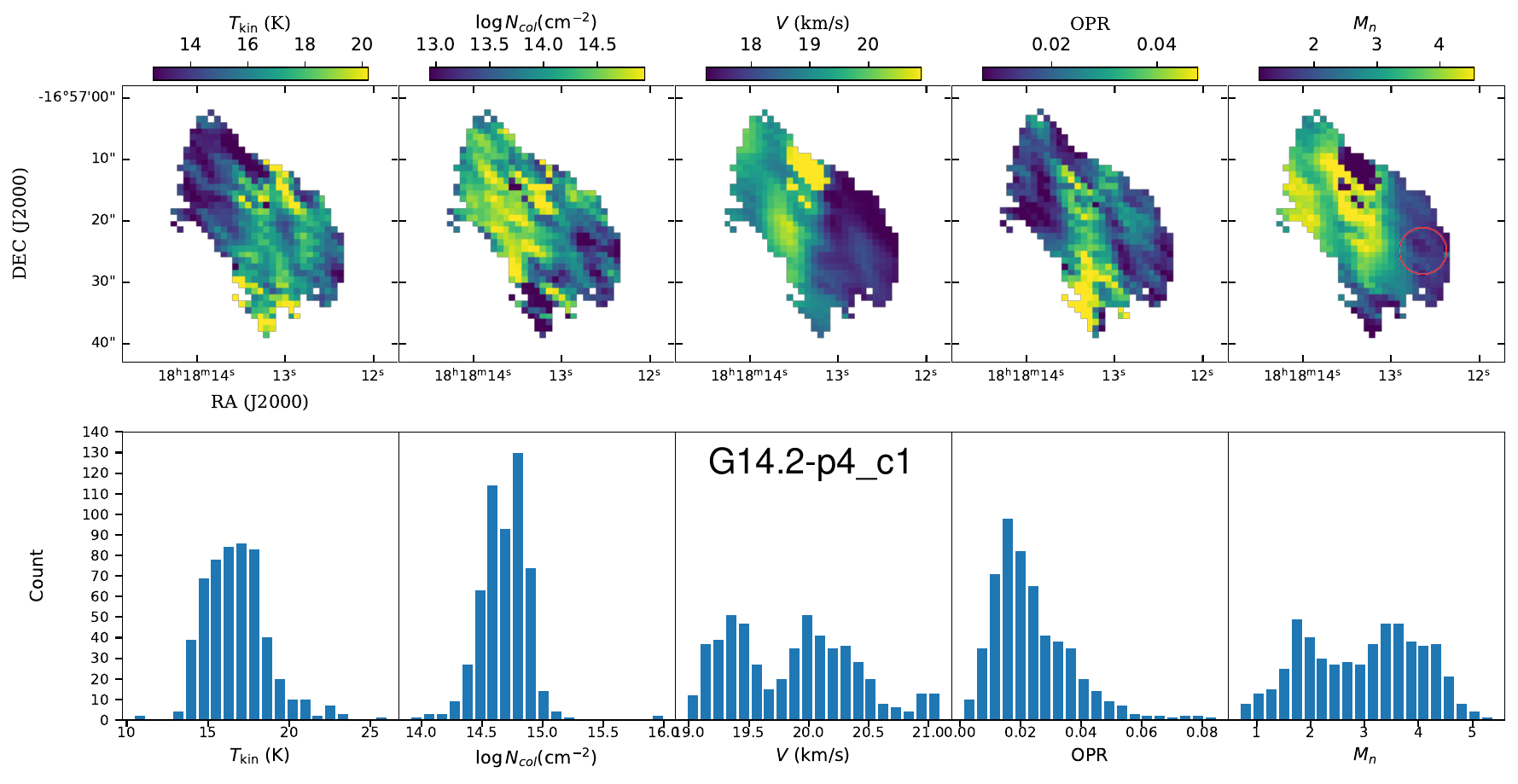}}
    \end{minipage}
    \caption{Maps and histograms of the fitted cores (continued)}
\end{figure*}

\begin{figure*}[htb!]
    \centering
    \includegraphics[width=0.7\textwidth]{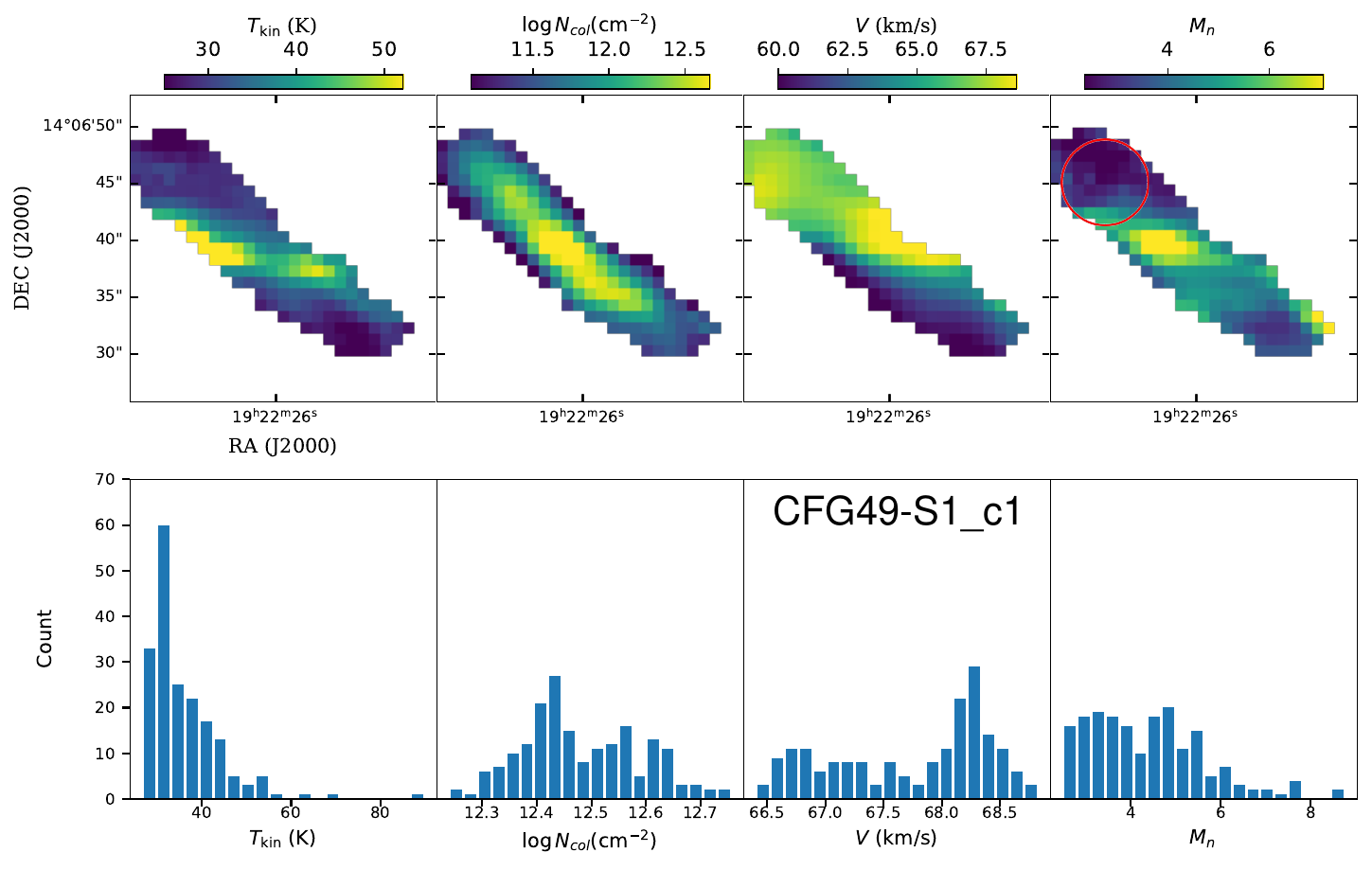}
    \includegraphics[width=0.7\textwidth]{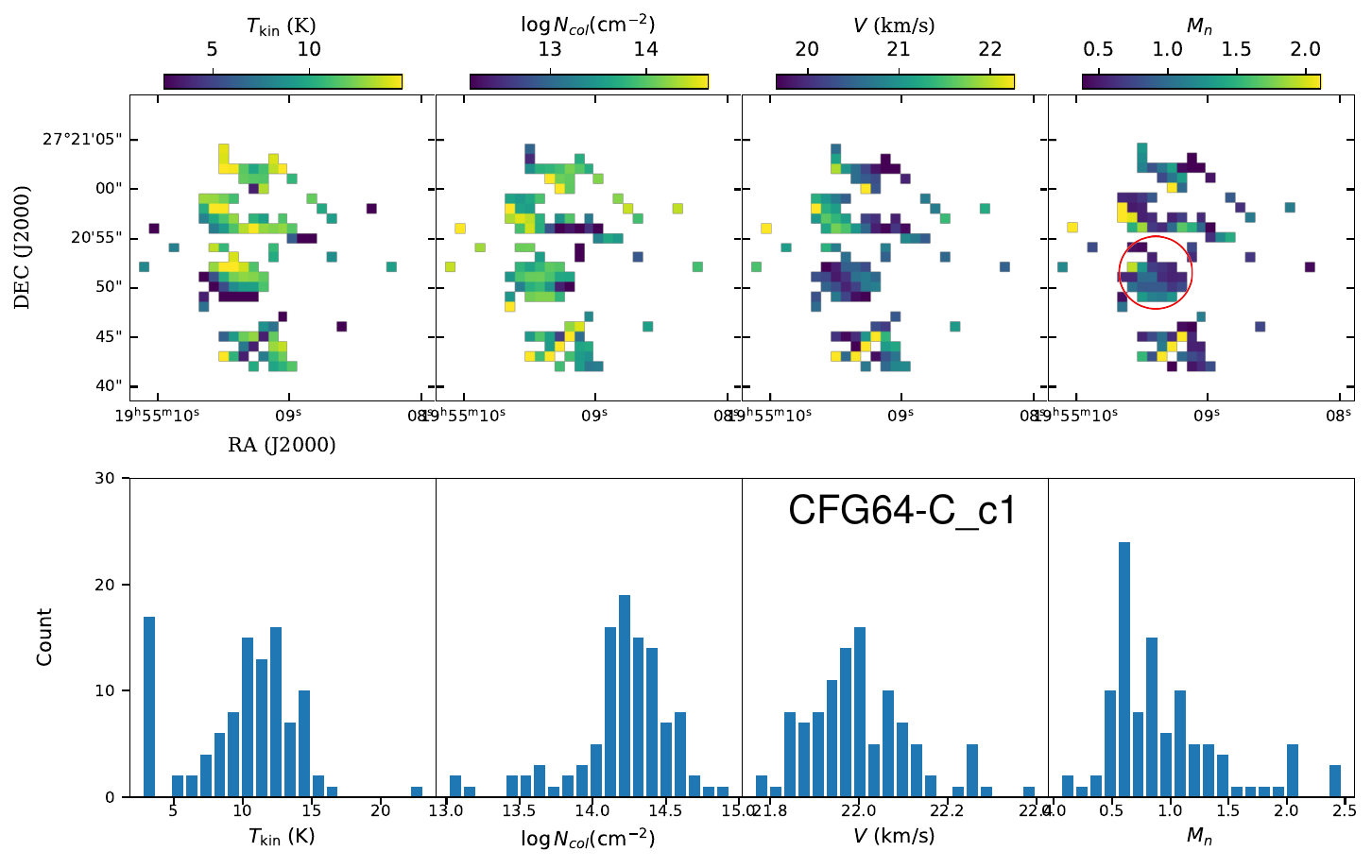}
    \includegraphics[width=0.7\textwidth]{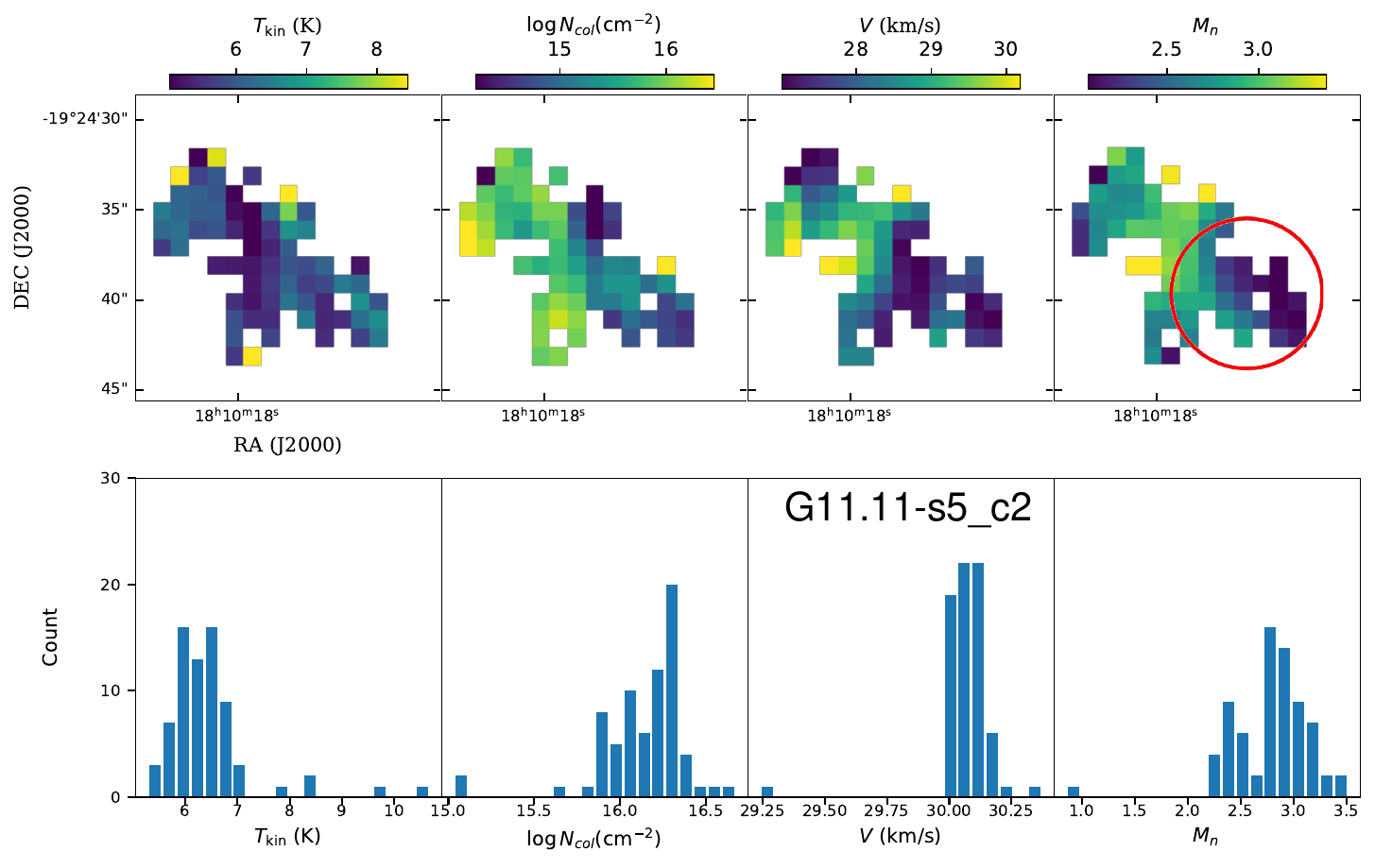}
    \caption{Maps and histograms of the fitted cores without the detection of the NH$_3$ (3,3). \textbf{First row:} Temperature (K), column density (in log scale), centroid velocity (km ${\rm{s}^{-1}}$), and Mach number (from left to right). The red circle is the sub-region we selected to study the core-averaged Mach number. \textbf{Second row:} Corresponding histograms of the parameter distribution.}
    \label{restresult2}
\end{figure*}
\addtocounter{figure}{-1}

\begin{figure*}[htb!]
    \centering
    \includegraphics[width=0.7\textwidth]{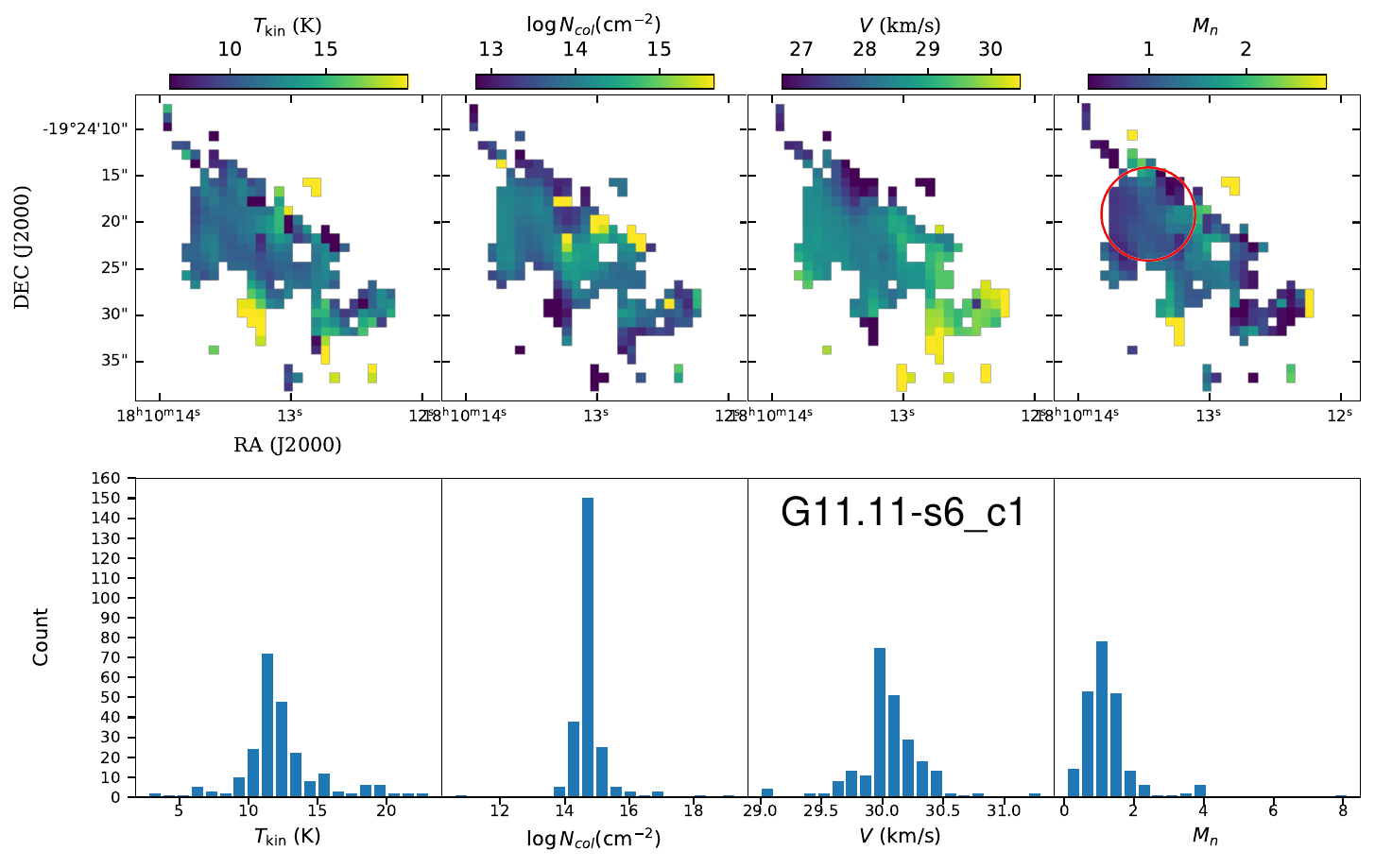}
    \includegraphics[width=0.7\textwidth]{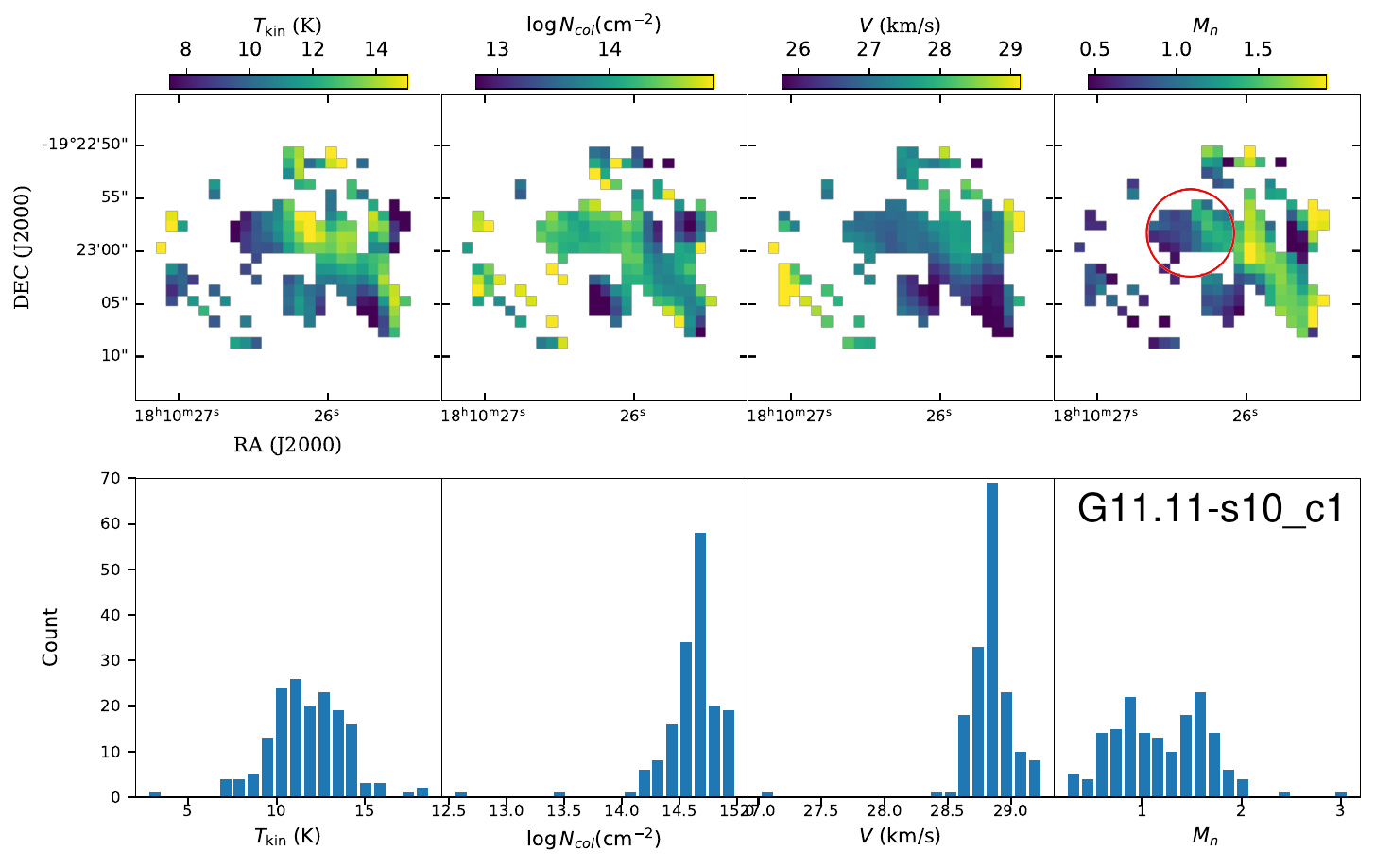}
    \includegraphics[width=0.7\textwidth]{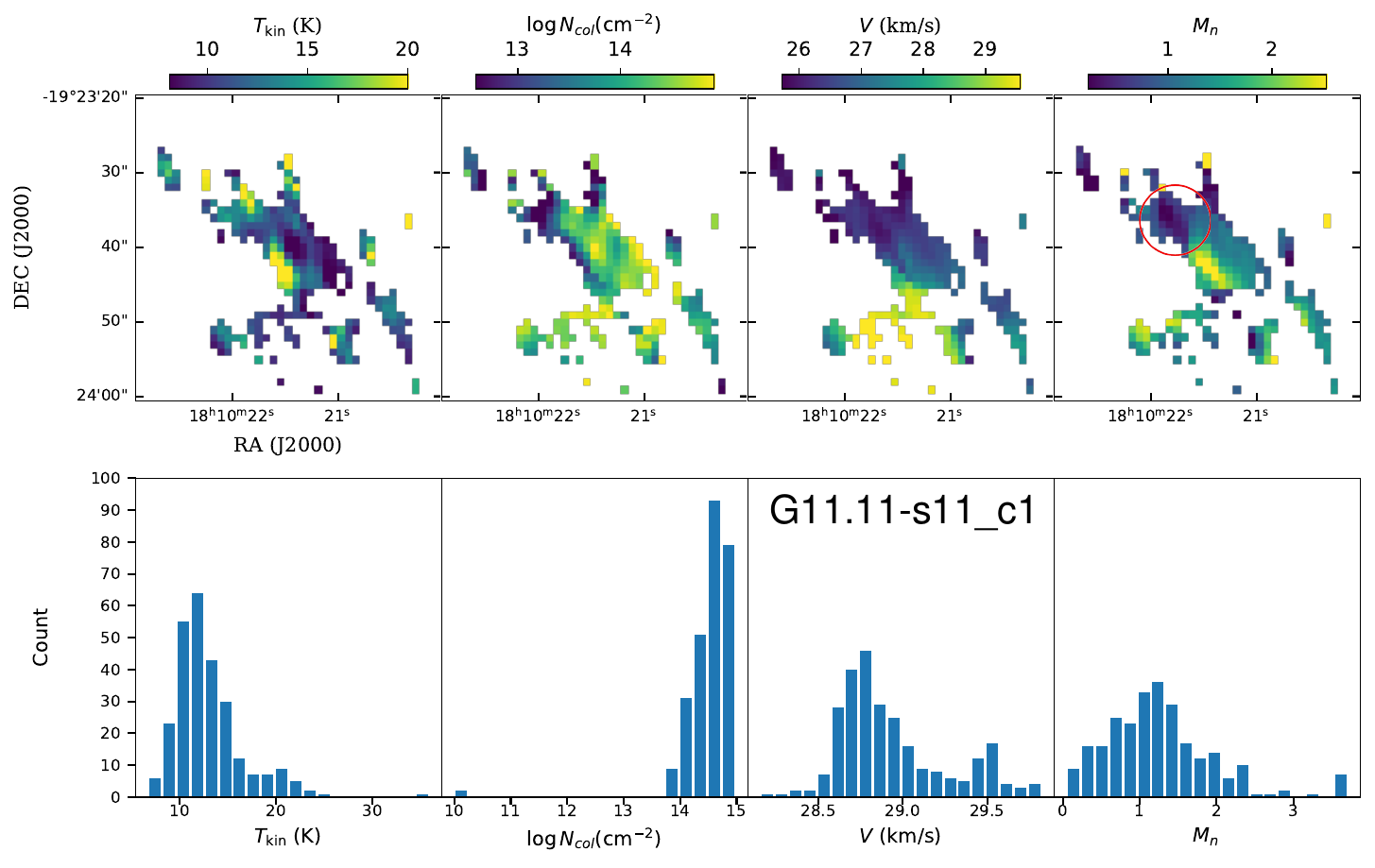}
    \caption{Maps and histograms of the fitted cores without the detection of the NH$_3$ (3,3) (continued)}
\end{figure*}
\addtocounter{figure}{-1}

\begin{figure*}[htb!]
    \centering
    \includegraphics[width=0.7\textwidth]{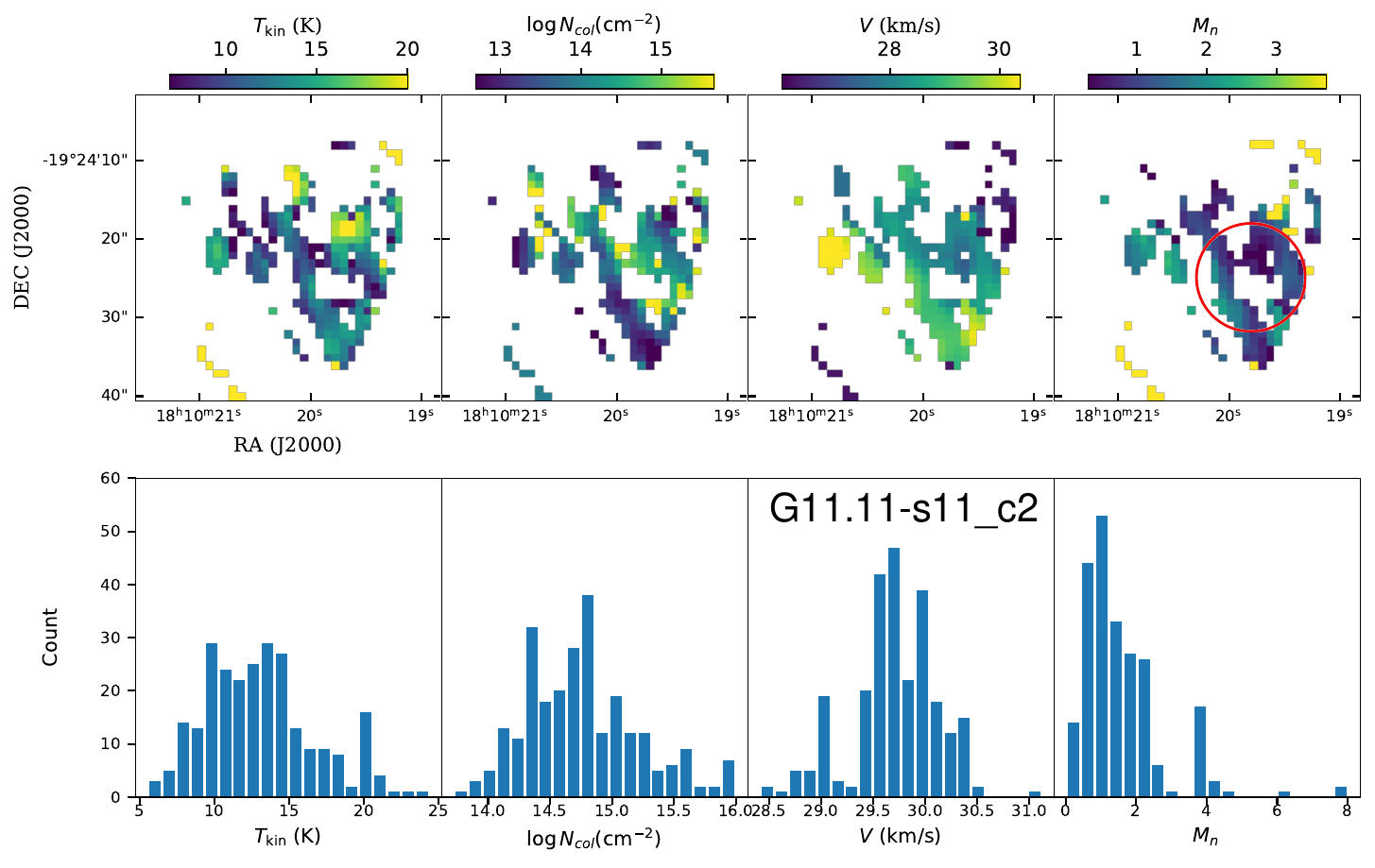}
    \includegraphics[width=0.7\textwidth]{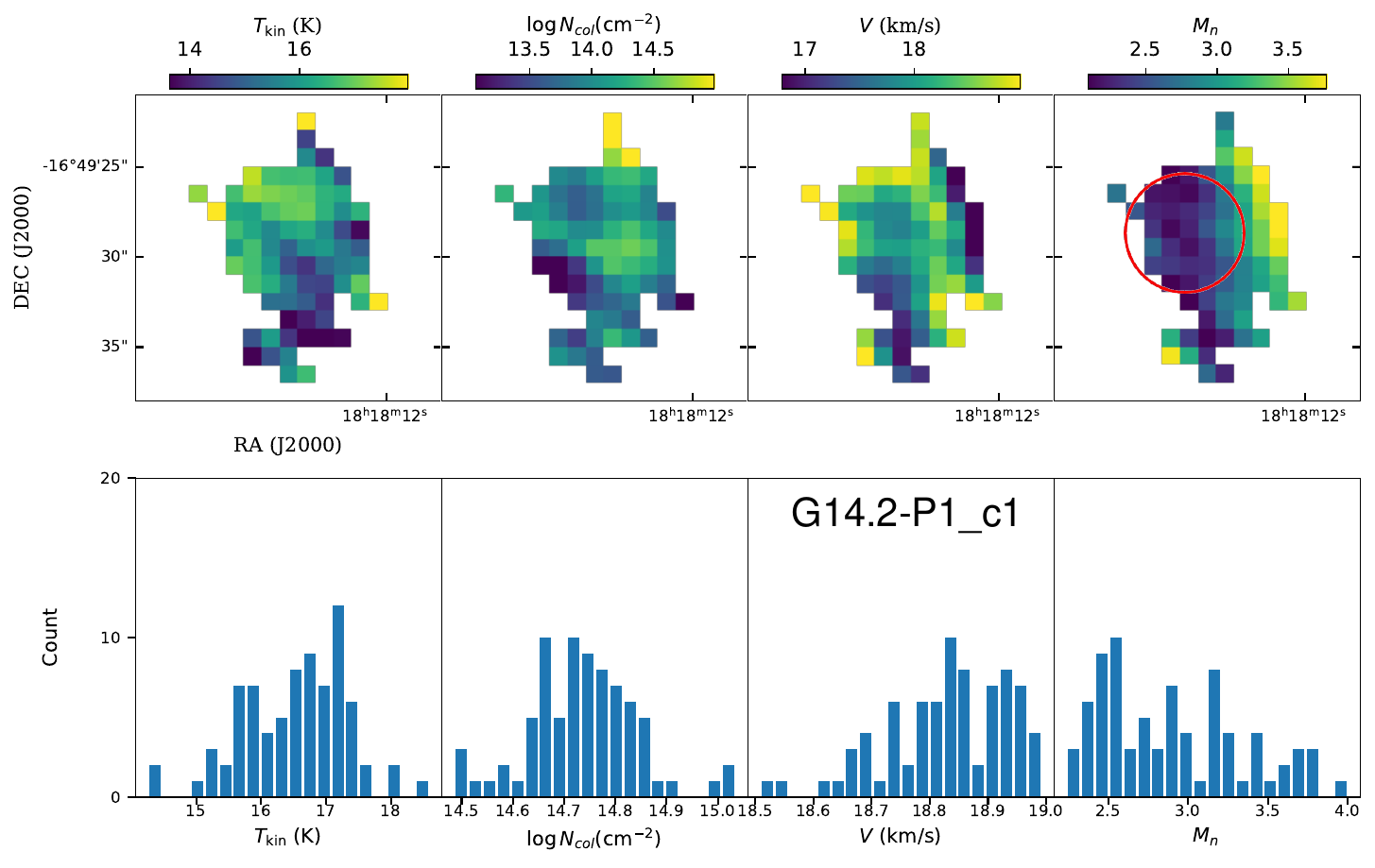}
    \includegraphics[width=0.7\textwidth]{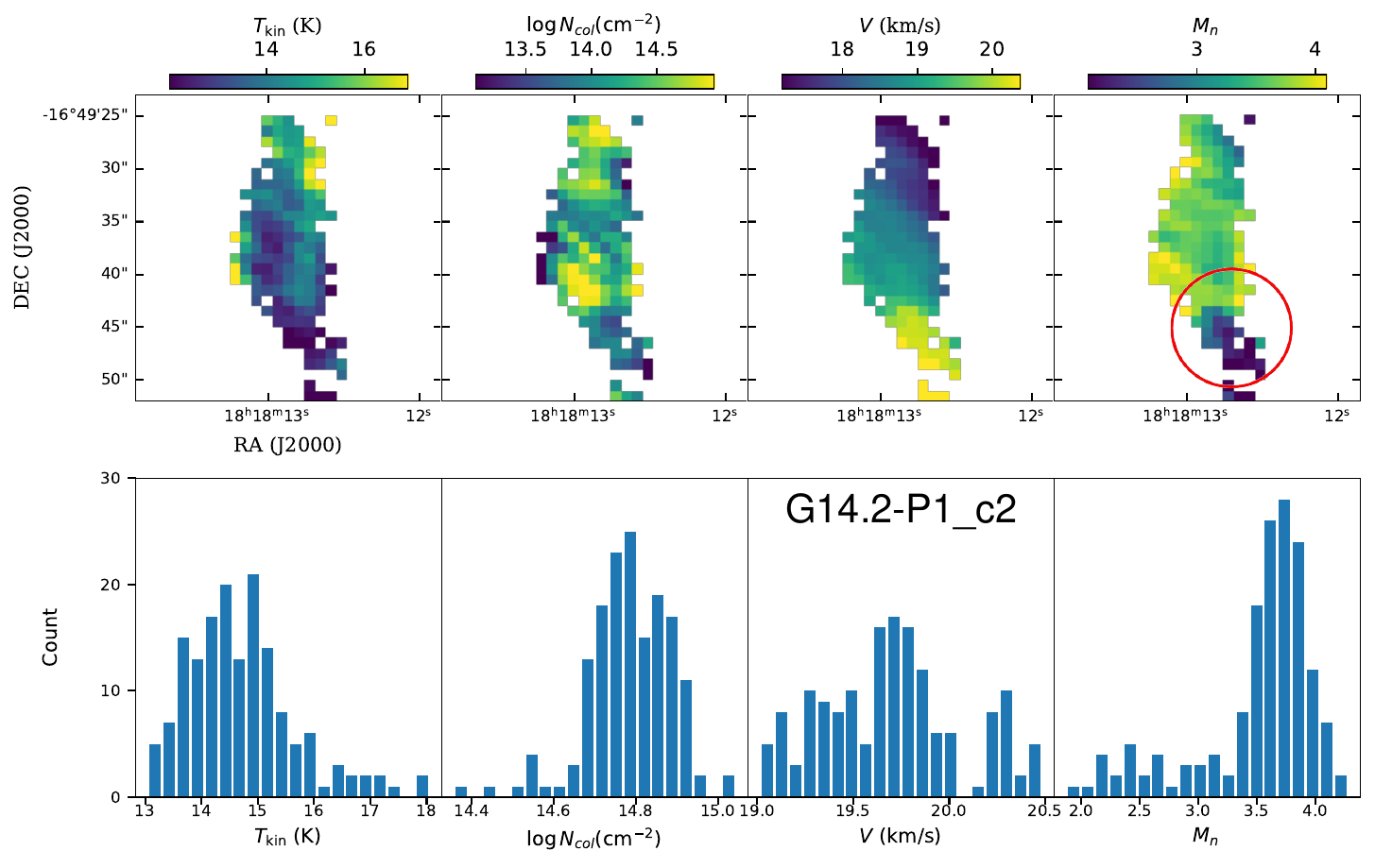}
    \caption{Maps and histograms of the fitted cores without the detection of the NH$_3$ (3,3) (continued)}
\end{figure*}
\addtocounter{figure}{-1}

\begin{figure*}[htb!]
    \centering
    \includegraphics[width=0.7\textwidth]{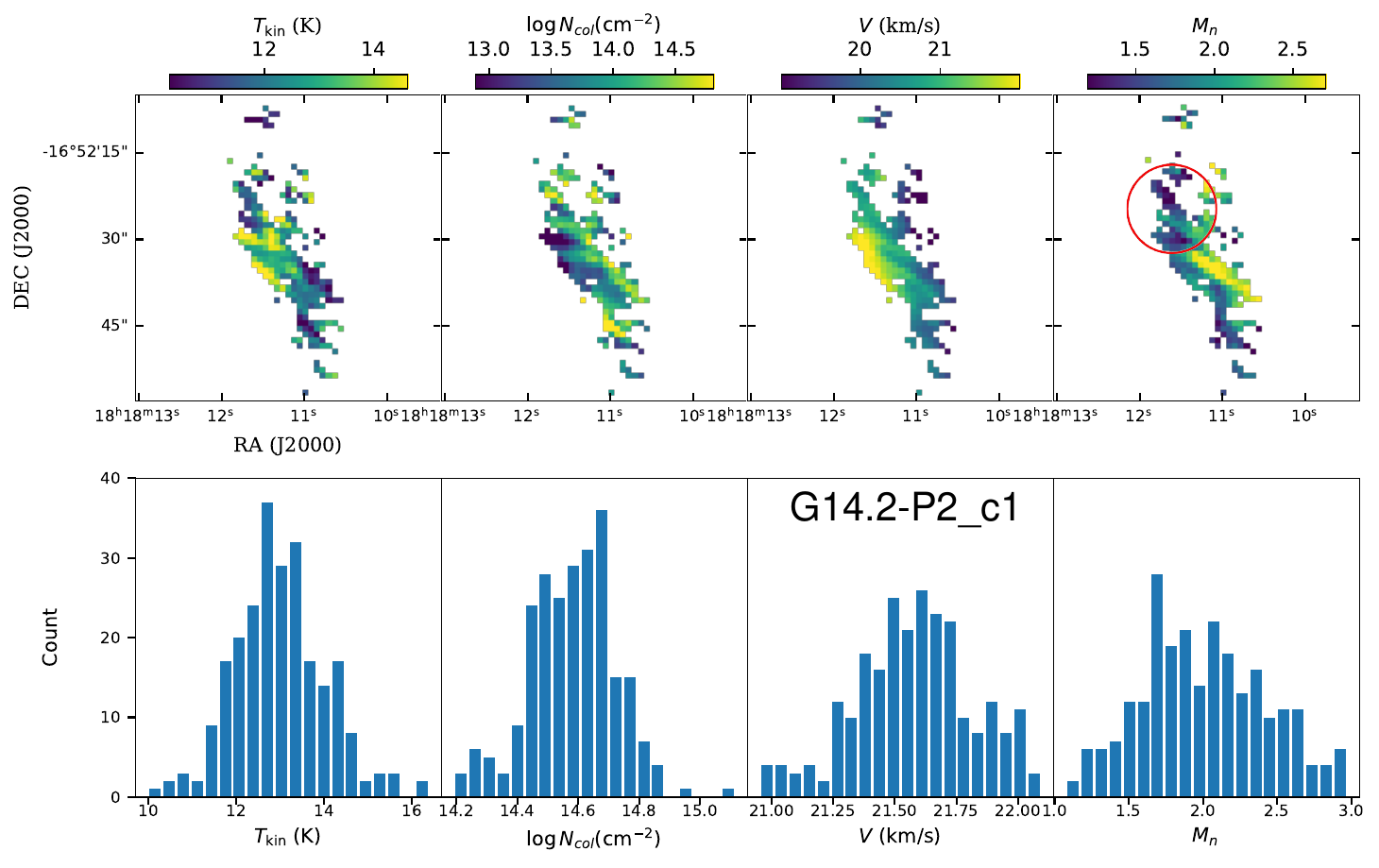}
    \includegraphics[width=0.7\textwidth]{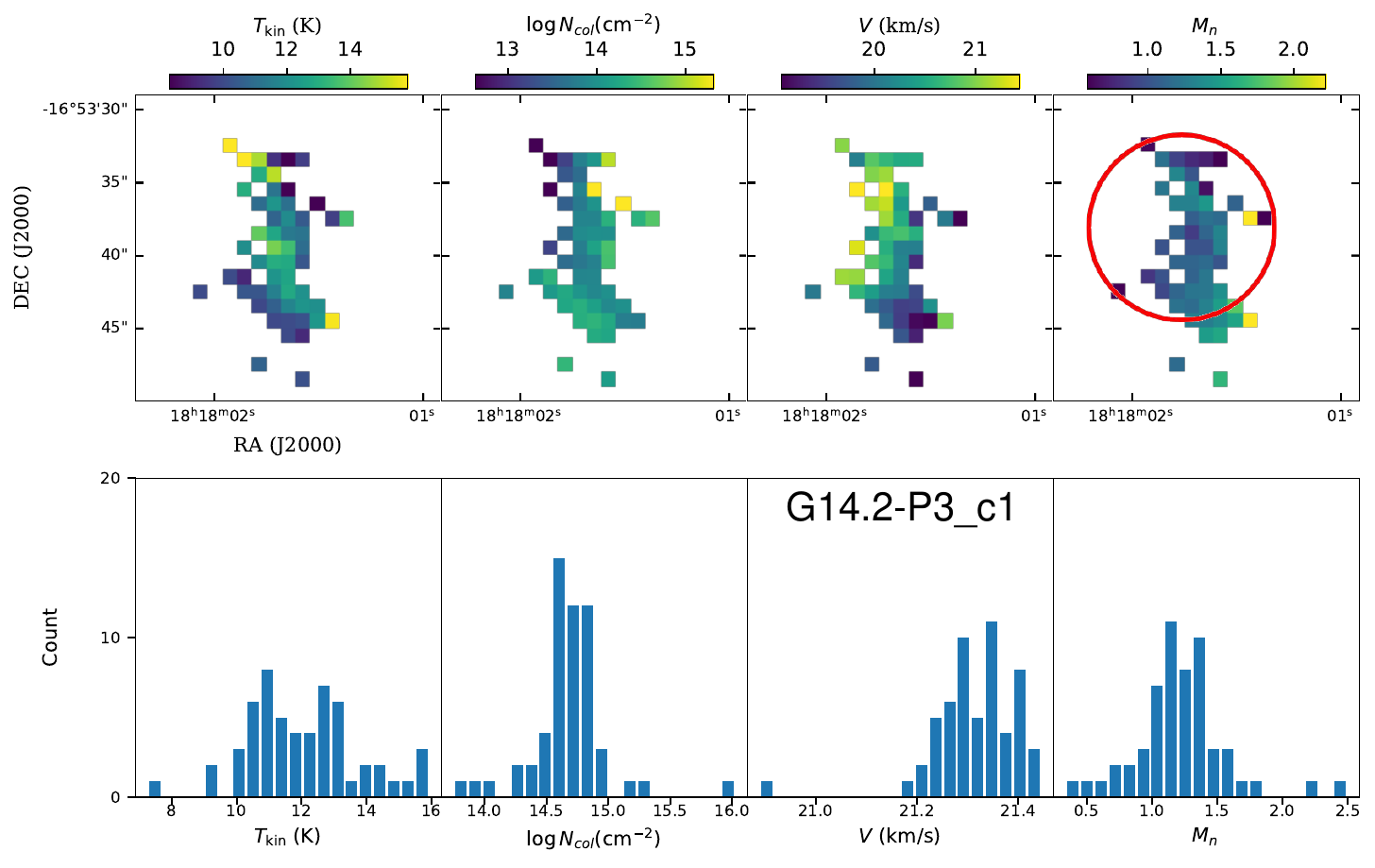}
    \includegraphics[width=0.7\textwidth]{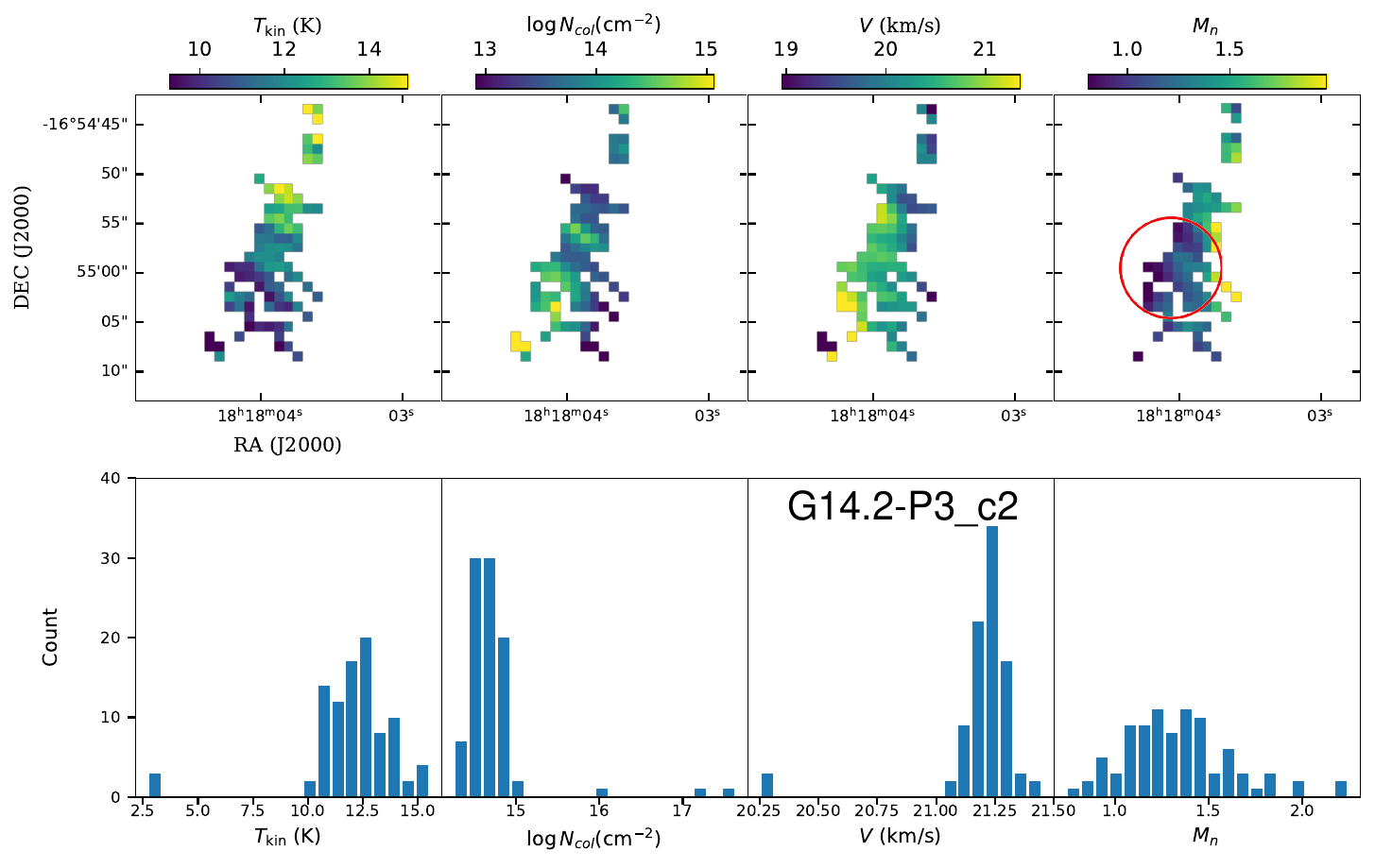}
    \caption{Maps and histograms of the fitted cores without the detection of the NH$_3$ (3,3) (continued)}
\end{figure*}
\addtocounter{figure}{-1}

\begin{figure*}[htb!]
    \centering
    \includegraphics[width=0.7\textwidth]{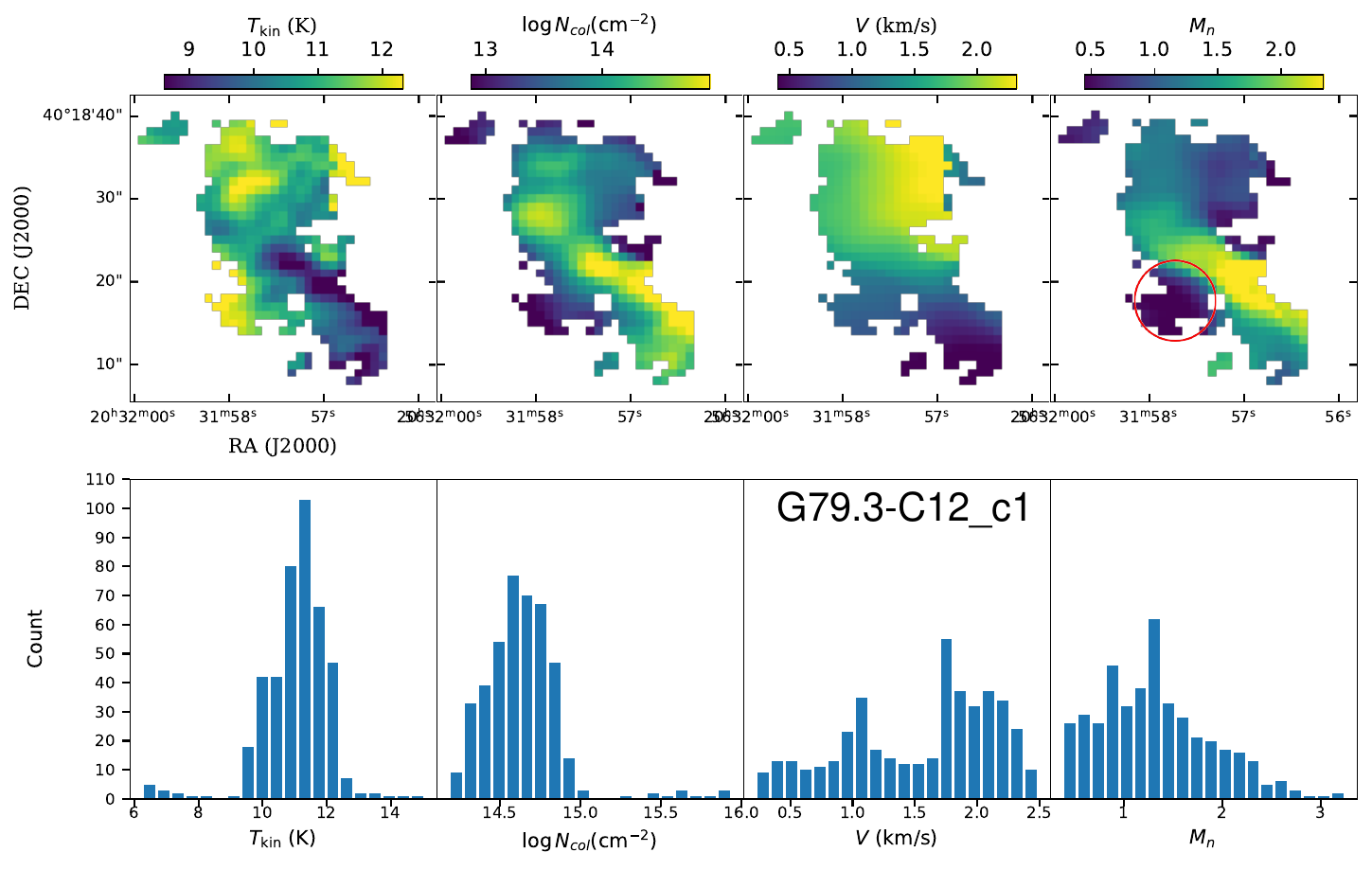}
    \includegraphics[width=0.7\textwidth]{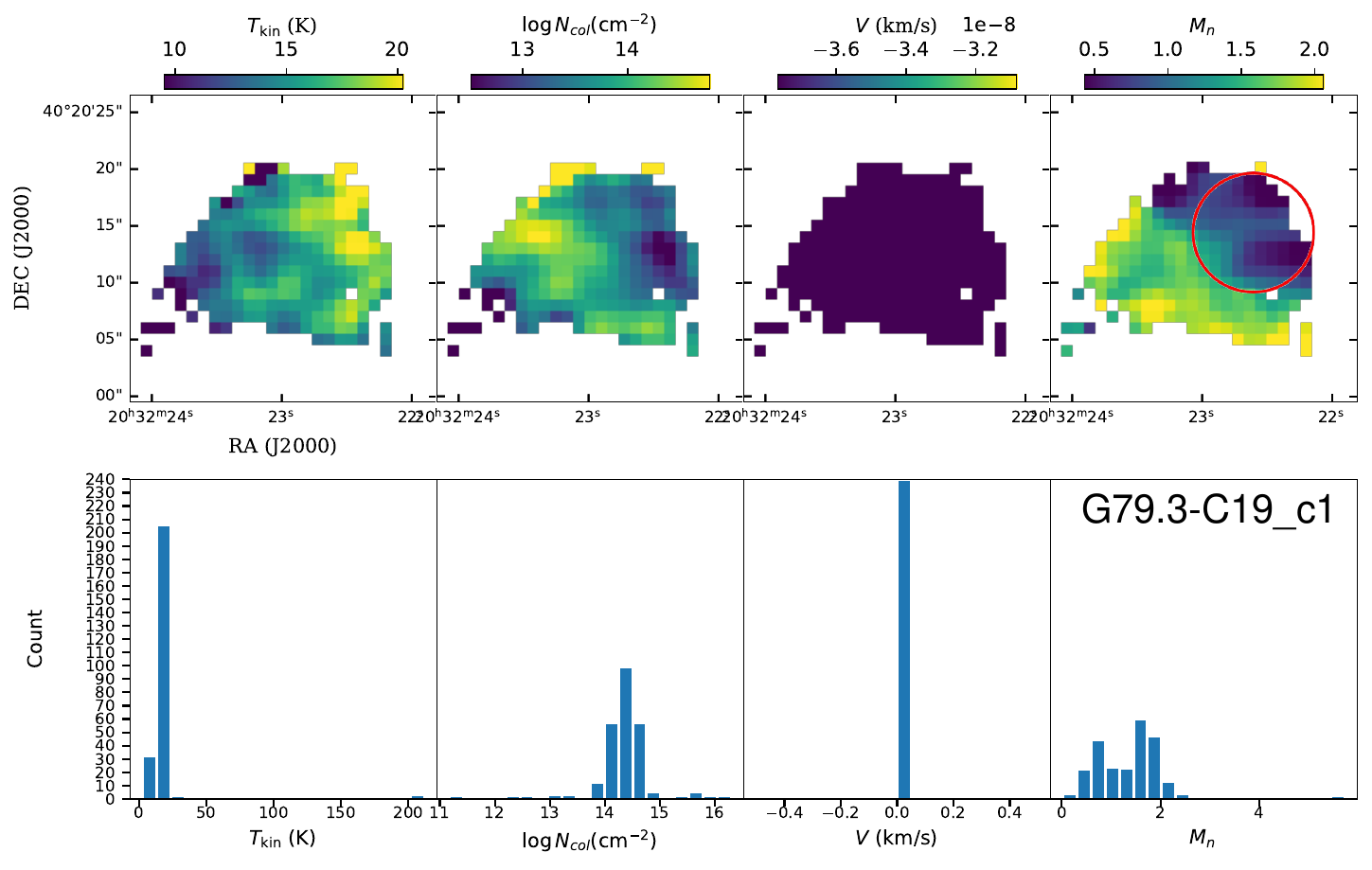}
    \includegraphics[width=0.7\textwidth]{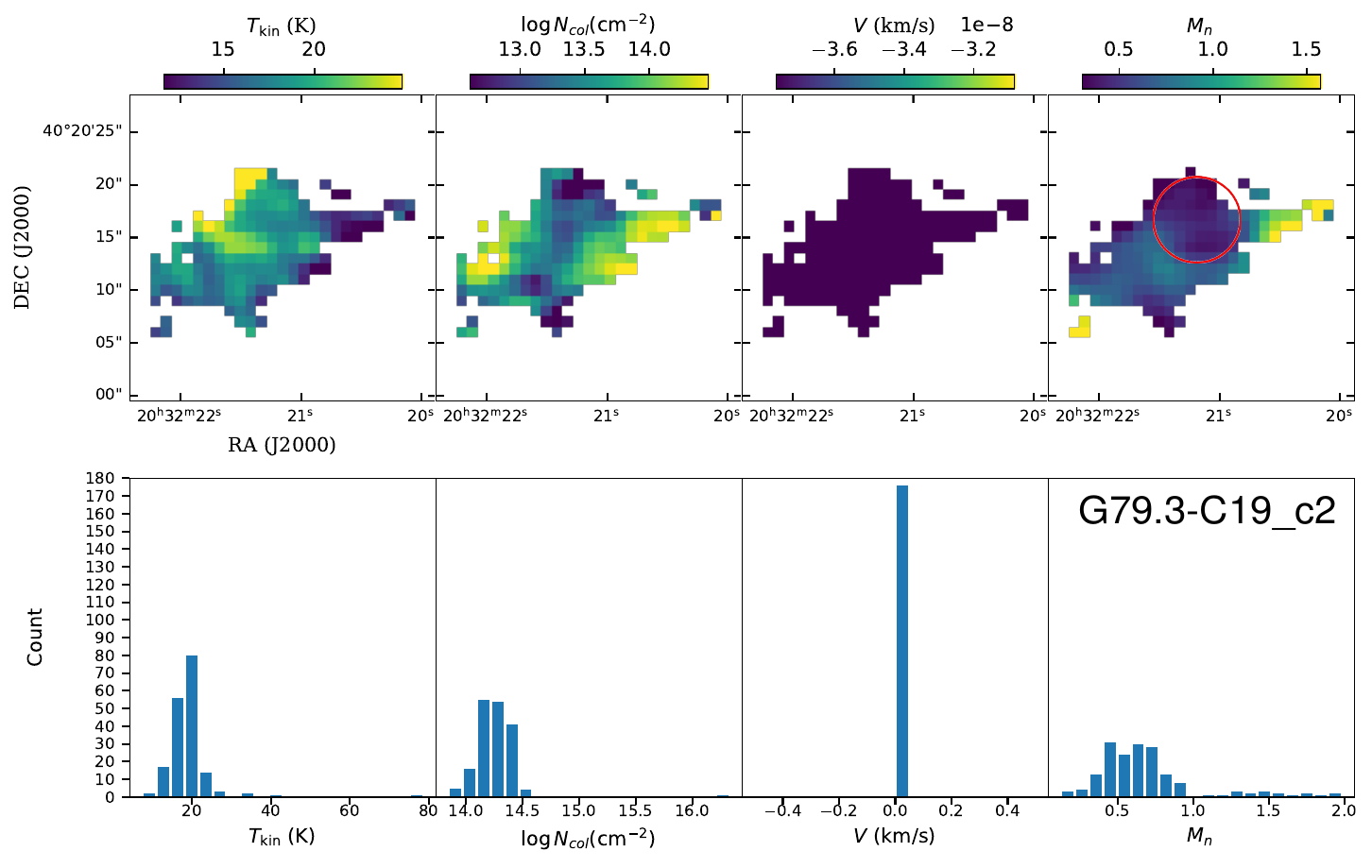}
    \caption{Maps and histograms of the fitted cores without the detection of the NH$_3$ (3,3) (continued).}
\end{figure*}
\addtocounter{figure}{-1}

\begin{figure*}[htb!]
    \centering
    \includegraphics[width=0.7\textwidth]{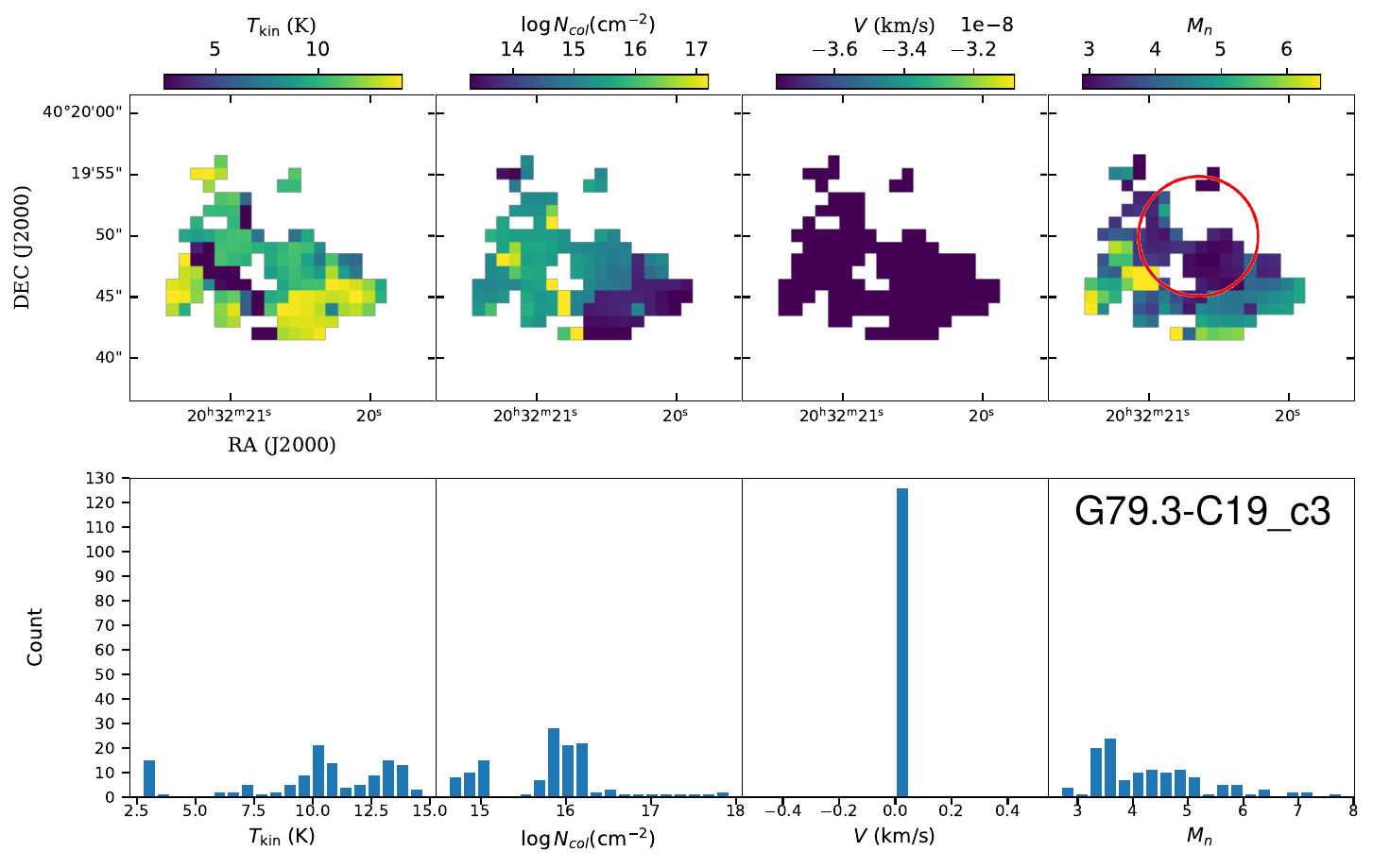}
    \includegraphics[width=0.7\textwidth]{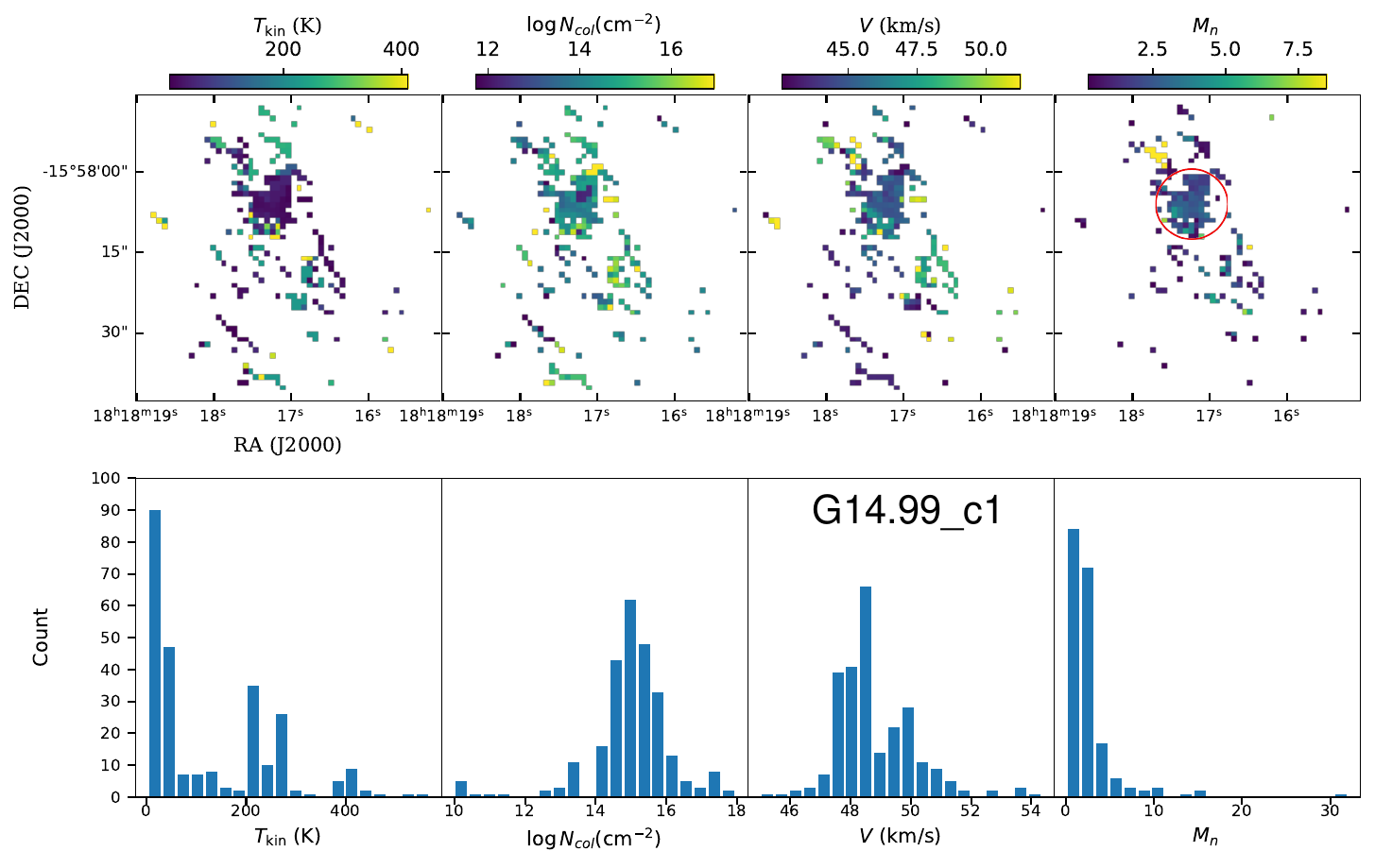}
    \includegraphics[width=0.7\textwidth]{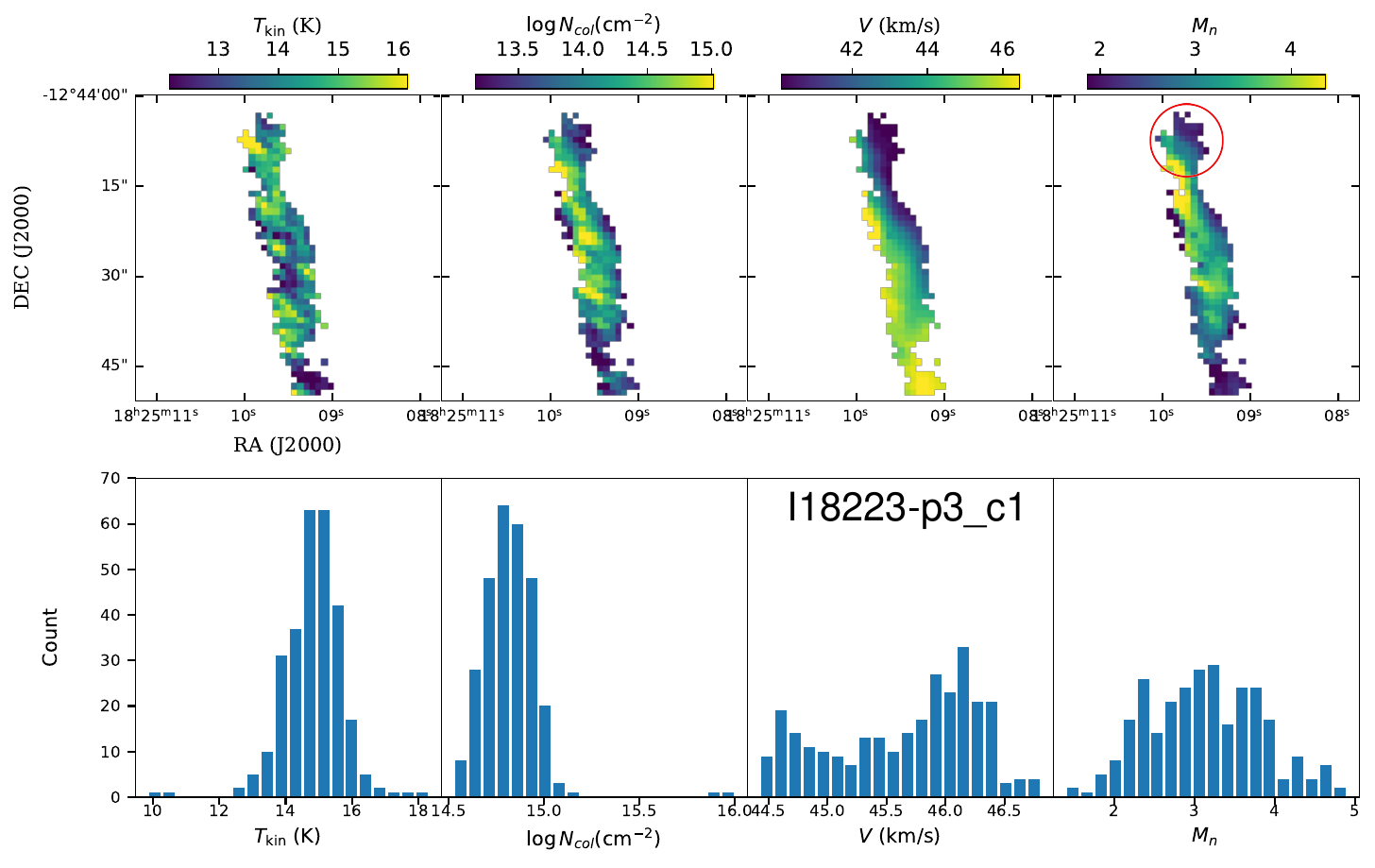}
    \caption{Maps and histograms of the fitted cores without the detection of the NH$_3$ (3,3) (continued).}
\end{figure*}
\addtocounter{figure}{-1}

\begin{figure*}[htb!]
    \centering
    \includegraphics[width=0.7\textwidth]{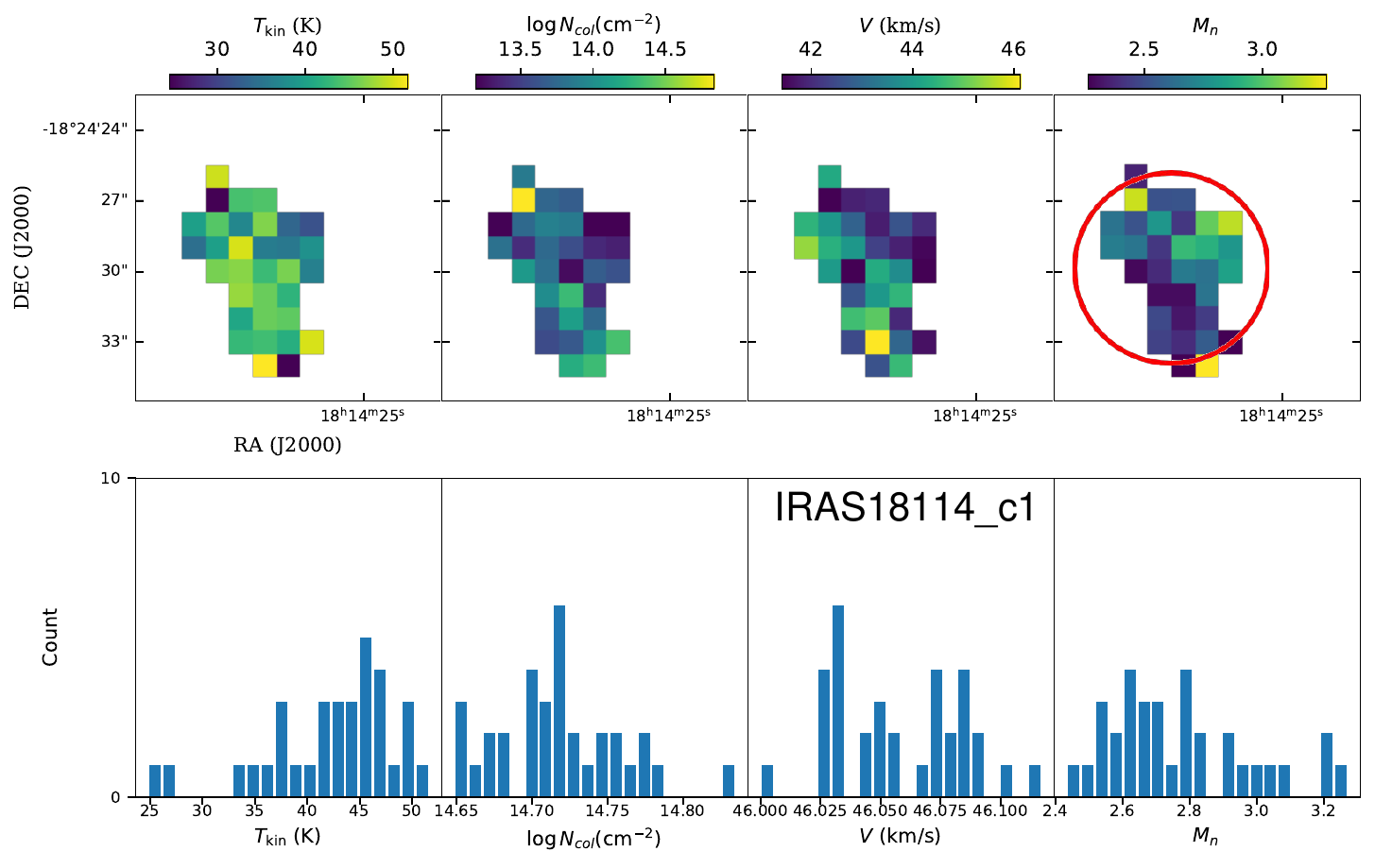}
    \includegraphics[width=0.7\textwidth]{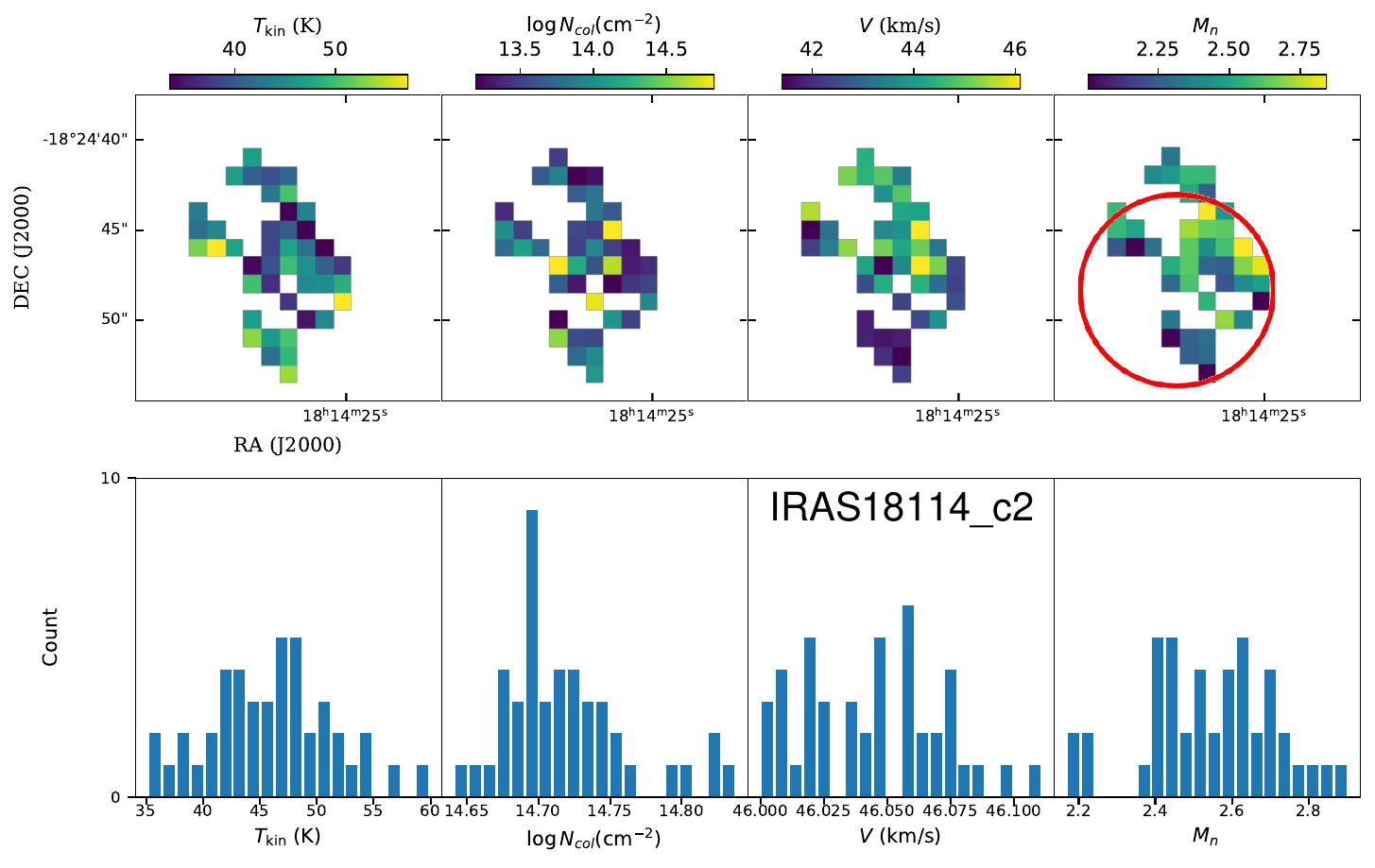}
    \includegraphics[width=0.7\textwidth]{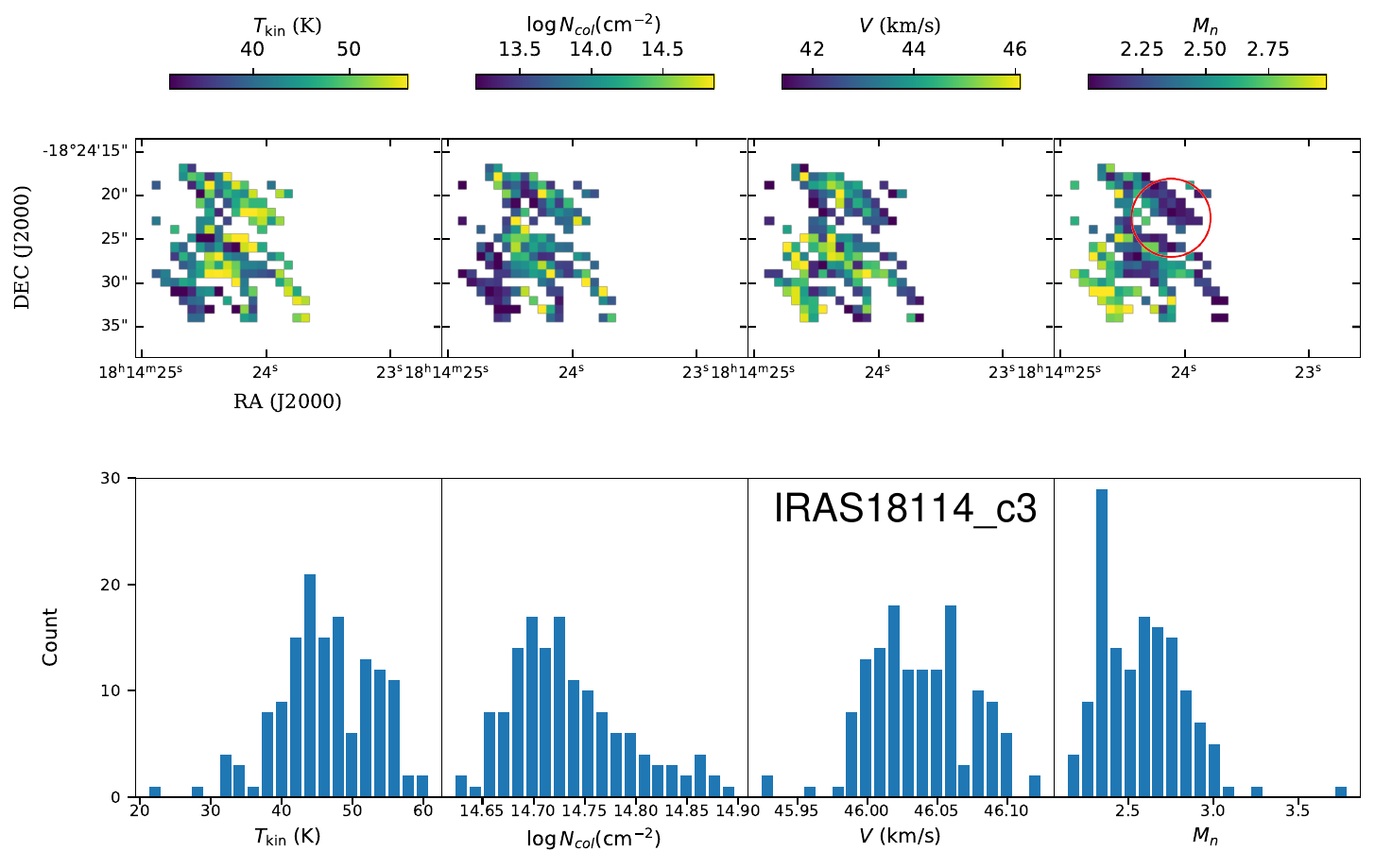}
    \caption{Maps and histograms of the fitted cores without the detection of the NH$_3$ (3,3) (continued).}
\end{figure*}
\addtocounter{figure}{-1}

\begin{figure*}[htb!]
    \centering
    \includegraphics[width=0.7\textwidth]{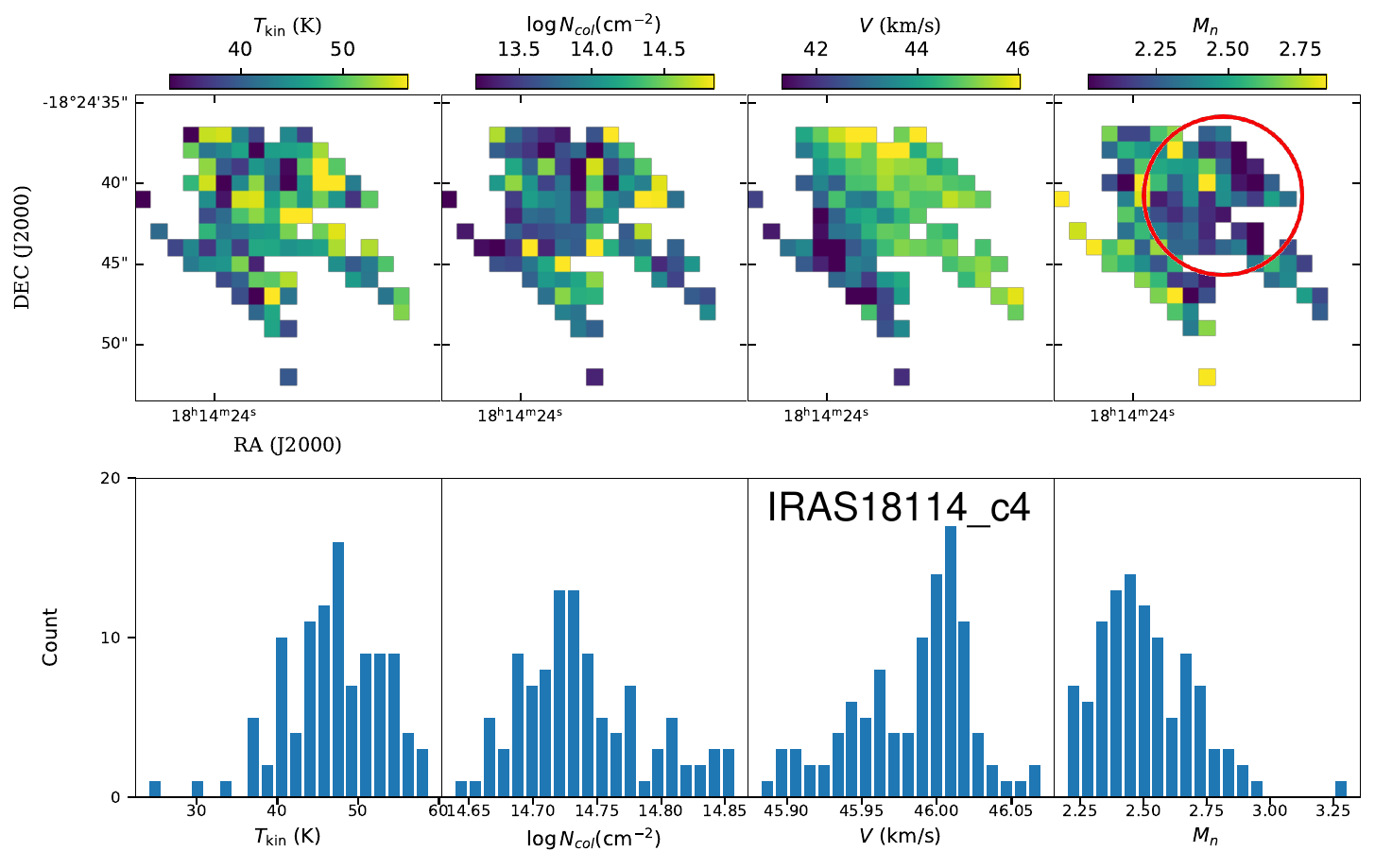}
    \caption{Maps and histograms of the fitted cores without the detection of the NH$_3$ (3,3) (continued).}
\end{figure*}
\addtocounter{figure}{-1}

\section{G15.19}
\begin{figure*}[htb!]
    \centering
    \includegraphics[width=0.4\textwidth]{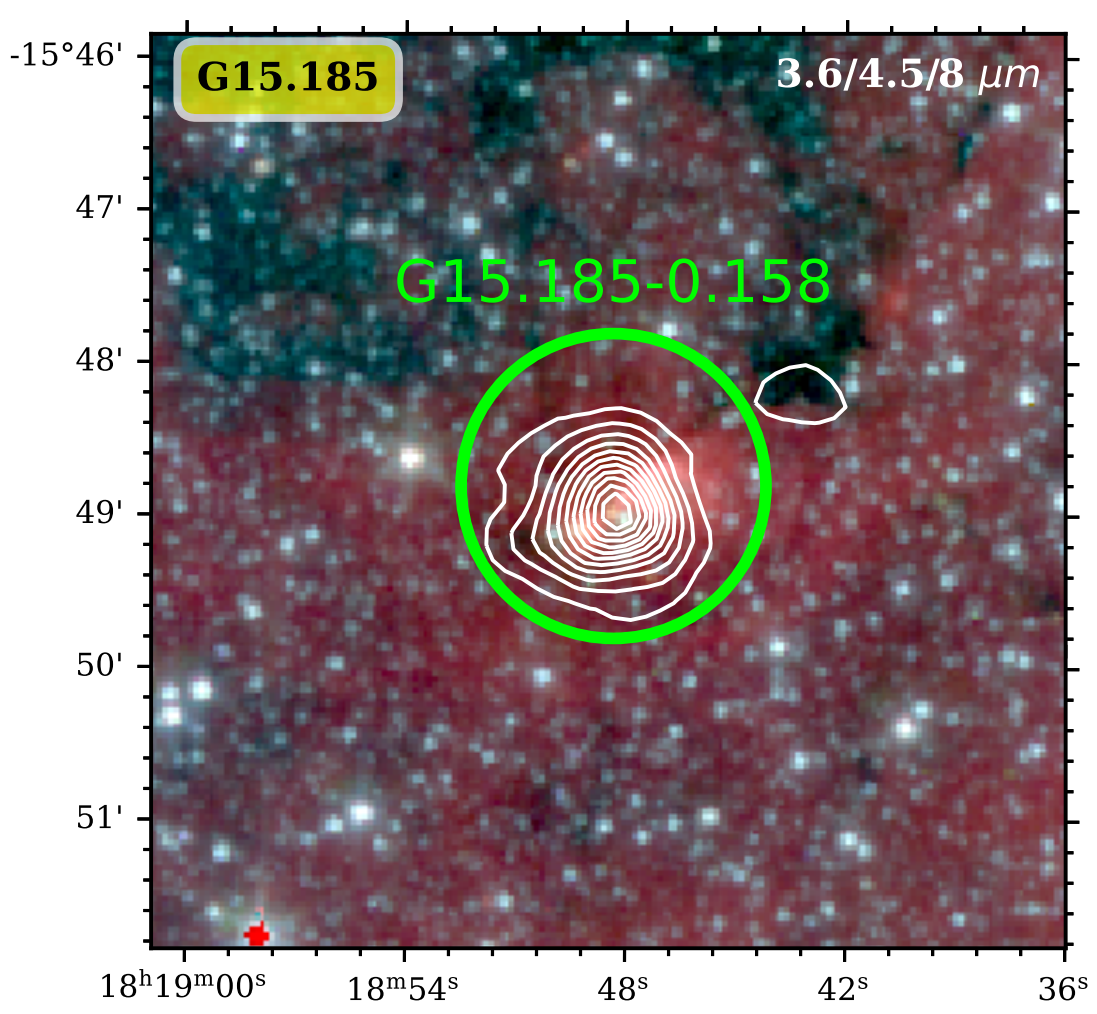}
    \caption{Infrared environment G15.19. Details are the same as in Fig. \ref{ob}.}
    \label{rgb:G15.19}
\end{figure*}

\begin{figure*}[htb!]
    \centering
    \includegraphics[width=0.7\textwidth]{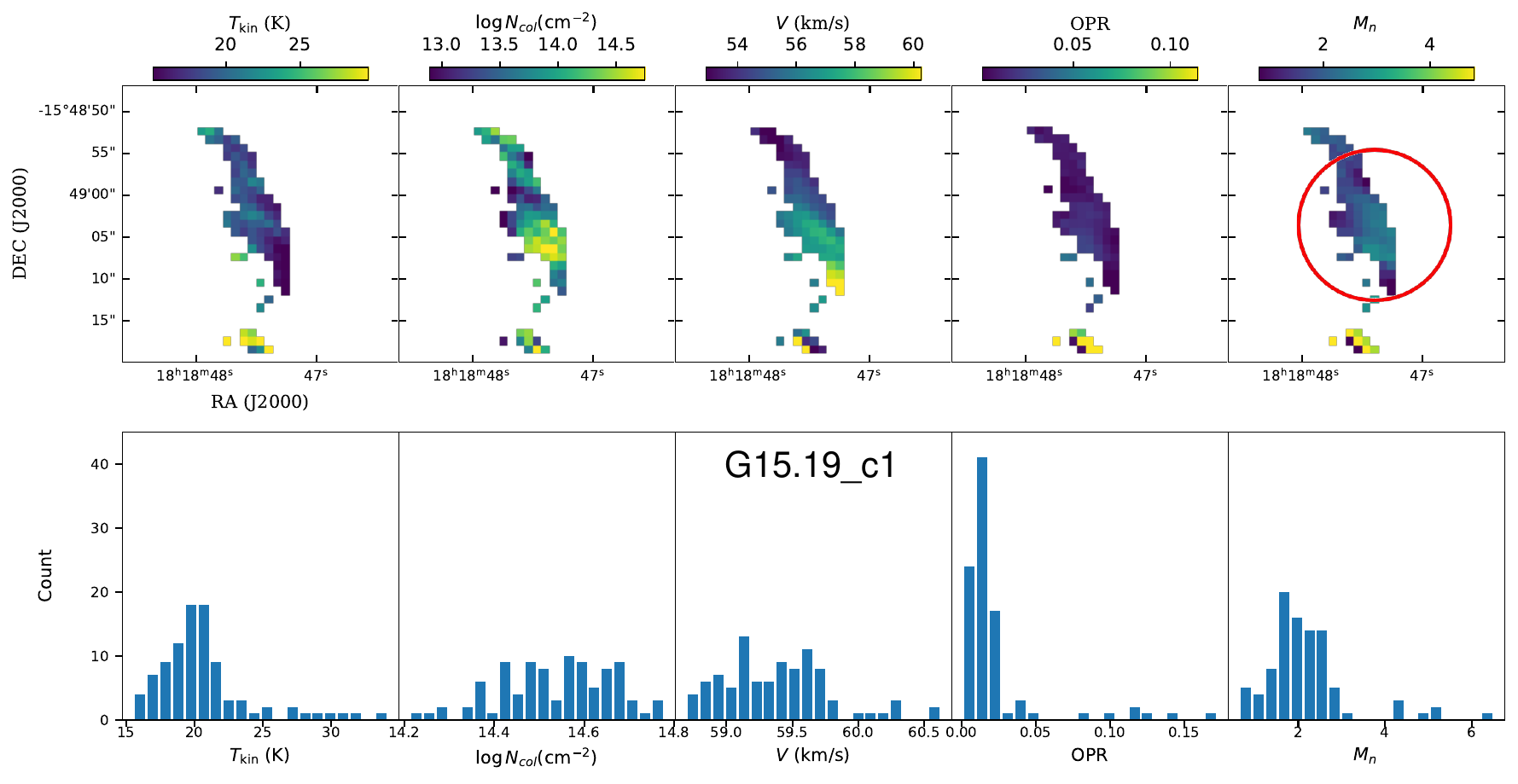}
    \caption{Maps and histograms of G15.19.}
    \label{fit:G15.19}
\end{figure*}

Beside the 13 sources reported in the main text, we have also observed G15.185-0.158 (G15.19 in short). G15.19 (18h18m48.2s, -15d48m36.0s) is located at 11.6 kpc and has 7000 $M_{\odot}$ within a radius of 16.6 pc \citep{2013A&A...549A..45C}. 
Due to its much larger distance than other sources, we have excluded analysis of G15.19, but kept its basic parameters and observing setup in Table 1 and 2, and present the fitting results here (Fig. \ref{rgb:G15.19}, \ref{fit:G15.19}).
As an IR-bright source, G15.19 is slightly warm (20.7 K) and relatively thin ($10^{14.5}$ cm$^{-2}$). The Mach number of G15.19 is 1.1 which means the turbulence is transonic. 

\end{appendix}








   
  



\end{document}